\documentclass[prd,preprintnumbers,showpacs,nofootinbib,superscriptaddress,notitlepage,floatfix]{revtex4-1}

\usepackage{amsmath}
\usepackage{amsfonts}
\usepackage{amssymb}
\usepackage{physics}
\usepackage{slashed}
\usepackage{simplewick}

\usepackage{graphicx}
\usepackage{float}
\usepackage[caption=false]{subfig}

\usepackage{tikz}
\usetikzlibrary{calc}
\usetikzlibrary{patterns}
\usetikzlibrary{positioning,decorations.pathmorphing,decorations.markings}
\usetikzlibrary{shapes,arrows}

\usepackage{feynmp-auto}
\usepackage{multirow}

\usepackage{enumitem}

\usepackage{hyperref}

\usepackage{color}
\usepackage{comment}
\def\str#1{{\it str} \{ #1 \}}

\def\ket#1{| #1 \ra }
\def\bra#1{\la #1}

\def\la{\langle}
\def\ra{\rangle}

\def\l{\left}
\def\r{\right}

\def\beq{\begin{equation}}
\def\eeq{\end{equation}}
\def\bea{\begin{eqnarray}}
\def\eea{\end{eqnarray}}
\def\overliner{\begin{array}}
\def\earr{\end{array}}

\def\sec#1{{sez.~\ref{#1}}}

\def\gev{\mbox{ GeV}}
\def\fm{\mbox{ fm}}

\def\ln#1{\log{\l( #1 \r)}}
\def\op{{\mathcal{O}}}

\def\div{\frac{1}{\hat{\epsilon}}}

\def\Im{\mathop{\mbox{Im}}}
\def\Re{\mathop{\mbox{Re}}}

\def\exponential{{\mathrm{e}}}

\usepackage{xcolor}

\newcommand{\myAffiliation}{Institute of Physics, National Yang Ming
  Chiao Tung University, Hsinchu 30010, Taiwan}

\begin{document}

\begin{abstract}
The pion light-cone distribution amplitude (LCDA) is a central non-perturbative object of interest for high-energy exclusive processes in quantum chromodynamics. 
In this article, the second Mellin moment of the pion LCDA is determined as a proof-of-concept calculation for the first numerical implementation of the heavy-quark operator product expansion (HOPE) method. The resulting value for the second Mellin moment, determined in quenched QCD at a pion mass of $m_\pi=550$ MeV at a factorization scale of 2 GeV is $ \expval{ \xi^2 }= 0.210 \pm 0.013\text{ (stat.)} \pm 0.034\text{ (sys.)}$.
This result is compatible with
those from previous determinations of this quantity.
\end{abstract}

\preprint{MIT-CTP/5330}

\title{
Parton physics from a heavy-quark operator product expansion:\\
  Lattice QCD calculation of the second moment\\
   of the pion distribution amplitude} 

\author{William~Detmold}
\email{wdetmold@mit.edu}
\affiliation{Center for Theoretical Physics, Massachusetts Institute of Technology,
Cambridge, MA 02139, USA}
\affiliation{The NSF AI Institute for Artificial Intelligence and Fundamental Interactions}

\author{Anthony~V.~Grebe}
\email{agrebe@mit.edu}
\affiliation{Center for Theoretical Physics, Massachusetts Institute of Technology,
  Cambridge, MA 02139, USA}

\author{Issaku~Kanamori}
\email{kanamori-i@riken.jp}
\affiliation{RIKEN Center for Computational Science, Kobe 650-0047, Japan}

\author{C.-J.~David~Lin}
\email{dlin@nycu.edu.tw}
\affiliation{\myAffiliation}
\affiliation{Centre for Theoretical and Computational Physics, National Yang Ming Chiao Tung University, Hsinchu 30010, Taiwan}
\affiliation{Centre for High Energy Physics, Chung-Yuan Christian University, Chung-Li, 32032, Taiwan}

\author{Santanu~Mondal}
\email{santanu.sinp@gmail.com}
\affiliation{Los Alamos National Laboratory, Theoretical Division T-2, Los Alamos, NM 87545, USA}

\author{Robert~J.~Perry}
\email{perryrobertjames@gmail.com}
\affiliation{\myAffiliation}
\affiliation{Centre for Theoretical and Computational Physics, National Yang Ming Chiao Tung University, Hsinchu 30010, Taiwan}

\author{Yong~Zhao}
\email{yong.zhao@anl.gov}
\affiliation{Physics Division, Argonne National Laboratory, Lemont, IL 60439, USA}

\collaboration{HOPE Collaboration}

\maketitle


%
\section{Introduction}
\label{sec:intro}
The pion light-cone distribution amplitude (LCDA) plays an important
role in parton physics.  It
is central to a description of a range of exclusive processes in high energy quantum chromodynamics~\cite{Lepage:1980fj}.   This LCDA, denoted as
$\phi_{\pi}$, is defined via its relation to the
matrix element for the transition between the vacuum and the
(charged\footnote{The isospin limit is used for the calculation
  presented in this article.})
pion state,
\beq
\label{eq:pion_DA_def}
\la 0 | \overline{\psi}_d(z) \gamma_{\mu} \gamma_{5} W[z, -z] \psi_u(-z) |
\pi^{+} ({\mathbf{p}}) \ra =
 i  p_{\mu} f_{\pi} \int_{-1}^{1} d \xi \mbox{ }
 {\mathrm{e}}^{-i \xi p\cdot z }\phi_{\pi}(\xi, \mu ) \, ,
\eeq
where
$\mu$ is the renormalization scale and ${\mathcal{W}}[-z,z]$ is a
light-like Wilson line between light-cone
coordinates $-z$ to $z$ ($z^{2} = 0$).  In the above equation,
$f_{\pi}$, ${\mathbf{p}}$, and $p_{\mu}$ are the decay constant, the
three-momentum, and the four-momentum of the pion.  In the light-cone gauge, $\phi_{\pi}(\xi, \mu )$ can
be interpreted as the probability amplitude for the pion to transition to the state of a quark and an antiquark that carry $(1+\xi)/2$ and $(1-\xi)/2$ fractions
of the pion momentum, respectively~\cite{Lepage:1980fj}. This LCDA plays an
important role in understanding exclusive
processes in
QCD~\cite{Radyushkin:1977gp,Lepage:1980fj}, in addition to being a crucial ingredient for 
extracting information regarding the Cabibbo-Kobayashi-Maskawa matrix in flavour physics via hadronic decays of
the $B$ meson~\cite{Dugan:1990de,Beneke:1999br,Keum:2000wi,Bauer:2005kd}.

Since $\phi_{\pi} (\xi,\mu)$ encodes non-perturbative physics from
strong interactions, it is natural to attempt to compute this quantity with lattice QCD (LQCD).  However, 
LQCD calculations are normally carried out in Euclidean
space because of the need to employ Monte-Carlo methods in evaluating
the path integrals.   Therefore it is challenging to adopt LQCD for
investigating parton physics which involves dynamics on the light
cone.  This leads to the conventional LQCD approach of determining the Mellin
moments for various parton
distribution functions (PDFs) and LCDAs.   In general these moments
can be extracted by computing matrix elements of local operators that result from an
operator product expansion (OPE)~\cite{Kronfeld:1984zv, Martinelli:1987si,Del_Debbio_2003,Arthur_2011,Bali:2019dqc}.   The analytic continuation between
Euclidean and Minkowski spaces for these matrix elements is
straightforward.   In principle, knowledge of the
relevant Mellin
moments enables the construction of PDFs and LCDAs.  Nevertheless, such
a strategy has been limited by the possibility of having reliable
LQCD results for only the first few
moments because the breaking of $O(4)$ Euclidean space-time
symmetry by the
lattice regularization requires the subtraction of power
divergences in the renormalization
of the relevant local operators~\cite{Kronfeld:1984zv}.  These power divergences
already appear in the computation of the first non-trivial (the
second) Mellin moment for $\phi_{\pi}(\xi,\mu)$.  They can be evaded
by using a method proposed in Ref.~\cite{Martinelli:1987si}, where one
chooses particular combinations of the Lorentz indices in the local
operator for computing the second moment.  However,
this method is not applicable for the extraction of higher moments,
making it desirable to have other approaches with which to extract
information of LCDAs and other light-cone quantities, including PDFs.

Alternative strategies for extracting information about LCDAs and PDFs
using LQCD have been proposed in the last two
decades~\cite{Aglietti:1998ur,Liu:1999ak,Detmold:2005gg,Braun:2007wv,Davoudi:2012ya,Ji:2013dva,Chambers:2017dov,Radyushkin:2017cyf,Ma:2017pxb},
and their implementations are being
intensively pursued, see
Refs.~\cite{Cichy:2018mum,Ji:2020ect,Constantinou:2020pek,Constantinou:2020hdm} for reviews.
To access information contained in higher Mellin moments while bypassing
the above-mentioned complication in renormalizing the local operators,
a generic character of these strategies is the computation for hadronic matrix elements of non-local
operators.  For instance, one much studied procedure, the quasi-PDF
approach~\cite{,Ji:2013dva}, features the calculation of matrix elements involving a
space-like Wilson line.
This work follows the method suggested in
Ref.~\cite{Detmold:2005gg}.   It relies on investigating the OPE analysis
for hadronic amplitudes in Euclidean space with the insertion of two local quark bilinears
that contain a fictitious, valence heavy quark.  For this reason, this
approach is termed the heavy-quark OPE (HOPE) method.   The
introduction of this heavy quark affords several advantages, such
the removal of certain higher-twist effects, the inclusion of an extra heavy scale to
control the OPE, and simple properties of analytic continuation to Minkowski space.
As pointed out in recent work~\cite{Detmold:2021uru}, this method
can be used to extract the $\xi$-dependence in $\phi_{\pi}(\xi,\mu)$
through a perturbative matching procedure.  The current
article presents its application to the numerical determination of
the (second) Mellin
and Gegenbauer moments\footnote{Preliminary results from this calculation have been presented in 
contributions to the proceedings of recent lattice
conferences~\cite{Detmold:2018kwu,Detmold:2020lev, Detmold:2021LattConf}.}.  Since this is the
first numerical implementation of the HOPE strategy, this calculation
bears the character of a feasibility study, which 
investigates the relatively well-known
second Mellin moment.  This allows a better quantification of the efficacy
of the method. The success of the current work paves the way for further studies using this technique. 
For this reason, the
quenched approximation and an unphysical pion mass of $\sim 550$ MeV are used.
Since the continuum extrapolation is an important ingredient in this
approach, the calculations are performed at four choices of the lattice
spacing, $a$, ranging from 0.04 fm to 0.08 fm.

Definitions relevant to this method and the conventions adopted in this article are declared in Sec.~\ref{sec:def}.
The HOPE method and
its particular features for the current work are explained in
Sec.~\ref{sec:strategy}, while Sec.~\ref{sec:numerical} describes the
details of the numerical implementation.    The analysis of the data
and the results are discussed in Sec.~\ref{sec:analysis}.   
The conclusion of this work and the outlook in this research
direction are then given in Sec.~\ref{sec:conclusion}.

\section{Definitions and conventions}
\label{sec:def}
Before proceeding to describe the numerical lattice determination for
the second Mellin and Gegenbauer moments of $\phi_{\pi} (\xi, \mu)$, a review of the definition of these moments is in order.   The discussion presented here also serves the purpose of setting the conventions and notation that will be used in subsequent calculations.

Conformal symmetry in QCD~\cite{Braun:2003rp} implies that the pion
LCDA can be conveniently studied using the Gegenbauer OPE.  $C$-parity imposes the constraint $\phi_{\pi}
(\xi,\mu) = \phi_{\pi} (-\xi, \mu)$ in the isospin limit, leading to
\beq
\label{eq:Gegen_OPE_phi_pi}
\phi_{\pi}(\xi, \mu) = \frac{3}{4} (1 - \xi^{2})
\sum_{n=0,{\mathrm{even}}}^{\infty} \phi_{n}(\mu)
{\mathcal{C}}_{n}^{3/2} (\xi) \, ,
\eeq
where ${\mathcal{C}}_{n}^{3/2} (\xi)$ are the Gegenbauer polynomials, with $\mathcal{C}_0^{3/2}(\xi) = 1$ and $\mathcal{C}_2^{3/2}(\xi) = (-3+15\xi^2)/2$.  The
Gegenbauer moments, $\phi_{n}(\mu)$, are defined as
\beq
\label{eq:Gegen_moment_def}
 \phi_{n}(\mu) = \frac{2 (2n+3)}{3 (n+1)(n+2)} \int_{-1}^{1} d \xi
 \mbox{ } {\mathcal{C}}_{n}^{3/2} (\xi) \phi_{\pi}(\xi, \mu) \, .
\eeq
Because of conformal symmetry, the $\phi_{n}(\mu)$ do not mix under the
renormalization group (RG) evolution at one loop.  To this order,
their renormalization scale dependence is~\cite{Efremov:1978rn}
\beq
\label{eq:Gegen_moment_one_loop_running}
 \phi_{n} (\mu_{2}) = \phi_{n} (\mu_{1}) \left ( \frac{\alpha_{s}(\mu_{2})}{\alpha_{s}(\mu_{1})} \right
 )^{\gamma_{n}/\beta_{0}} \, ,
\eeq
where $\alpha_{s}$ is the strong coupling, $\beta_{0} = 11 - 2N_{f}/3$ ($N_{f}$ being the number of
flavours) is the coefficient of the leading-order (LO) QCD
$\beta$-function, and $\gamma_{n}$ is the anomalous dimension,
\beq
\label{eq:Gegen_moment_anomalous_dim}
 \gamma_{n} = \frac{4}{3} \left [ 1 - \frac{2}{(n+1)(n+2)} +4
 \sum_{j=2}^{n+1}j^{-1} \right ] \, .
\eeq
Since $\gamma_{n}$ increases monotonically with $n$, one expects that
a truncated version of the OPE in Eq.~(\ref{eq:Gegen_OPE_phi_pi})
can be a good approximation to $\phi_{\pi}(\xi,\mu)$ at large enough
renormalization scales.
In the regime where $\mu \gg \Lambda_{{\mathrm{QCD}}}$ (with
$\Lambda_{{\mathrm{QCD}}}$ being the QCD dynamical scale),  the pion LCDA is dominated by the
zeroth Gegenbauer moment.  
Equations~(\ref{eq:Gegen_moment_one_loop_running}) and (\ref{eq:Gegen_moment_anomalous_dim})  lead to the asymptotic form the pion LCDA,
\beq
\label{eq:Phi_pi_asymptotic}
 \lim_{\mu\rightarrow\infty}\phi_{\pi} (\xi, \mu) = \frac{3}{4}
 \left ( 1 - \xi^{2} \right ) \, ,
\eeq
where the normalization $\phi_{0} = 1$ has been imposed.

\begin{figure}
\centering
\includegraphics[scale=0.6]{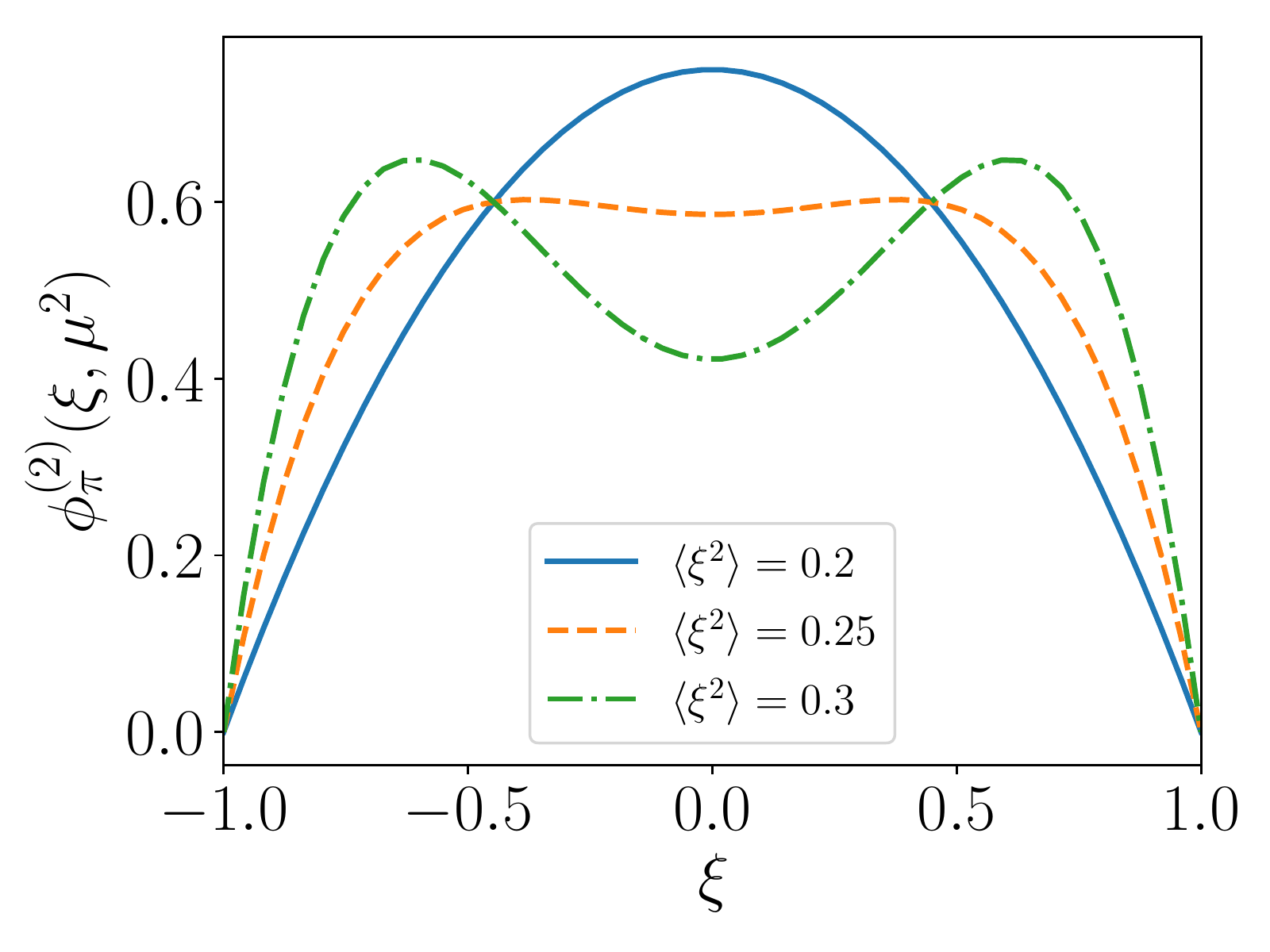}
\caption{Dependence of $\phi_{\pi}^{(2)}(\xi,
\mu)$, defined in Eq.~(\ref{eq:phi_pi_OPE_to_xi2}), on the values of
$\la \xi^{2} \ra$.  The range of $\la \xi^{2} \ra$ values shown in the
plot covers typical results for this second Mellin moment at $\mu
\sim 2$ GeV from modern lattice
computations.  They lead to consistency with a single-humped or double-humped structure of the DA, and more precise measurements would resolve this.
}
\label{fig:Xi2_pheno}
\end{figure}

The Gegenbauer moments in Eqs.~(\ref{eq:Gegen_OPE_phi_pi}) and (\ref{eq:Gegen_moment_def}) can be expressed as linear
combinations of the Mellin moments $\la \xi^{n} \ra$, which are defined as
\beq
\label{eq:Mellin_moments_def}
 \la \xi^{n} \ra (\mu) = \int_{-1}^{1} d\xi \mbox{ }\xi^{n}
 \phi_{\pi}(\xi,\mu) \, .
\eeq
For instance, from Eqs.~(\ref{eq:Gegen_OPE_phi_pi}) and (\ref{eq:Mellin_moments_def}) it is straightforward to obtain
\begin{equation}
\label{eq:G_to_M_moments}
 \phi_{0}  = \la \xi^{0} \ra = 1\, , \,\,  \phi_{2} = \frac{7}{12}
 \left ( 5 \la \xi^{2} \ra - \la \xi^{0} \ra \right ) \, , \,\,
 \phi_{4} = \frac{11}{24} \left ( 21 \la \xi^{4}\ra - 14 \la
 \xi^{2}\ra + \la \xi^{0}\ra\right ) \, , \ldots \,  .
\end{equation}
In general, knowledge of $\la \xi^{0} \ra$, $\la \xi^{2}\ra$, ...,
$\la \xi^{n} \ra$ is equivalent to that of $\phi_{0}$, $\phi_{2}$, ...,
$\phi_{n}$.  These Mellin moments, $\la \xi^{n} \ra$, can be related to matrix elements of local,
twist-two, operators, 
\beq
\label{eq:Mellin_moments_to_ME}
 \la 0 |  \big [ \overline{d}
  \gamma^{\{\mu_0}\gamma_5(i\overset{\leftrightarrow}{D}\vphantom{D}^{\,\mu_1})\dots(i\overset{\leftrightarrow}{D}\vphantom{D}^{\,\mu_n\}})
  u  - {\mathrm{traces}} \big  ] | \pi^{+} (\mathbf{p})
\ra
 =f_{\pi} \expval{\xi^{n}}(\mu^2) \left [ p^{\mu_0}p^{\mu_1}\dots p^{\mu_n}-
 {\mathrm{traces}} \right ] \, ,
\eeq
where the Lorentz indices are totally symmetrized, with 
\beq
\label{eq:two_way_derivative_def}
\overset{\leftrightarrow}{D}\vphantom{D}^{\,\mu}=\frac{1}{2}(\overset{\rightarrow}{D}\vphantom{D}^{\,\mu}-\overset{\leftarrow}{D}\vphantom{D}^{\,\mu})
\, ,
\eeq
and the traces are taken in all possible pairs amongst the Lorentz
indices, $\mu_{0}, \mu_{1}, \ldots, \mu_{n}$.  As discussed in the
last paragraph, from the leading-order result of QCD perturbation
theory, it is natural to expect that knowledge of the first few Gegenbauer
moments allows one to construct $\phi_{\pi} (\xi, \mu)$ reliably at
sufficiently large $\mu$.  This also implies
that obtaining important information about the LCDA at $\mu \gg \Lambda_{{\mathrm{QCD}}}$ is possible from the first few Mellin
moments\footnote{The qualitative feature of low-moment dominance in
  $\phi_{\pi}$ was also argued using QCD sum rules~\cite{Chernyak:1981zz}.}.
This point
can be illustrated by investigating an extreme scenario where one
truncates the Gegenbauer OPE in Eq.~(\ref{eq:Gegen_OPE_phi_pi}) at
$n=2$.  The pion LCDA constructed with this truncated OPE is denoted $\phi_{\pi}^{(2)}(\xi, \mu) $.  Using
Eq.~(\ref{eq:G_to_M_moments}), one obtains
\bea
\label{eq:phi_pi_OPE_to_xi2}
\phi_{\pi}^{(2)}(\xi, \mu) &=& \frac{3}{4} (1 - \xi^{2}) \left [
\phi_{0}(\mu) {\mathcal{C}}^{3/2}_{0} (\xi) +\phi_{2}(\mu)
{\mathcal{C}}^{3/2}_{2} (\xi) \right ]\nonumber\\
 &=& \frac{15}{32} \left [ 3 - 7 \la \xi^{2}
\ra (\mu) \right ] + \frac{16}{15} \left [ -5 + 21 \la \xi^{2} \ra (\mu)\right ]
\xi^{2} + \frac{105}{32} \left [ 1 - 5 \la \xi^{2} \ra (\mu)\right ] \xi^{4}
\, .
\eea
Figure~\ref{fig:Xi2_pheno} shows the result with $\phi_{\pi}^{(2)}(\xi,
\mu)$ at $\la \xi^{2} \ra (\mu) = 0.2$, 0.25 and 0.3.  
Note that these
are typical values for this second Mellin moment at $\mu
\sim 2$ GeV from modern lattice
computations~\cite{Del_Debbio_2003, Arthur_2011, 
  Bali:2017gfr, Zhang:2017bzy, Bali:2019dqc}.   This figure demonstrates
that the shape of the pion LCDA can depend strongly upon $\la
\xi^{2}\ra (\mu)$.  Naturally, the inclusion of higher moments will likely reduce the sensitivity to the second moment, but nevertheless, this exercise shows that the second Mellin moment is a phenomenologically interesting quantity.

\section{Strategy and correlation functions}
\label{sec:strategy}
To present the calculation for the second Mellin and Gegenbauer
moments of $\phi_{\pi}(\xi, \mu)$, it is first necessary to describe the strategy and the
correlators that have to be computed using LQCD.  
To extract moments for the pion LCDA employing the HOPE method, 
the hadronic amplitudes
\begin{equation}
\label{eq:antisymm_hadronic_tensor}
V^{[ \mu\nu ]} (q,p) \equiv \frac{V^{\mu\nu}(q,p) - V^{\nu\mu}(q,p)}{2} \, ,
\end{equation}
are computed, where
\begin{equation}
\label{eq:hadronic_tensor}
V^{\mu\nu} (q,p) = \int d^{4}z \mbox{ } \exponential^{i q \cdot z} \mbox{ }
 \la 0 | {\mathcal{T}} \{ J_{A}^{\mu} (z/2) J_{A}^{\nu}(-z/2) \} | \pi ({\mathbf{p}})\ra \, ,
\end{equation}
with $J_{A}^{\mu}$ being the axial-vector current involving a light
quark $\psi$ and a fictitious valence heavy quark $\Psi$ with mass $m_{\Psi}$,
\begin{equation}
\label{eq:axial_current}
 J^{\mu}_{A} = \overline{\Psi} \gamma^{\mu}\gamma^{5}\psi + \overline{\psi} \gamma^{\mu}\gamma^{5} \Psi \, .
\end{equation}
The antisymmetrization in Eq.~(\ref{eq:antisymm_hadronic_tensor}) is performed explicitly to reduce statistical noise, and the axial-axial correlator in Eq.~(\ref{eq:hadronic_tensor}) was empirically found to have less excited state contamination than the analogous correlator with two vector current insertions.

\subsection{Relevant results from the HOPE strategy}
\label{sec:HOPE_strategy}

Conformal symmetry in QCD~\cite{Braun:2003rp} implies that it is natural to
proceed with the Gegenbauer OPE for studying the twist-two contribution to the
hadronic amplitude defined in
Eq.~(\ref{eq:antisymm_hadronic_tensor}).  This OPE allows one to
express the twist-two component of $V^{[\mu\nu]}(q,p)$ in terms of the
same Gegenbauer
moments, $\phi_{n}(\mu)$, defined in Eq.~(\ref{eq:Gegen_moment_def}). 
These Gegenbauer moments do not mix under renormalization group (RG)
evolution at one-loop~\cite{Efremov:1979qk,Lepage:1979zb}.
At this order, generically the Gegenbauer OPE leads to~\cite{Detmold:2021uru}
\begin{equation}
\label{eq:Gegen_OPE_had_amp}
 V^{[ \mu\nu ]} (q,p) = - \frac{2 i  \epsilon^{\mu\nu\rho\sigma}
 q_{\rho} p_{\sigma}}{\tilde{Q}^{2}}  f_{\pi}
 \sum_{n=0,{\mathrm{even}}}^{\infty} {\mathcal{F}}_{n} (\tilde{Q}^{2},
 \mu, \tilde{\omega}, m_\Psi) \phi_{n} (\mu)  + 
  \text{higher-twist terms}\, ,
\end{equation}
where $\mu$ is the renormalization scale, $m_{\Psi}$ is the mass of
the fictitious valence heavy quark, and ${\mathcal{F}}_{n}$  are coefficients that can be computed in QCD
perturbation theory and can be expressed as functions of the kinematic variables
\begin{equation}
\label{eq:kin_var_def}
 \tilde{Q}^{2} = Q^{2} + m_{\Psi}^{2} \, , ~~~~~
 \tilde{\omega} = \frac{2 p\cdot q}{\tilde{Q}^{2}} \, ,
\end{equation}
with $Q^{2} = -q^{2}$.  For simplicity,
higher-twist contributions
to $V^{[ \mu\nu ]} (q,p)$ will be discarded below in this
section.  They will be
discussed in detail in Sec.~\ref{sec:analysis}.

Employing results in Ref.~\cite{Detmold:2021uru}, it can be
demonstrated that
\begin{equation}
\label{eq:G_nm_def}
  {\mathcal{F}}_{n} (\tilde{Q}^{2},
 \mu, \tilde{\omega}, m_{\Psi}) = \sum_{m=n, {\mathrm{even}}}^{\infty}  {\mathcal{F}}_{n,m} (\tilde{Q}^{2},
 \mu, m_{\Psi}) \left ( \frac{\tilde{\omega}}{2} \right )^{m} .
\end{equation}
As an example, the leading-order (tree-level) result of
${\mathcal{F}}_{n}$ is independent of $\mu$ and  gives, up to ${\mathcal{O}}(\tilde{\omega}^{2})$,
\bea
\label{eq:Gegen_OPE_tree}
V^{[ \mu\nu ]}_{{\mathrm{tree}}} (q,p) &=& - \frac{2 i \epsilon^{\mu\nu\rho\sigma}
 q_{\rho} p_{\sigma}}{\tilde{Q}^{2}}  f_{\pi} \left \{ \left [  1 + \frac{1}{20}
 \tilde{\omega}^{2} +
 \op(\tilde{\omega}^{4})  \right
 ]\phi_{0} + \left [ \frac{3}{35} \tilde{\omega}^{2} +
 \op(\tilde{\omega}^{4}) \right ] \phi_{2}
 +
 \op(\tilde{\omega}^{4}) \right \} \, \nonumber\\
 &=& - \frac{2 i \epsilon^{\mu\nu\rho\sigma}
 q_{\rho} p_{\sigma}}{\tilde{Q}^{2}}  f_{\pi} \left \{ 1 + \la \xi^{2} \ra \left ( \frac{\tilde{\omega}}{2}\right )^{2}  +
 \op(\tilde{\omega}^{4}) \right \} \, ,
\eea
where the relation between the leading two Mellin and 
Gegenbauer moments given in Eq.~(\ref{eq:G_to_M_moments}) is used.
In this work, the goal is to compute the second Mellin and Gegenbauer
moments of $\phi_{\pi}(\xi, \mu)$, working in the kinematic regime
where $\tilde{\omega} \ll 1$, such that the OPE for
$V^{[\mu\nu]}(p,q)$ can be truncated at the order of $\tilde{\omega}^{2}$.

Beyond the leading order (LO) in QCD perturbation theory, adopting Eqs.~(\ref{eq:Gegen_OPE_had_amp}),
(\ref{eq:G_nm_def}) and (\ref{eq:G_to_M_moments}), one obtains
\begin{equation}
\label{eq:Mellin_OPE_had_amp}
V^{[ \mu\nu ]} (q,p) = - \frac{2 i \epsilon^{\mu\nu\rho\sigma}
 q_{\rho} p_{\sigma}}{\tilde{Q}^{2}}  f_{\pi} \sum_{n=0,{\mathrm{even}}}^{\infty}
 C_{W}^{(n)} (\tilde{Q}^{2},
 \mu, m_\Psi)  \la \xi^{n}\ra  \left ( \frac{\tilde{\omega}}{2}\right )^{n}\, ,
\end{equation}
where the Wilson coefficients,
$C_{W}^{(n)}(\tilde{Q}^{2},\mu,m_\Psi)$, are
linear combinations of ${\mathcal{F}}_{n,m} (\tilde{Q}^{2},\mu,
m_\Psi)$~\cite{Detmold:2021uru}. 
Since this work uses a relatively heavy pion ($m_{\pi} \sim 550$ MeV), it is beneficial to resum higher-twist target mass effects proportional to $m_\pi$.  The resummation prescription given in~\cite{Nachtmann:1973mr} is to replace 
$\tilde{\omega}^{n}$ by $\zeta^{n}{\mathcal{C}}_{n}^{2}
(\eta)/(n+1)\tilde{Q}^{2}$, where $\zeta=\sqrt{p^{2}
q^{2}}/\tilde{Q}^{2}$, $\eta=p\cdot q/\sqrt{p^{2}q^{2}}$, and ${\mathcal{C}}_{n}^{2}(\eta)$
is a Gegenbauer polynomial.
In other words,
\begin{equation}
\label{eq:Mellin_OPE_had_amp_target_mass}
V^{[ \mu\nu ]} (q,p) = - \frac{2 i \epsilon^{\mu\nu\rho\sigma}
 q_{\rho} p_{\sigma}}{\tilde{Q}^{2}}  f_{\pi} \sum_{n=0,{\mathrm{even}}}^{\infty}
 C_{W}^{(n)} (\tilde{Q}^{2},
 \mu, m_\Psi)  \la \xi^{n}\ra  \left [ \frac{\zeta^{n} {\mathcal{C}}_{n}^{2}
(\eta)}{2^{n}(n+1)\tilde{Q}^{2}}\right ]\, .
\end{equation}

Truncating at the order of $\tilde{\omega}^{2}$,
\begin{equation}
\label{eq:2nd_Mellin_OPE_had_amp_target_mass}
V^{[ \mu\nu ]} (q,p) \approx - \frac{2 i \epsilon^{\mu\nu\rho\sigma}
 q_{\rho} p_{\sigma}}{\tilde{Q}^{2}}  f_{\pi} \left \{
 C_{W}^{(0)} (\tilde{Q}^{2},
 \mu, m_\Psi)  +  C_{W}^{(2)} (\tilde{Q}^{2},
 \mu, m_\Psi)  \la \xi^{2}\ra  \left [ \frac{\zeta^{2} \mathcal{C}_{2}^{2}
(\eta)}{12\tilde{Q}^{2}} \right ] \right \} \, ,
\end{equation}
where the explicit one-loop expressions for $C_{W}^{(0)}
(\tilde{Q}^{2},\mu, m_\Psi)$ and $C_{W}^{(2)} (\tilde{Q}^{2},\mu,
m_\Psi)$ are given in Ref.~\cite{Detmold:2021uru}. 
Equation~(\ref{eq:2nd_Mellin_OPE_had_amp_target_mass}) is used
in this analysis to extract $\la \xi^{2} \ra$.  As described in
Refs.~\cite{Detmold:2005gg,Detmold:2021uru},  in addition to $\la
\xi^{2} \ra$, $f_{\pi}$ and $m_{\Psi}$ are also fit parameters in the
analysis procedure that will be presented in detail in Sec.~\ref{sec:analysis}. Note that while the hadronic matrix element is renormalization scheme and scale independent, the factorization of this matrix element into short-distance Wilson coefficients and long-range Mellin moments are dependent on the renormalization scheme and scale. The calculation of the Wilson coefficients was determined in the $\overline{\text{MS}}$ scheme and thus the fitted heavy-quark masses and Mellin moments are directly extracted in this scheme.
\subsection{The correlation functions}
\label{sec:correlators}
The power of the hadronic tensor lies in its amenability to lattice QCD calculations.
The pion LCDA defined in Eq.~(\ref{eq:pion_DA_def}) cannot be computed directly in Euclidean-space LQCD due to the light-like separation vector $z$.  In contrast, the hadronic tensor $V^{\mu\nu}$ can be written in terms of quantities calculable on the lattice.
Defining
\begin{align}
  R^{\mu\nu}(\tau; \mathbf{p}, \mathbf{q})
 &= \int d^3 \mathbf{z} \,e^{i\mathbf{q}\cdot \mathbf{z}} \langle 0 |\mathcal{T}[ J_A^\mu(\tau/2,\mathbf{z}/2) J_A^\nu(-\tau/2,-\mathbf{z}/2)] | \pi(\mathbf{p}) \rangle
 \nonumber \\
 &= \langle 0 | J_A^\mu(\tau/2; (\mathbf{p}+\mathbf{q})/2) J_A^\nu(-\tau/2; (\mathbf{p}-\mathbf{q})/2) | \pi(\mathbf{p}) \rangle \, ,
  \label{ratio}
\end{align}
then the hadronic tensor is the Fourier transform of $R^{\mu\nu}$ in the temporal direction:
\begin{equation}
  V^{\mu\nu}(q,p) = \int d\tau \, e^{iq_4 \tau} R^{\mu\nu}(\tau; \mathbf{p}, \mathbf{q}) \, .
  \label{fourier-transform-temporal}
\end{equation}

\begin{figure}[h]
  \centering
  \includegraphics[scale=0.7]{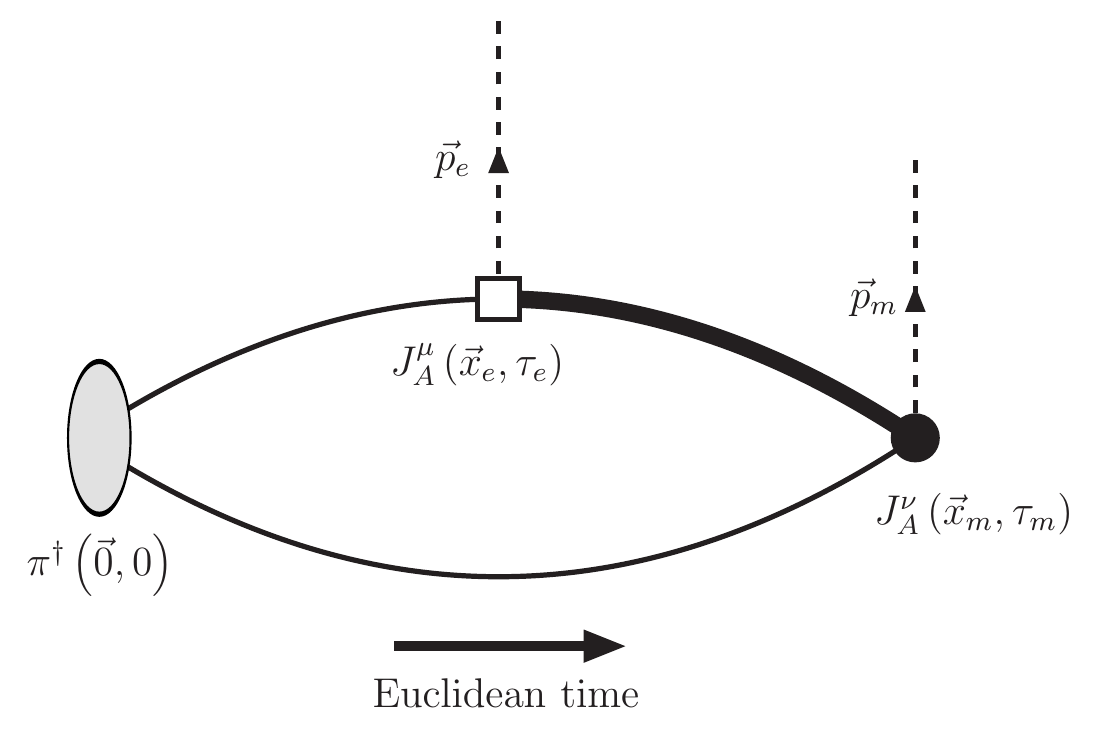}
  \caption{A diagram of the three-point correlator used in this calculation with current insertions at times $\tau_e$ and $\tau_m$.}
  \label{3-pt-figure}
\end{figure}

Using lattice methods, one can compute two-point and three-point correlation functions
\begin{align}
  C_2 (\tau, \mathbf{p})
  &= \int d^3 \mathbf{x} \, e^{i\mathbf{p}\cdot \mathbf{x}}  \langle 0|\mathcal{O}_\pi(\tau,\mathbf{x}) \mathcal{O}^\dagger_\pi (0, \mathbf{0}) |0 \rangle \nonumber \\
  &= \langle 0 | \mathcal{O}_\pi(\tau, \mathbf{p})
    \mathcal{O}^\dagger_\pi(0, \mathbf{p}) | 0 \rangle \, ,
  \label{2pt-corr}
\end{align}
and
\begin{align}
  C^{\mu\nu}_3 (\tau_e, \tau_m; \mathbf{p}_e, \mathbf{p}_m)
  &= \int d^3x_e \, d^3x_m\, e^{i\mathbf{p}_e\cdot \mathbf{x}_e}e^{i\mathbf{p}_m\cdot \mathbf{x}_m}\langle 0 | \mathcal{T}\left[ J_A^\mu(\tau_e, \mathbf{x}_e) J_A^\nu(\tau_m, \mathbf{x}_m) \mathcal{O}^\dagger_\pi(\mathbf{0}) \right] | 0 \rangle
  \nonumber \\
  &= \langle 0 | J_A^\mu(\tau_e, \mathbf{p}_e) J_A^\nu(\tau_m, \mathbf{p}_m) \mathcal{O}^\dagger_\pi(0, \mathbf{p}) | 0 \rangle  \, .
  \label{3pt-corr}
\end{align}
The three-point correlator is shown diagrammatically in Figure~\ref{3-pt-figure}.

For $0 \ll \tau \ll T$, the 2-point correlator is saturated with the contribution of the lowest-lying hadronic state and can be written as
\begin{equation}
  C_2 (\tau, \mathbf{p}) \sim \frac{|Z_\pi(\mathbf{p})|^2}{2E_\pi(\mathbf{p})} \left[ e^{-E_\pi(\mathbf{p}) \tau} + e^{-E_\pi(\mathbf{p}) (T - \tau)} \right] \, ,
  \label{2-pt-func}
\end{equation}
which allows determination of the overlap factor\footnote{$Z_\pi(\mathbf{p})$ is taken to be real and positive in this analysis.}
$Z_\pi(\mathbf{p}) = \langle 0 | \mathcal{O}_\pi | \pi(\mathbf{p}) \rangle$ and the pion energy $E_\pi(\mathbf{p})$.

Similarly, for $0 \ll \tau_e, \tau_m \ll T/2$, the 3-point correlation function takes the form
\begin{equation}
  C_3^{\mu\nu}(\tau_e, \tau_m; \mathbf{p}_e, \mathbf{p}_m) \sim
  R^{\mu\nu}(\tau; \mathbf{p}, \mathbf{q})
  \frac{Z_\pi(\mathbf{p})}{2E_\pi(\mathbf{p})}
  e^{-E_\pi(\mathbf{p})(\tau_e + \tau_m)/2} \, ,
\end{equation}
with $\mathbf{p} = \mathbf{p}_e + \mathbf{p}_m$ and $\tau = \tau_e - \tau_m$, allowing one to extract 
\begin{equation}\label{eq:r12}
  R^{\mu\nu}(\tau; \mathbf{p}, \mathbf{q}) = \frac{2E_\pi(\mathbf{p}) C_3^{\mu\nu}(\tau_e, \tau_m; \mathbf{p}_e, \mathbf{p}_m)}{Z_\pi(\mathbf{p}) e^{-E_\pi(\mathbf{p})(\tau_e + \tau_m)/2}}
\end{equation}
from the two- and three-point correlators.  From this and Eq.~(\ref{eq:2nd_Mellin_OPE_had_amp_target_mass}) and (\ref{fourier-transform-temporal}), one can extract the second Mellin moment on the lattice.

\section{Details of numerical implementation}
\label{sec:numerical}
\subsection{Lattice Action and $\mathcal{O}(a)$ Improvements}\label{sec:order-a}
Order-$a$ corrections to correlation functions arise from both the action and the interpolating operators~\cite{Symanzik:1983dc,Symanzik:1983gh,Luscher:1996sc}. Thus in order to remove these effects, one must in general improve both.
The ensembles were generated with the standard Wilson gauge action
\begin{align}
  S_G[\psi, \overline \psi, U] &= \frac{\beta}{3} \sum_n \sum_{\mu < \nu} \text{Re}\left\{ \text{Tr}\left[ 1-P_{\mu\nu}(n) \right] \right\} \, ,
  \label{gauge-action}
\end{align}
where $\beta = 6/g^2$ is inverse coupling and $P_{\mu\nu}$ is the Wilson plaquette.  The gauge action is automatically $\mathcal{O}(a)$ improved.

The Wilson fermion action, which is omitted from ensemble generation in the quenched approximation but is used in propagator construction, is given by
\begin{align}
  S_F[\psi, \overline \psi, U] &= a^4 \sum_{f=1}^{N_f} \sum_{m,n} \overline \psi^{(f)}_{\alpha a}(n) \left[ \frac{1}{2a\kappa^{(f)}} \delta_{\alpha\beta} \delta_{ab} \delta_{nm} - \frac{1}{2a} \sum_{\mu = \pm 1}^{\pm 4} (1-\gamma_\mu)_{\alpha\beta} U_{\mu}(n)_{ab} \delta_{n+\hat \mu, m} \right] \psi^{(f)}_{\beta b}(m) \, ,
  \label{fermion-action}
\end{align}
where $\kappa^{(f)}$ is the hopping parameter for flavour $f$~\cite{gatringer}.  The fermion action can be improved by addition of the clover term
\begin{equation}
  \delta S[\overline{\psi},\psi,U]=\frac{1}{4} a^5 c_{\text{sw}} \sum_n \sum_{\mu\nu} \overline{\psi}(n)i\sigma^{\mu\nu} \hat F_{\mu\nu}(n)\psi(n)\,,
\end{equation}
where $\sigma_{\mu\nu} = \left[ \gamma_\mu, \gamma_\nu \right]/2i$ and $\hat F_{\mu\nu}$ is a discretized version of the field strength tensor corresponding to a sum over plaquettes.
The clover coefficient $c_\text{sw}$ is taken from the non-perturbative tuning in Ref.~\cite{Luscher_1997}. 

The $\mathcal{O}(a)$-improved, renormalized axial current operator is given by
\begin{equation}\label{eq:order-a_current}
  J_A^\mu = Z_A^{(0)} (1 + \tilde{b}_A a \tilde{m}_{ij})\left[\overline\psi \gamma_\mu \gamma_5 \Psi + a c_A \partial_\mu \overline \psi \gamma_5 \Psi - a \frac{c_A'}{4} \left( \overline \psi \gamma_\mu \gamma_5 (\overrightarrow D + m_\Psi) \Psi - \overline \psi (\overleftarrow D + m_\psi) \gamma_\mu \gamma_5 \Psi \right) + \left( \psi \leftrightarrow \Psi \right)\right] \, ,
\end{equation}
where $Z_A^{(0)}$ is the axial-vector renormalization constant calculated in the chiral limit, $\tilde{m}_{ij}=(\tilde m_i+\tilde m_j)/2$ is the average value of the masses of the two quark fields and $\tilde b_A, c_A, c_A'$ are couplings which must be tuned to remove the $\mathcal{O}(a)$ corrections \cite{Bhattacharya_2006}.  However, only $b_A$ is required for the $\mathcal{O}(a)$ improvement of the hadronic matrix element considered in this work.

This work studies the antisymmetric correlator $R^{[\mu\nu]}(\tau,\mathbf{p},\mathbf{q})$, which may be obtained by taking the antisymmetric combination of Eq.~(\ref{ratio}). For this specific matrix element, it is possible to show that the term proportional to $ac_A$ vanishes by symmetry (see Appendix~\ref{app:ca} for details). 
The terms proportional to $c_A'$ can also be shown to vanish; details of this argument are given in Appendix~\ref{app:ca-prime}.
This means that both renormalization and $\mathcal{O}(a)$ improvement of the current are, for the purposes of this work, purely multiplicative effects.  $Z_A^{(0)}$ is given in Ref.~\cite{Luscher_1997} and $\tilde{b}_A$ is given in Ref.~\cite{Bhattacharya_2006}.  However, in the time-momentum representation analysis procedure described below, any multiplicative factors only affect the fitted pion decay constant $f_\pi$ and not the second moment $\langle \xi^2 \rangle$, and thus the second moment is independent of any uncertainties in these $\mathcal{O}(a)$ improvement parameters.  As a result, lattice artifacts in $\langle \xi^2 \rangle$ only enter at $\mathcal{O}(a^2)$ or higher order.
\subsection{Lattice Parameters}

Despite the $\mathcal{O}(a)$-improvement scheme described above, the method in this work requires very fine lattice spacings because of the large mass used for the intermediate heavy quark propagator.
With renormalized heavy quark masses ranging from about 2 GeV to 4.5 GeV, lattice spacings between 0.08 fm and 0.04 fm are needed to keep $am_\Psi \lesssim 1$.  At larger values of $am_\Psi$, lattice artifacts were uncontrolled and could not be reliably removed.

Due to critical slowing down, generating dynamical configurations at such fine spacings is expensive and beyond the scope of this preliminary study.  As a result, this analysis uses gauge configurations generated in the quenched approximation following the multiscale procedure of Ref.~\cite{Detmold:2018zgk}.  
The lattice spacings for the ensembles with $L/a = $ 24, 32, and 48 had previously been determined in Ref.~\cite{Detmold:2018zgk} using Wilson flow with a reference scale of $w_{0.4} = 0.193$ fm \cite{asakawa2015determination}.  This scale-setting procedure was repeated for the ensemble with $L/a = 40$.
The lattice geometries were tuned to a constant physical volume of 1.92 fm, which was kept small to reduce computational costs.  

Finite volume effects were suppressed by using light quark masses tuned to give $m_\pi \approx 550$ MeV so that $m_\pi L \approx 5.3$.  The heavy quark masses were chosen to give approximately constant masses of the heavy-heavy pseudoscalar meson across the four lattices. 
Details on the lattices and quark masses used are listed in Table \ref{lattices-used}.
The required 2- and 3-point functions were generated using the software package \textsc{Chroma} with the \textsc{QPhiX} inverters \cite{chroma, qphix}.

\begin{table}
  \begin{tabular}{ c c c c c c c c c } \hline \hline
    $(L/a)^3 \times T/a$ & $\beta$ & $a$ (fm) & $\kappa_\text{light}$ & $\kappa_\text{heavy}$ & $c_\text{sw}$ & Configurations Used & Sources/Config & Total Sources Used \\ \hline
    $24^3 \times 48$ & ~6.10050~ & 0.0813 & ~0.134900~ & $\begin{matrix} 0.1200 \\ 0.1100 \end{matrix}$ & ~1.6842~ & 650 & 12 & 7800 \\ \hline
    $32^3 \times 64$ & 6.30168 & 0.0600 & 0.135154 & $\begin{matrix} 0.1250 \\ 0.1184 \\ 0.1095 \end{matrix}$ & 1.5792 & 450 & 10 & 4500 \\ \hline
    $40^3 \times 80$ & 6.43306 & 0.0502 & 0.135145 & $\begin{matrix} 0.1270 \\ 0.1217 \\ 0.1150 \\ 0.1088 \end{matrix}$ & 1.5292 & 250 & 6 & 1500 \\ \hline
    $48^3 \times 96$ & 6.59773 & 0.0407 & 0.135027 & $\begin{matrix} 0.1285 \\ 0.1244 \\ 0.1192 \\ 0.1150 \\ 0.1100 \end{matrix}$ & 1.4797 & 341 & 10 & 3410 \\ \hline
  \end{tabular}
  \caption{Details of the ensembles used in this numerical study.}
  \label{lattices-used}
\end{table}

\begin{figure}
  \includegraphics[scale=0.5]{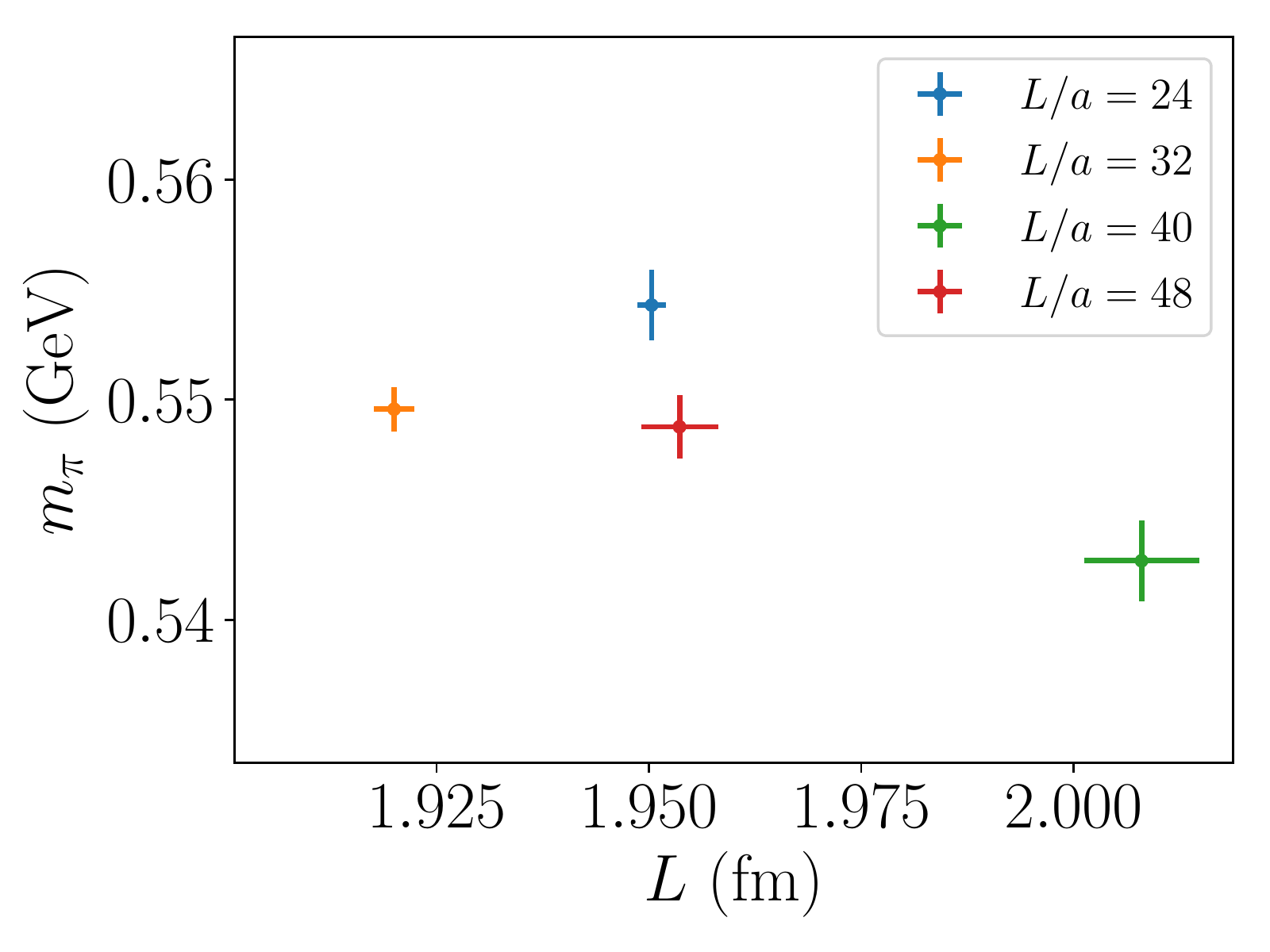}
\caption{A comparison of tuning of pion mass and lattice spatial extent $L$ across the used ensembles.}\label{fig:tuning}
\end{figure}

\subsection{Choice of Heavy-Quark Masses}

The operator product expansion in Eq.~(\ref{eq:2nd_Mellin_OPE_had_amp_target_mass}) will require higher-twist corrections that scale like $\Lambda_\text{QCD} / \tilde{Q}$ or $m_\pi / \tilde{Q}$.  Unlike light-quark operator product expansions that rely on large momenta or small distances to suppress higher-twist effects~\cite{Bali_2018}, this work relies on the heavy intermediate mass for this suppression.  With $m_\Psi \gtrsim q_4 \gg |\mathbf{q}|$, higher-twist effects will scale as $\Lambda_\text{QCD}/m_\Psi$ or $m_\pi/m_\Psi$, so
this analysis requires $\Lambda_\text{QCD}, m_\pi \ll m_\Psi$.  Separately, lattice artifacts enter as powers of either $a \Lambda_\text{QCD}$ or $a m_\Psi$.  Suppressing these will require $a m_\Psi \lesssim 1$, so altogether it is required that
\beq
\label{eq:scale_hierarchy}
\Lambda_\text{QCD}, m_\pi \ll m_\Psi \lesssim a^{-1} \, .
\eeq

\begin{figure}[h]
  \centering
  \includegraphics[scale=0.5]{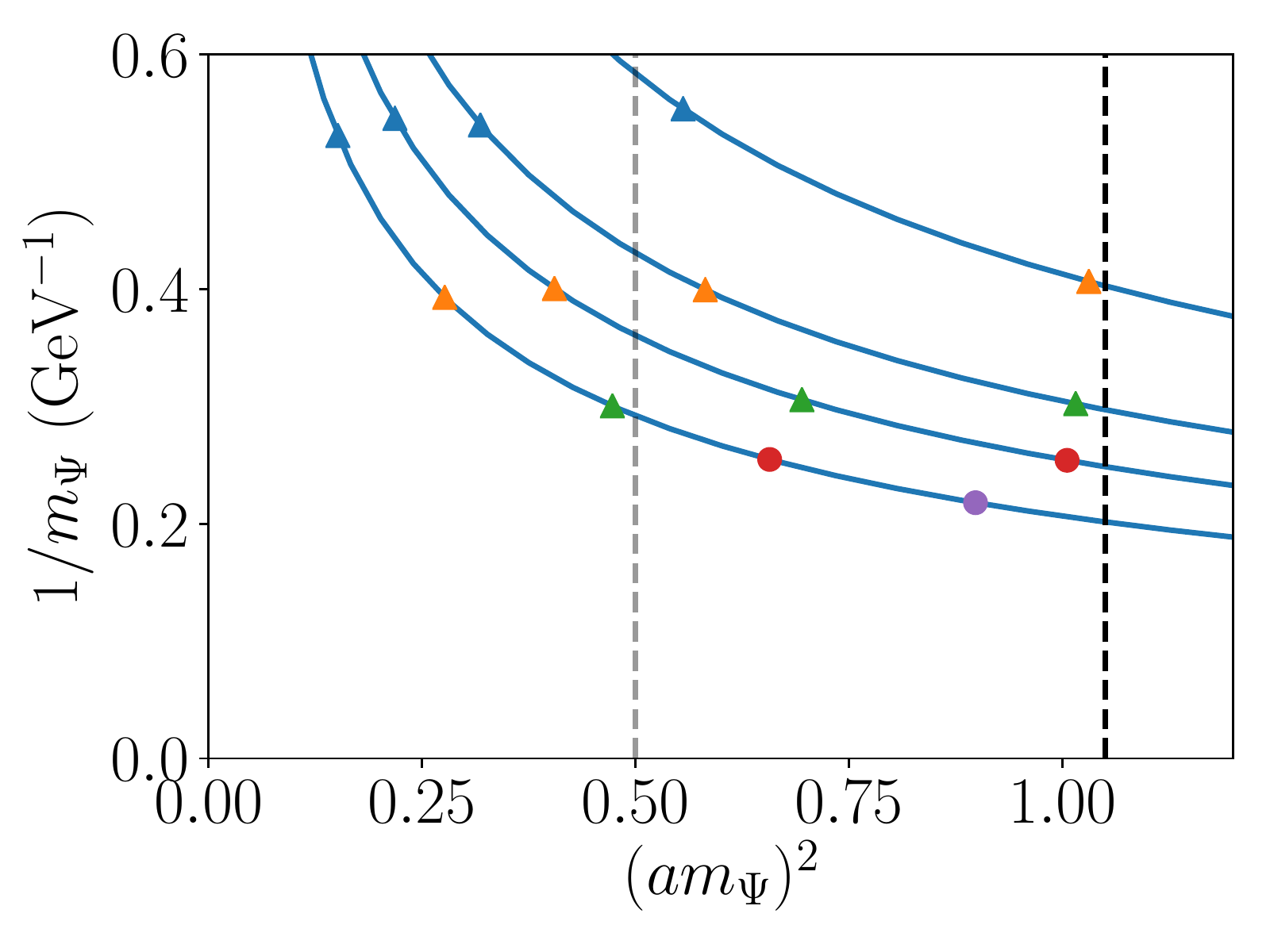}
  \caption{The four lattice spacings and the heavy quark masses used.  The plot shows the trade-off between discretization effects (which can depend on $(am_\Psi)^2$) and higher-twist effects (which scale as $1/m_\Psi$).  At fixed lattice spacing, one can decrease the higher twist effects at the cost of increasing discretization errors, and the available trade-offs at the four lattice spacings studied here are shown by the blue curves.  The coloured points show the masses actually used in this study.  The black dashed line at $am_\Psi = 1.05$ shows the cutoff beyond which discretization effects were no longer found to be well controlled, and the gray line at $am_\Psi = 0.7$ shows a more conservative threshold used for analysing systematic errors arising from lattice artifacts.}
  \label{fig:masses-used}
\end{figure}

With $\Lambda_\text{QCD} \sim 250$ MeV and $m_\pi \approx 550$ MeV, $m_\Psi > 1.8$ GeV provides some suppression of higher twist effects.  To fit the residual higher-twist effects away will require a range of heavy quark masses, and therefore this analysis will consider $m_\Psi$ as large as 4.5 GeV.  The lattice spacing is small enough for the finest discretization to accommodate such a heavy mass while maintaining $a m_\Psi < 1$; coarser lattices out of necessity have a smaller range of heavy quark masses.  Figure \ref{fig:masses-used} shows the quark masses used on each of the four ensembles considered here.

\subsection{Choice of Kinematics}

The heavy-quark OPE is given by
\begin{equation}
V^{[\mu\nu]} = -\frac{2if_\pi \varepsilon_{\mu\nu\rho\lambda}q^\rho p^\lambda}{\tilde{Q}^2}\left[ C_W^{(0)} + \langle \xi^2 \rangle \tilde{\omega}^2 \left( 1 - \frac{p^2 q^2}{6(p\cdot q)^2} \right) C_W^{\left( 2 \right)}+ \cdots +\mathcal{O}\left( \frac{\Lambda_\text{QCD}}{\tilde{Q}} \right) \right] \, ,
\label{ope-second-order}
\end{equation}
where $p$ is the momentum of the incoming pion and $q$ is the difference in momenta between the two outgoing currents, and where the ellipsis represents the contributions of higher moments that are negligibly small in this analysis.
The exact form of the higher-twist effects suppressed by $\Lambda/\tilde{Q}$ is not known, but symmetries (see Appendix~\ref{app:lor_invar_decomp}) constrain it to be proportional to $\varepsilon_{\mu\nu\rho\lambda}q^\rho p^\lambda$.  

\begin{figure}
\includegraphics[scale=0.5]{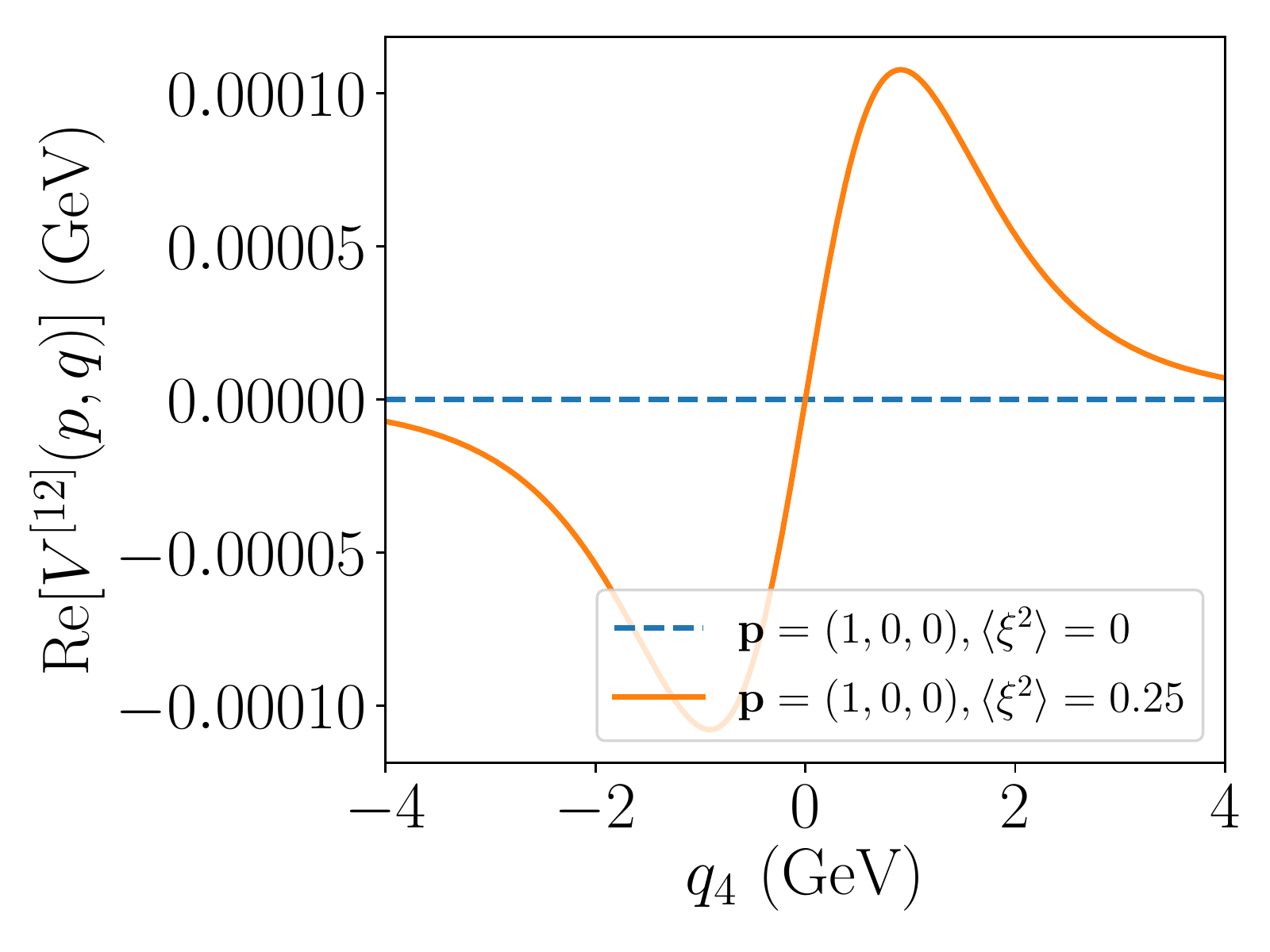}
\includegraphics[scale=0.5]{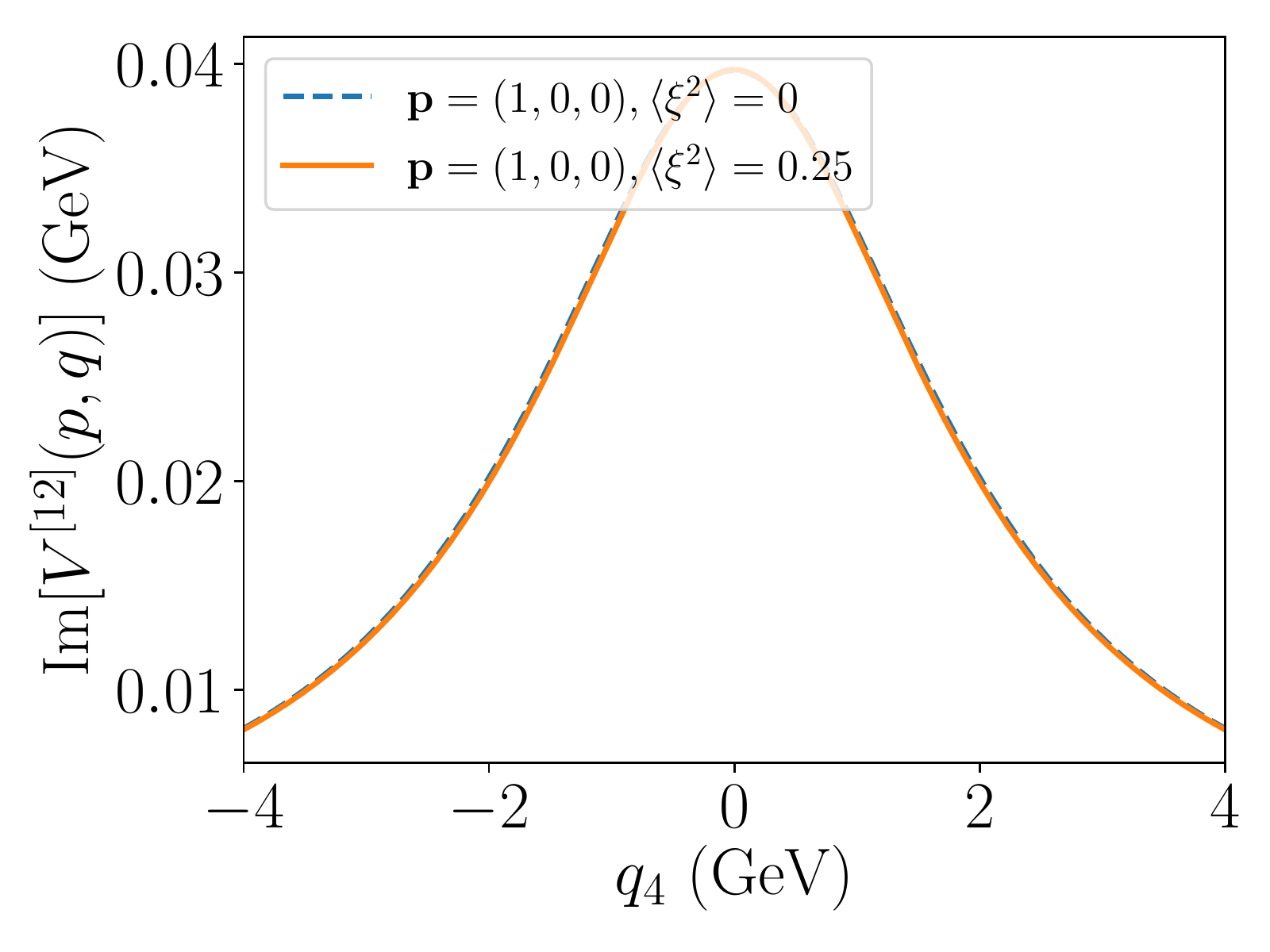}
\caption{Investigation of optimal kinematics for the HOPE procedure. Feasible lattice simulations are restricted to small lattice momentum, and thus for the HOPE method, small $\tilde{\omega}$. Generally speaking, this results in moments beyond the leading zeroth moment being highly suppressed. This behaviour may be seen in the imaginary part of the amplitude where the variation of $\expval{\xi^2}$ has minimal effect on the matrix element. By choosing $\hat{\mathbf{p}}=(1,0,0)$ and $\hat{\mathbf{q}}=(1/2,0,1)$, it is possible to show that the leading contribution to the real component of the amplitude arises from $\expval{\xi^2}$. Thus this choice of kinematics offers improved access to the second Mellin moment.
}
\label{fig:special-kinematics}
\end{figure}

In order to enhance the contribution of the second moment, one would like its prefactor to be as large as possible.  However, $\tilde{Q}^2$ must be large to suppress higher twist effects, and $\mathbf{p}$ is limited by noise that grows exponentially with the pion energy on the lattice.  In this work, $\mathbf{\hat p} \equiv \frac{\mathbf{p}}{2\pi/L}$ was constrained to be one unit of momentum, which for the volumes used in this work corresponds to $|\mathbf{p}| = $ 640 MeV.  At these kinematics, the second moment is a small contribution to the hadronic tensor.
As such, it is desirable to isolate its effect from the much larger contribution of the zeroth moment.  In this study, the axial current indices are fixed to be $\mu=1, \nu=2$, the prefactor on the right-hand side of Eq.~(\ref{ope-second-order}) becomes
\beq
\label{eq:prefactor_our_kinematics}
i\varepsilon_{\mu\nu\rho\sigma}q^\rho p^\sigma = i\left(q^0 p^3 - p^0 q^3\right) = - q^4 p^3 - i E_\pi q^3 \, .
\eeq
If the kinematics are chosen such that $p^3 = 0$, then this prefactor is purely imaginary.  At tree level, the entire contribution of the zeroth moment will be pure imaginary as well.  However, $p\cdot q = iE_\pi q^4 - \mathbf{p}\cdot \mathbf{q}$ is generically complex (as long as $\mathbf{p}\cdot \mathbf{q} \neq 0$), so the contribution of the second moment to the hadronic tensor will have nonzero real part. The effect of these special kinematics is shown in Fig.~\ref{fig:special-kinematics}. This work met these criteria by choosing 
\begin{equation}
\mathbf{\hat p} = (1,0,0)~~\text{and}~~\mathbf{\hat q} = (1/2,0,1)
\end{equation}
in units of $2\pi/L$, as well as all combinations of $\mathbf{\hat p}, \mathbf{\hat q}$ that are equivalent under lattice symmetries.  With these choices, $\langle \xi^2 \rangle$ can be extracted as the leading contribution in the real part of $V^{\mu\nu}$.  (Note that $\mathbf{\hat q} = (\mathbf{\hat p}_m - \mathbf{\hat p}_e)/2$, so it is quantized in half-integers rather than integers.)

There are two caveats to this argument:
\begin{enumerate}
  \item The Wilson coefficient of the zeroth moment $C_W^{(0)}$, while real at tree level, becomes complex at 1-loop order and also contributes to the real part of $V^{\mu\nu}$.  However, this contribution is suppressed by $\alpha_s$, and it is known analytically, so this small correction can be subtracted.
  \item Higher-twist contributions may also be complex, but their contribution to $\text{Re}[V^{\mu\nu}]$ must also contain powers of $(p\cdot q)^2/(\tilde{Q}^2)^2$, just like the second moment contribution.  They are further suppressed by $\Lambda_\text{QCD}/\tilde{Q}$, so it is expected that they are smaller than the second moment contribution, particularly for large $\tilde{Q}^2$.
\end{enumerate}
However, assuming the second Mellin moment is $\langle \xi^2 \rangle \sim 0.25$, both of the above contributions are subdominant at the special kinematics chosen here.

\subsection{Computing Real and Imaginary Parts of $V^{\mu\nu}$}
To compute the hadronic tensor, values of $R^{\mu\nu}(\tau) \propto C_3^{\mu\nu}(\tau_e, \tau_m = \tau_e + \tau)$ are needed at both positive and negative values of $\tau$.  In particular, using the fact that $R^{\mu\nu}(\tau)$ is pure imaginary\footnote{In terms of the Minkowski 4-momentum $q = (q_0, \mathbf{q})$, the hadronic tensor is pure imaginary and can be related to $R^{\mu\nu}(\tau)$ via Laplace transform, which has a purely real kernel.}, the imaginary and real parts of $V^{\mu\nu}$ can be written in terms of symmetric and antisymmetric combinations of $R^{\mu\nu}(\pm \tau)$:

\begin{align}
  \text{Re}[V^{\mu\nu}(\mathbf{p}, q)] &= \int_0^\infty d\tau \, \left[ R^{\mu\nu}(\tau; \mathbf{p}, \mathbf{q}) - R^{\mu\nu}(-\tau; \mathbf{p}, \mathbf{q}) \right] \sin (q_4 \tau) \, ,
  \label{antisymmetry-R} \\ 
  \text{Im}[V^{\mu\nu}(\mathbf{p}, q)] &= \int_0^\infty d\tau \, \left[ R^{\mu\nu}(\tau; \mathbf{p}, \mathbf{q}) + R^{\mu\nu}(-\tau; \mathbf{p}, \mathbf{q}) \right] \cos (q_4 \tau) \, .
  \label{symmetry-R}
\end{align}

At the kinematics of interest, the real part of $V^{\mu\nu}$ is about two orders of magnitude smaller than the imaginary part, so computing the difference in Eq.~(\ref{antisymmetry-R}) requires a delicate cancellation between $R^{\mu\nu}(\pm \tau)$.  The computation becomes more tractable if the two terms are highly correlated, as this increases the statistical power of the correlated difference.  These correlations are substantially enhanced if values of $C_3^{\mu\nu}$ for $\tau < 0$ are obtained using the identity\footnote{
This identity can be proven by writing $C_3^{\mu\nu}$ in terms of the quark propagators
\begin{equation*}
  C_3^{\mu\nu}(\tau_e, \tau_m; \mathbf{p}_e, \mathbf{p}_m) = \int d^3 \mathbf{x}_e \, d^3 \mathbf{x}_m \, e^{i\mathbf{p}_e \cdot \mathbf{x}_e + i \mathbf{p}_m \cdot \mathbf{x}_m} \text{Tr}\left[ \gamma_5 D^{-1}_\psi(0|x_m) \gamma_5 \gamma_\nu D^{-1}_\Psi (x_m|x_e) \gamma_5 \gamma_\mu D^{-1}_\psi (x_e|0) \right] \, ,
\end{equation*}
and applying $\gamma_5$-hermiticity to each of the propagators.
}
\begin{equation}
  C_3^{\mu\nu}(\tau_e, \tau_m; \mathbf{p}_e, \mathbf{p}_m)^* = C_3^{\nu\mu}(\tau_m, \tau_e; -\mathbf{p}_m, -\mathbf{p}_e) \, .
\label{c3-identity}
\end{equation}
Then Eq.~(\ref{antisymmetry-R}) and (\ref{symmetry-R}) can be written as
\begin{align}
  \text{Re}[V^{\mu\nu}(\mathbf{p}, q)] &= \int_0^\infty d\tau \, \left[ R^{\mu\nu}(\tau; \mathbf{p}, \mathbf{q}) + R^{\mu\nu}(\tau; -\mathbf{p}, \mathbf{q}) \right] \sin (q_4 \tau) \, , \label{antisymmetry-R-improved} \\
  \text{Im}[V^{\mu\nu}(\mathbf{p}, q)] &= \int_0^\infty d\tau \, \left[ R^{\mu\nu}(\tau; \mathbf{p}, \mathbf{q}) - R^{\mu\nu}(\tau; -\mathbf{p}, \mathbf{q}) \right] \cos (q_4 \tau) \, . \label{symmetry-R-improved}
\end{align}

\begin{figure}
  \centering
  \includegraphics[scale=0.5]{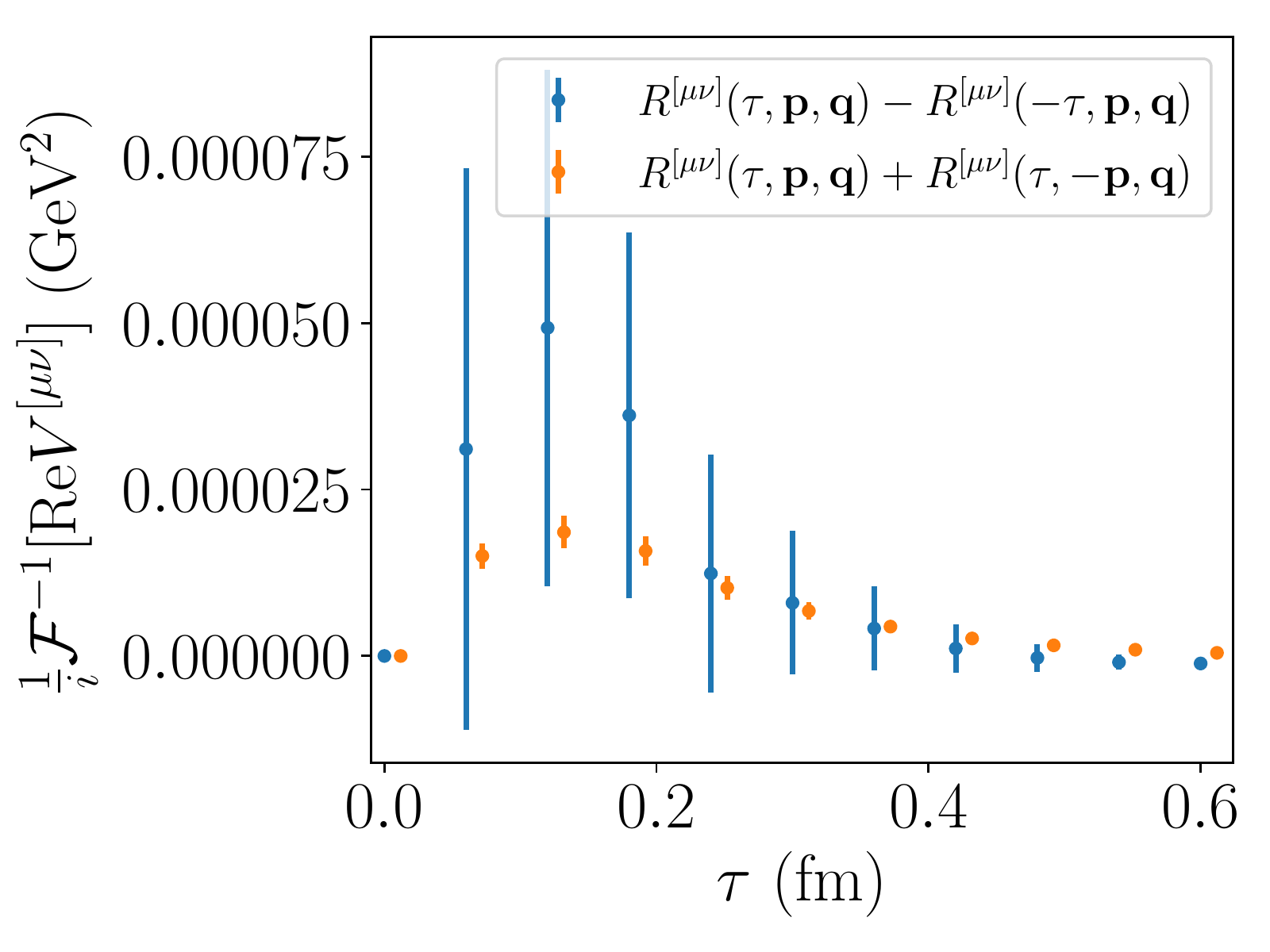}
  \caption{A comparison of the real part of the hadronic tensor computed using Eq.~(\ref{antisymmetry-R}) versus Eq.~(\ref{antisymmetry-R-improved}), the latter of which has been manipulated to reduce the statistical error.  While the manipulations do not change the expectation of the correlator, they reduce statistical uncertainties by about an order of magnitude.  For comparison purposes, both quantities were measured on 2 sources on each of 450 configurations.}
  \label{error-cancellation}
\end{figure}

Consequently, one can obtain both $\tau > 0$ and $\tau < 0$ at the same sets of current insertion times, which will enhance the correlations.  A demonstration of this reduction in statistical error is shown in Figure~\ref{error-cancellation}.

\subsection{Excited State Contamination and Choice of $\tau_e$}
\label{sec:excited-states}

\begin{figure}
\subfloat[][]{
\includegraphics[scale=0.5]{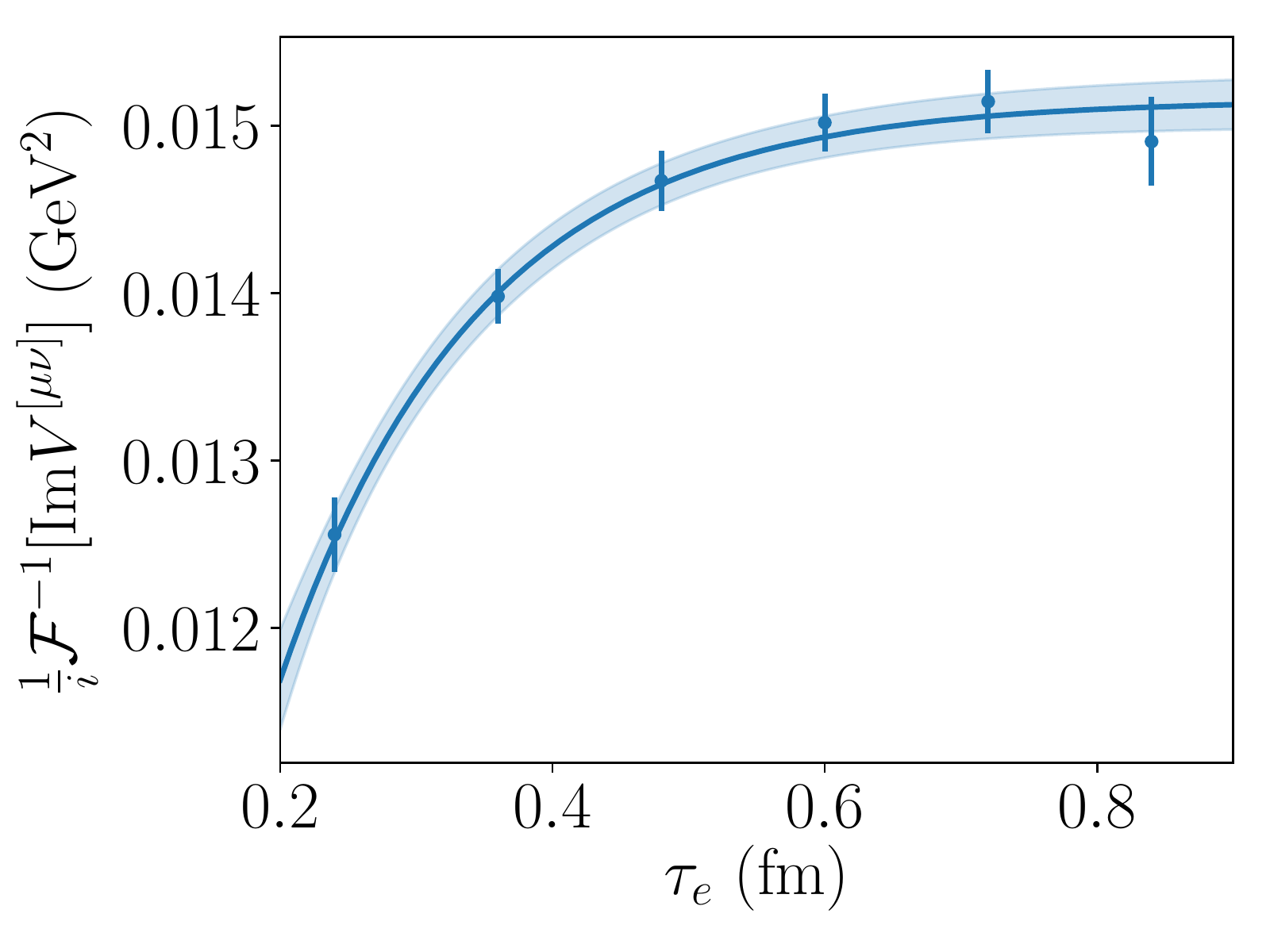}
}
\hspace{20pt}
\subfloat[][]{
\includegraphics[scale=0.5]{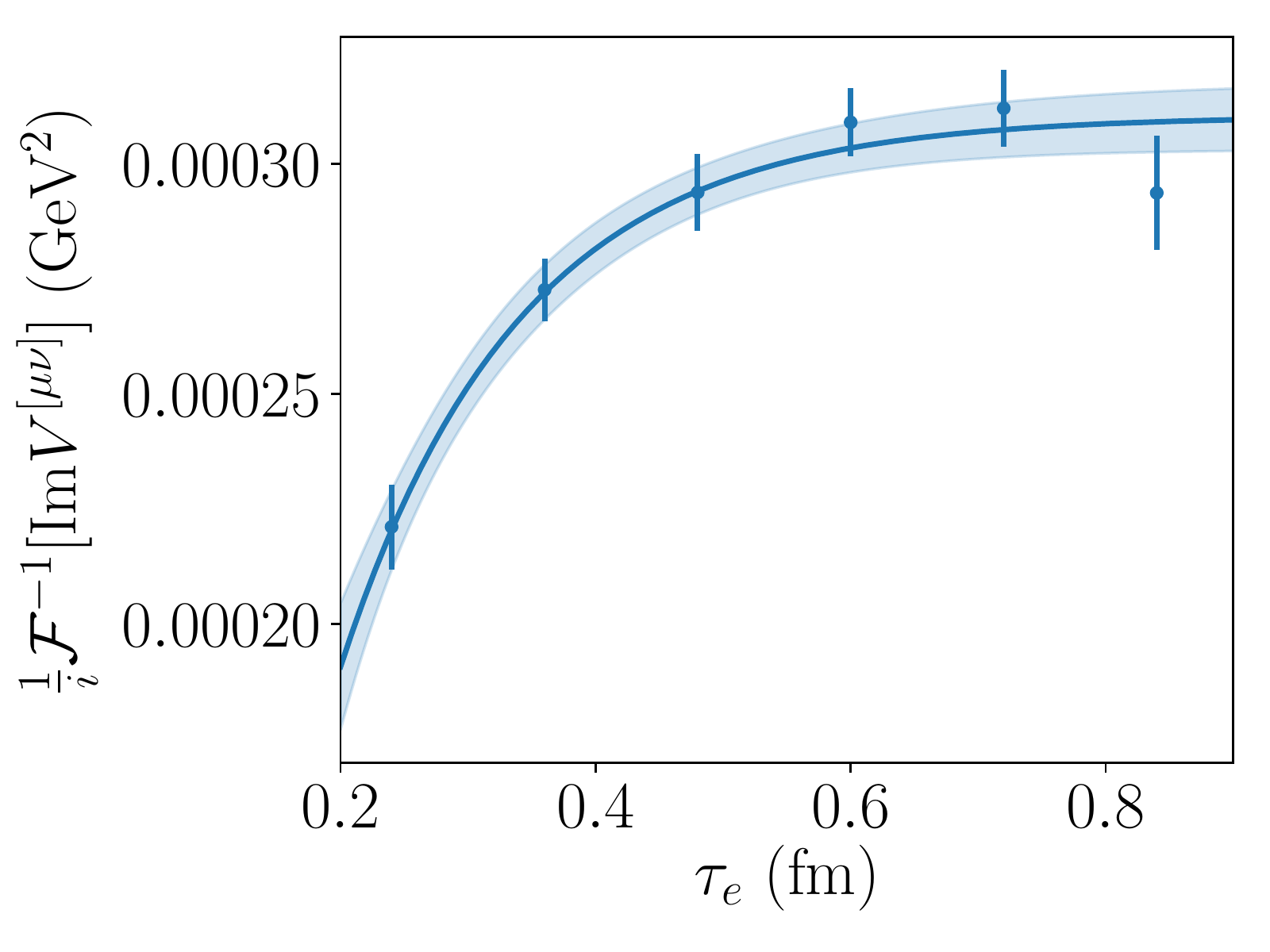}
}
\caption{Excited state contamination at operator separations (a) $\tau = 0.06$ fm and (b) $\tau = 0.36$ fm.}
  \label{fig:excited-states}
\end{figure}

The three point correlator $C_3^{\mu\nu} (\tau_e,\tau_m) = \langle J_A^\mu(\tau_e) J_A^\nu (\tau_m) \mathcal{O}_\pi^\dagger(0) \rangle$ is computed by creating a pion source, propagating one of the quarks forward to $\tau_e$, creating a sequential source, and then tying together the sequential heavy-quark propagator and the other light quark propagator at the sink.  Since $\tau_e \leq \tau_m$ is chosen in this work, excited state effects arise from the fact that the combination of states created by the pion interpolator have not fully relaxed to the ground state before $\tau_e$, so they are suppressed exponentially in $\tau_e$.  Excited-state effects are reduced by using a Gaussian-smeared pion source~\cite{GUSKEN1990361} with smearing radius equal to the inverse pion mass ($aw_\text{smear} = \left\{ 4.5, 6.0, 8.0, 9.0 \right\}$ for $L/a = \left\{ 24, 32, 40, 48 \right\}$, respectively).  With this smearing, numerical study on the $L/a = 32$ lattices showed that excited state contamination is estimated to be about 1\% for a source-operator separation $\tau_e$ of about 0.7 fm, as shown in Figure~\ref{fig:excited-states}.

\begin{figure}
\centering
\includegraphics[scale=0.7]{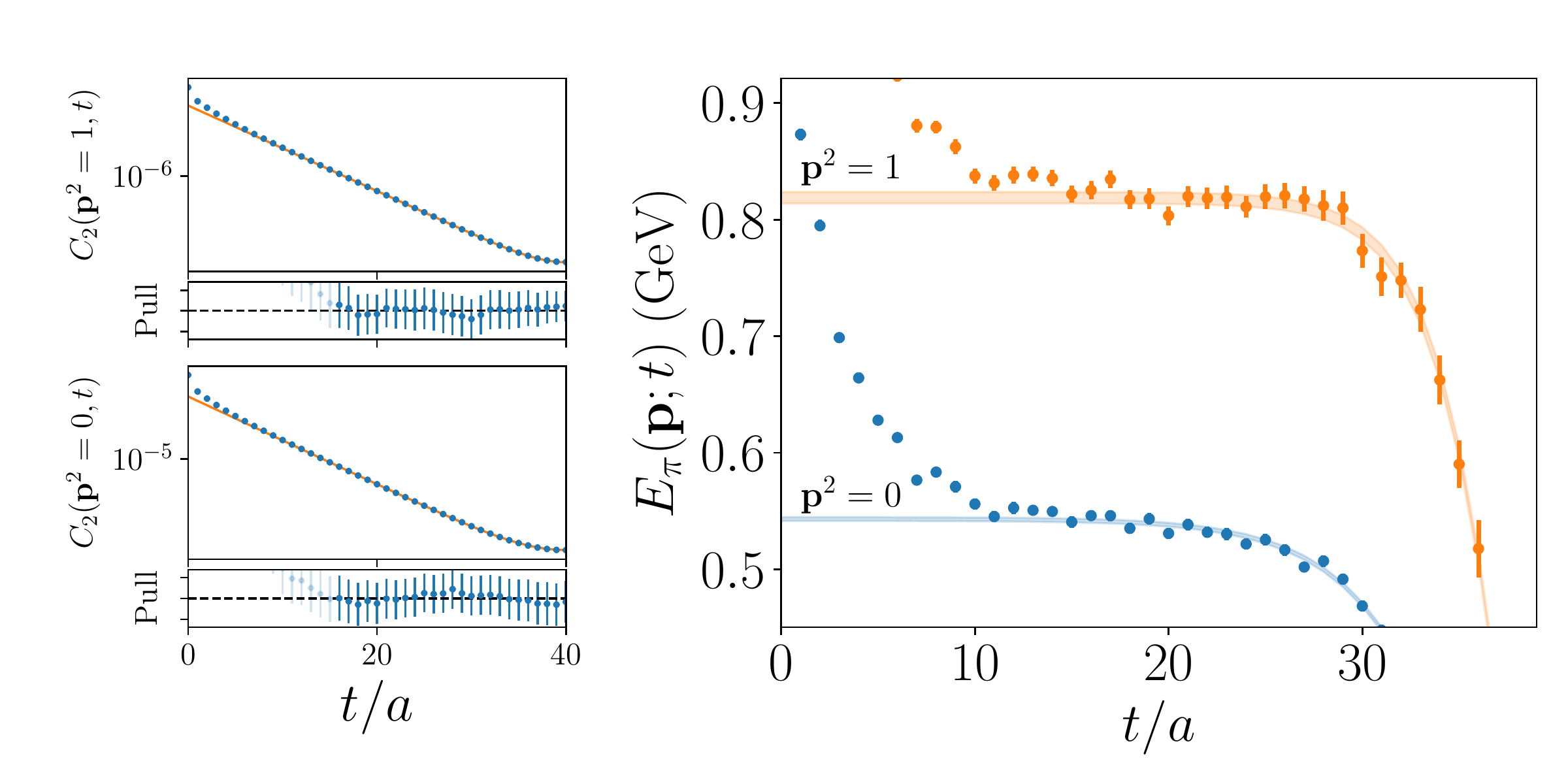}
\caption{Validation plots for the single-state \textit{ansatz} used in this study. The left panels show fits to the $L/a=40$ pseudoscalar correlator for $|\mathbf{p}|=1$  and $|\mathbf{p}|=0$, respectively, while the right panel shows the resulting effective mass plateaus.  The ``pull'' in the left-hand plot is the residual of each data point normalized to the its standard deviation and is a measure of that point's effect on the fit.}\label{fig:2pt_fit}
\end{figure}

Since $\tau_e$ must be fixed at runtime, $\tau_e \sim 0.7$ fm is chosen, leading to  a $\sim 1\%$ systematic error due to excited-state contamination.  Excited state contamination in the 2-point function is better controlled since one does not need to choose the source-sink separation at runtime, and one can afford the very conservative fit range of $\left[ T/4, 3T/4 \right]$ since the statistical errors on the 2-point function are smaller than those on the 3-point function (for an example of the goodness of fit, see Fig.~\ref{fig:2pt_fit}).
\section{Analysis, results and discussion}
\label{sec:analysis}

Extraction of the second moment from the 2- and 3-point correlation functions measured here is nontrivial due to signal contamination by both lattice artifacts and higher-twist effects.  The extraction of $Z_\pi(\mathbf{p})$, $E_\pi(\mathbf{p})$ from fitting the 2-point correlation function $C_2(\tau, \mathbf{p})$ at late $\tau$ and the construction of $R^{\mu\nu}(\tau)$ in Eq.~(\ref{ratio}) are relatively straightforward.  However, comparing the 3-point data to the OPE of the hadronic amplitude can be done in multiple ways, which can lead to somewhat different systematics.  As this is the first numerical study of the HOPE method, two analysis methods, called the \textit{time-momentum analysis} and the \textit{momentum-space analysis}, are performed.  This enables a cross-check of the results and ensures that they are robust against systematics in the analysis procedure.  These analysis methods are as follows:
\begin{enumerate}
  \item Time-momentum analysis
  \begin{enumerate}[label=(\roman*)]
    \item Fit $f_\pi$ and the heavy quark mass $m_\Psi$ by comparing the symmetric part of the data to the inverse Fourier transform of the OPE of $\text{Im}[V(\mathbf{p},q)]$, that is, from the inverse Fourier transform of Eq.~(\ref{symmetry-R-improved}).
    \item Use the fitted results of $f_\pi, m_\Psi$ and the antisymmetric part of the data to fit the second moment $\langle \xi^2 \rangle$ from the inverse Fourier transform of $\text{Re}[V(\mathbf{p},q)]$, using Eq.~(\ref{antisymmetry-R-improved}), at each heavy quark mass and lattice spacing.
    \item Perform a combined fit to $\langle \xi^2 \rangle(a, m_\Psi)$ to remove both lattice spacing and higher-twist effects.
  \end{enumerate}
  \item Momentum-space analysis
  \begin{enumerate}[label=(\roman*)]
    \item Perform a Fourier transform of $R^{\mu\nu}(\tau)$ in the temporal direction.
    \item Extrapolate the momentum-space hadronic amplitude to the continuum.
    \item Fit $f_\pi$, $m_\Psi$ and $\expval{\xi^2}$ to the hadronic amplitude in the continuum limit using the momentum-space, continuum HOPE formula presented in Sec.~\ref{sec:HOPE_strategy}. 
  \end{enumerate}
\end{enumerate}

\begin{figure}
\centering
\tikzstyle{decision} = [diamond, draw, fill=blue!20, 
     text width=4.5em, text badly centered, node distance=3cm, inner sep=0pt]
\tikzstyle{block} = [rectangle, draw, 
     text width=20em, text centered, rounded corners, minimum height=6em] 
\tikzstyle{smallblock} = [rectangle, draw, 
     text width=10em, text centered, rounded corners, minimum height=4em] 
\tikzstyle{line} = [draw, -latex']
\tikzstyle{cloud} = [draw, ellipse,fill=red!20, node distance=3cm,
     minimum height=2em]
\begin{tikzpicture}
\node [block] at (0,0){Construct ratio from bare correlators:\\
$
R^{\mu\nu}(\tau,\mathbf{p},\mathbf{q};a)=\int d^3z e^{i\mathbf{q}\cdot \mathbf{z}}\bra{\Omega} |\mathcal{T}\{J_A^\mu(z/2)J_A^\nu(-z/2)\}\ket{\pi(\mathbf{p})}
$
};
\node [block] at (4,-3.5){Perform Fourier transform to momentum space:\\
$
V^{\mu\nu}(p,q;a)=a\sum_{\tau} e^{i\tau q_4}R^{\mu\nu}(\tau,\mathbf{p},\mathbf{q};a)
$
};
\node [block] at (4,-6){Extrapolate data to continuum:\\$V^{\mu\nu}(p,q;a)=V^{\mu\nu}(p,q)+a^2V_{(2)}^{\mu\nu}(p,q)$};
\node [block] at (4,-8.5){Fit continuum matrix element to HOPE formula:\\$V^{\mu\nu}(p,q)=\frac{2if_\pi\epsilon^{\mu\nu\alpha\beta}q_\alpha p_\beta}{\tilde{Q}^2}\times \sum_{\substack{n=0\\\text{even}}} \frac{\mathcal{C}_n^{(2)}(\eta)}{2^n(n+1)}C_W^{(n)}(\tilde{Q}^2/\mu^2)\expval{\xi^n}(\mu^2)\zeta^n$};

\node [block] at (-4,-3.5){Fit lattice matrix element to inverse Fourier transform of HOPE formula:\\$R^{\mu\nu}(\tau,\mathbf{p},\mathbf{q})=\int dq_4 \, e^{-iq_4 \tau}\frac{2if_\pi\epsilon^{\mu\nu\alpha\beta}q_\alpha p_\beta}{\tilde{Q}^2}\times \sum_{\substack{n=0\\\text{even}}} \frac{\mathcal{C}_n^{(2)}(\eta)}{2^n(n+1)}C_W^{(n)}(\tilde{Q}^2/\mu^2)\expval{\xi^n}(\mu^2)\zeta^n$};
\node [block] at (-4,-6){Extrapolate fixed lattice spacing parameters $f_\pi(a,m_\Psi)$, and $\expval{\xi^2}(a,m_\Psi)$ to continuum, $m_\Psi\to\infty$ limit.};
\path [line] (0,-1) -- (-4,-2.5);
\path [line] (-4,-4.5) -- (-4,-5);

\path [line] (0,-1) -- (4,-2.5);
\path [line] (4,-4.5) -- (4,-5);
\path [line] (4,-7) -- (4,-7.5);

\node[align=left] at (5,-1.8) {Momentum-space analysis};
\node[align=left] at (-5,-1.8) {Time-momentum analysis};

\end{tikzpicture}
\caption{Comparing the two analysis strategies. Both approaches utilize the same bare correlators, but note that the order of the continuum extrapolation and HOPE fit are reversed.}\label{fig:analysis_flow}
\end{figure}
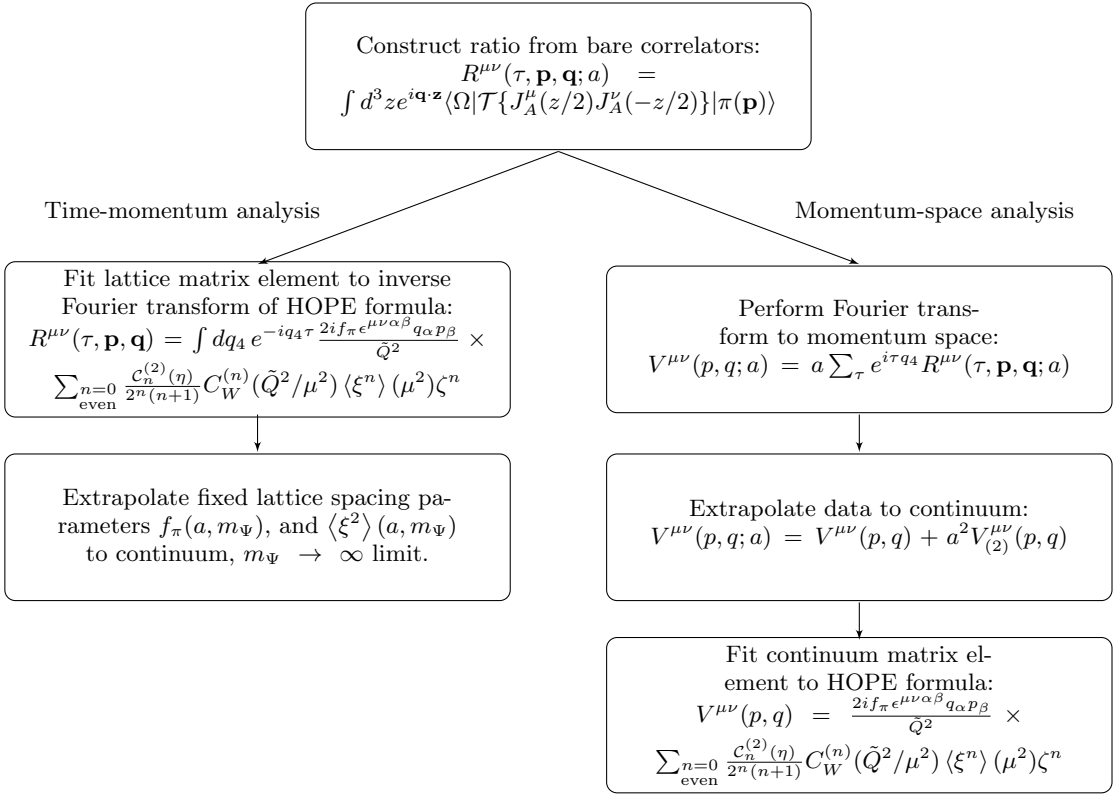

These alternative procedures are shown diagrammatically in Fig.~\ref{fig:analysis_flow}. The following sections will detail both analysis strategies. 

\subsection{Time-Momentum Analysis}
\label{time-momentum}

The ratio $R^{\mu\nu}\left( \tau; \mathbf{p}, \mathbf{q} \right)$ was
constructed for $0 < \tau \leq \tau_\text{max} \approx 0.6$ fm.
The statistical quality of the signal
deteriorates with time, and large-$\tau$ data may be more susceptible to
higher-twist contamination, motivating the cut at $\tau_\text{max}$.  The
symmetric and anti-symmetric components of $R^{\mu\nu}(\tau)$ are
constructed as described in Section II-B.

\begin{figure}
\subfloat[][]{
    \includegraphics[scale=0.5]{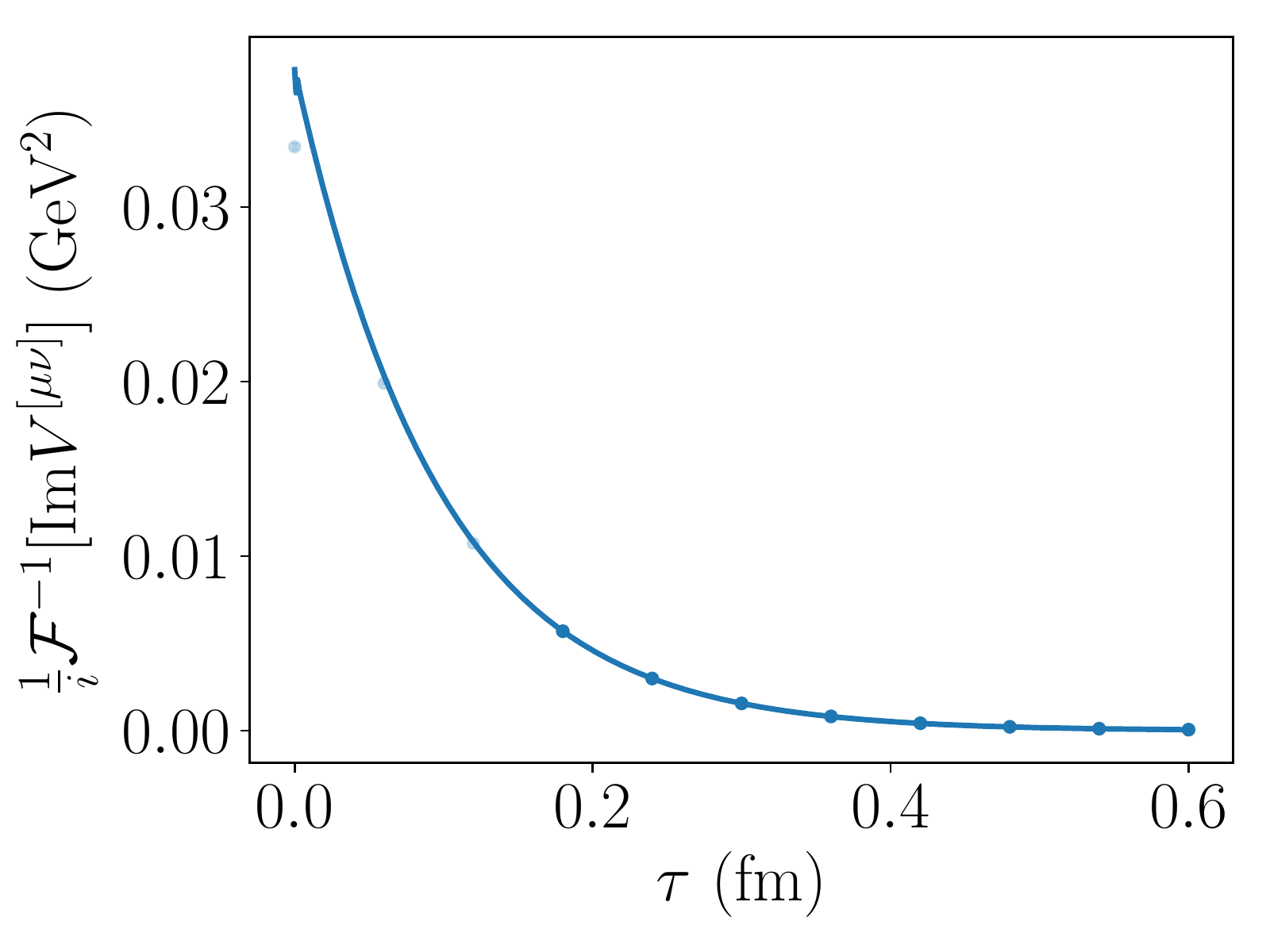}
}
\hspace{20pt}
\subfloat[][]{
    \includegraphics[scale=0.5]{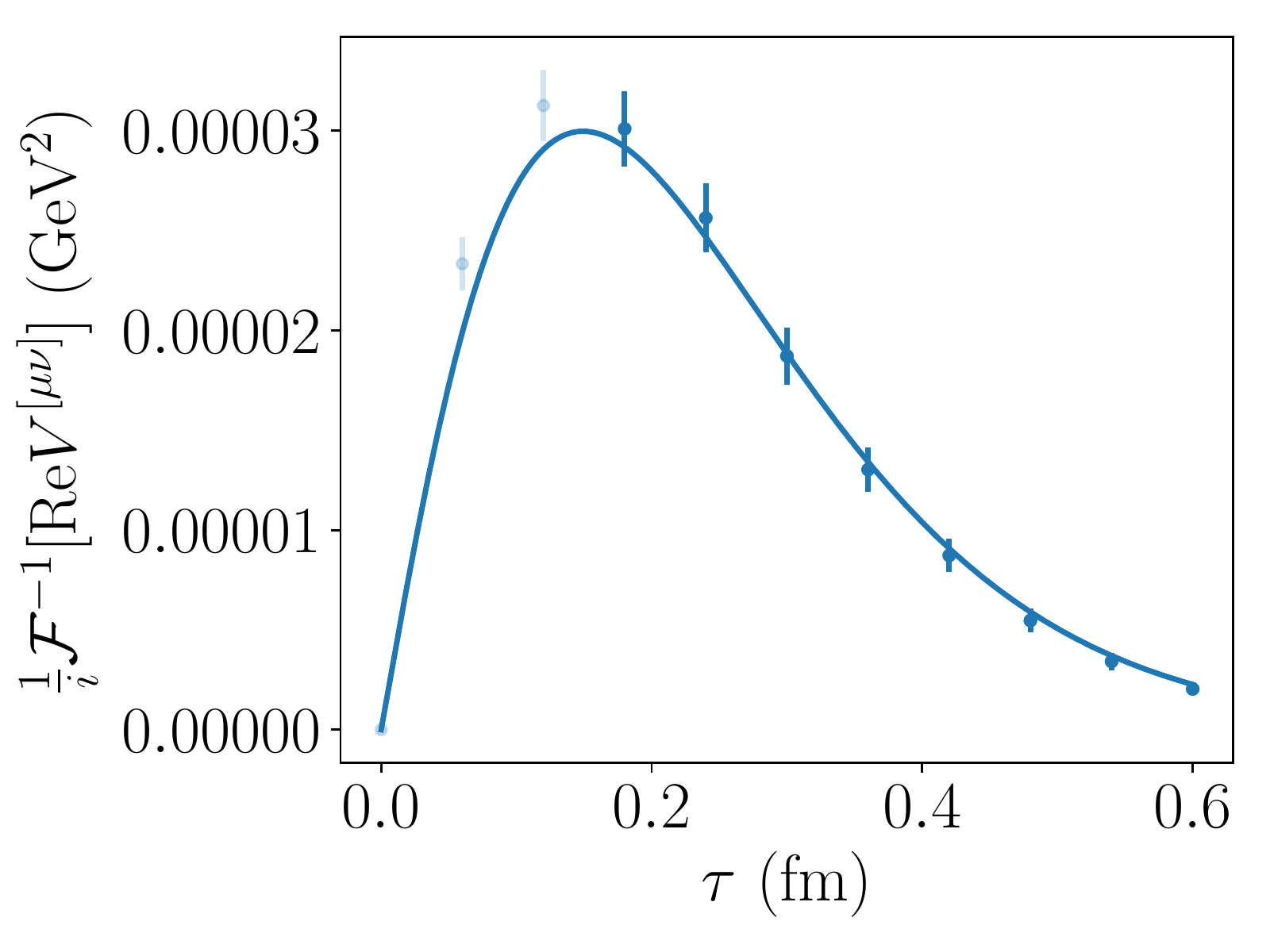}
}
\caption{The real and imaginary parts of the hadronic tensor $V^{\mu\nu}$ can be inverse Fourier transformed (indicated by $\mathcal{F}^{-1}$) to produce the antisymmetric and symmetric components of the ratio of correlators $R^{\mu\nu}$.  The symmetric part of $R^{\mu\nu}(\tau)$ (corresponding to $\mathcal{F}^{-1}[\text{Im}(V^{\mu\nu})]$) is dominated by the zeroth moment contribution, allowing extraction of $f_\pi$ and $m_\Psi$.  These are then used as inputs to the fit of $\mathcal{F}^{-1}[\text{Re}(V^{\mu\nu})]$ to extract the second moment $\langle \xi^2 \rangle$.  In order to avoid contamination with UV divergences near $\tau = 0$, points with $\tau < 3a$ (grayed out in the plots) are excluded from the fit.}
  \label{fig:single-mass}
\end{figure}

An example fit to $V^{\mu\nu}$ for a single heavy quark mass at a single lattice spacing is shown in Fig.~\ref{fig:single-mass}.  At the chosen kinematics, the second moment provides a negligible contribution to the imaginary part of the hadronic tensor (see Fig.~\ref{fig:special-kinematics}), so the fitting procedure can be split into two steps: one in which $f_\pi$ and $m_\Psi$ are fit to the imaginary part of $V^{\mu\nu}$ and a second step that consists of a single-parameter fit of $\langle \xi^2 \rangle$ to the real part of $V^{\mu\nu}$, where $f_\pi$ and $m_\Psi$ are used as inputs.
At values of $\tau$ comparable to the lattice spacing, uncontrolled discretization effects are to be expected.  Additionally, if the two current insertions are close in space-time relative to the lattice spacing, they may mix with lower-dimensional operators and lead to UV divergences.  Both of these effects suggest that small-$\tau$ data should be removed from the fits.
Empirically, the $\chi^2$ values for the fits to the various heavy-quark masses became reasonable if the $\tau \leq 2a$ data are excluded, so all fits will only use data with $\tau \geq 3a$.


\begin{figure}
  \centering
  \includegraphics[scale=0.5]{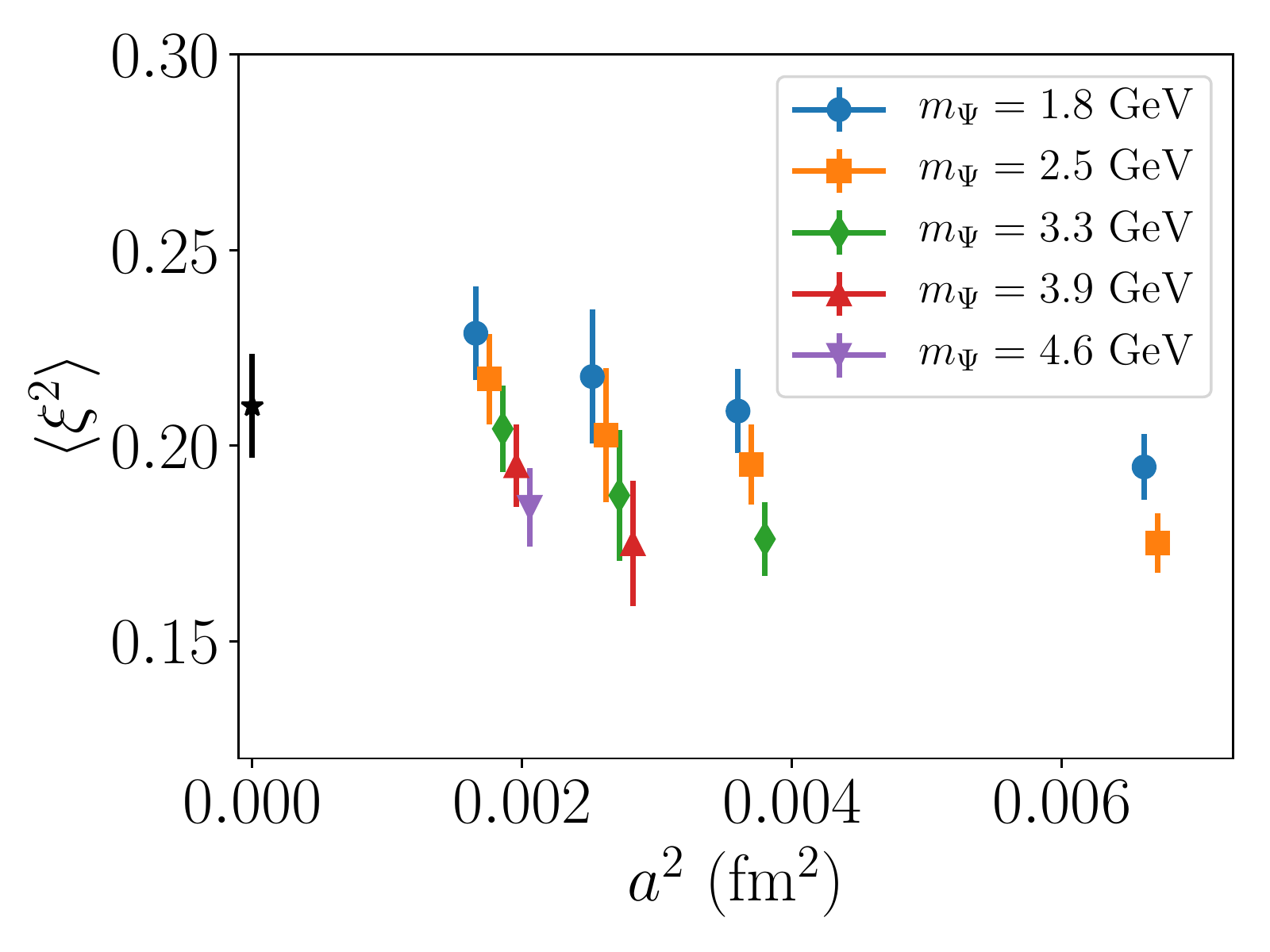}
\caption{The values of $\langle \xi^2 \rangle(a,m_\Psi)$ at the gauge couplings and heavy quark masses listed in Table \ref{lattices-used}, plotted as a function of $a^2$ with the heavier masses at each lattice spacing displaced slightly to the right for visual clarity.  The black star at $a^2 = 0$ represents the extrapolated value from the fit to Eq.~(\ref{eqn:global-fit}).}
  \label{fig:global-fit}
\end{figure}

This fitting procedure compares lattice data to a continuum, twist-2 OPE.  As a result, the extracted second moment $\langle \xi^2 \rangle (a, m_\Psi)$ will be contaminated by both lattice artifacts and higher-twist corrections.  The lattice artifacts enter at $O(a^2)$ (see Appendices~\ref{app:ca},~\ref{app:ca-prime} for details), and by dimensional analysis, $a^2$ must be accompanied by two factors of a mass scale, either the typical momentum scale of $\Lambda_\text{QCD}$ or the heavy quark mass $m_\Psi$, so there may be discretization effects proportional to $a^2$, $a^2 m_\Psi$, or $a^2 m_\Psi^2$.  With $a m_\Psi < 1.05$, these terms were sufficient to describe lattice artifacts without need for additional $O(a^3)$ terms.
Higher-twist effects scale as powers of $\Lambda_\text{QCD}/\tilde{Q}$ or $m_\pi/\tilde{Q}$, and $\Lambda_{\text{QCD}} \sim m_\pi$ in this analysis.  The fitting procedure effectively integrates over the $q_4$ dependence, and $m_\Psi \gg |\mathbf{q}|$, so the twist-3 contribution can be approximated by a $\Lambda/m_\Psi$ term.  Therefore, to extract $\langle \xi^2 \rangle$ in the continuum limit without higher-twist contamination, $\langle \xi^2 \rangle (a, m_\Psi)$ is fit to the formula
\begin{equation}
  \langle \xi^2 \rangle(a,m_\Psi) = \langle \xi^2 \rangle + \frac{A}{m_\Psi} + B a^2 + C a^2 m_\Psi + D a^2 m_\Psi^2 \, ,
  \label{eqn:global-fit}
\end{equation}
where $\langle \xi^2 \rangle$, $A$, $B$, $C$, and $D$ are the fit parameters.
At an intermediate mass scale of $m_\Psi = 3$ GeV and a lattice spacing of $a = 0.06$ fm, the magnitudes of the various terms are
\begin{align}
  \langle \xi^2 \rangle &= 0.210 \pm 0.013, \\
  \frac{A}{m_\Psi} &= 0.009 \pm 0.005, \\  
  B a^2 &= -0.004 \pm 0.013, \\  
  C a^2 m_\Psi &= -0.004 \pm 0.013, \\ 
  D a^2 m_\Psi^2 &= -0.027 \pm 0.006, 
\end{align}
where the renormalization scale for $\langle \xi^2 \rangle$ is taken to be $\mu = 2$ GeV and the error bars are purely statistical.  Neither the higher-twist nor the discretization effects can be neglected at the precision considered in this work.
The fit result for $\langle \xi^2 \rangle$ is shown in Figure \ref{fig:global-fit}.

\subsection{Estimation of Systematic Uncertainties for Time-Momentum Analysis}
The analysis procedure described in the previous subsection contains several systematic errors.
Excited state contamination in the 3-point function was estimated in Sec.~\ref{sec:excited-states} to be a $\sim 1\%$ effect (and contamination in the 2-point function is much smaller).  Finite-volume effects are expected to scale as $\frac{1}{m_\pi L}e^{-m_\pi L} < 10^{-3}$ and are negligible compared to both statistical and other systematic errors.

This work uses an unphysically heavy pion mass of $m_\pi \sim 550$ MeV.
Previous studies \cite{Braun_2015} have indicated that $\langle \xi^2 \rangle$ at such a pion mass differs from its physical value by about $5\%$.  Therefore, this is taken as a systematic effect arising from the unphysical pion mass.

Other systematic errors can be estimated by studying the effects of changing input parameters or varying the analysis procedure.
\begin{itemize}
  \item The global fit described in Eq.~(\ref{eqn:global-fit}) restricted the heavy quark masses to those satisfying $am_\Psi < 1.05$.  To test whether this cut is sufficient to exclude lattice artifacts of $O(a^3)$ or higher, one could choose a more conservative cut, using only data satisfying $am_\Psi < 0.7$.  Refitting with this more limited data set results in a fit value of $\langle \xi^2 \rangle = 0.226 \pm 0.043$.  Although these two results are compatible within one standard deviation, the difference between the central values (0.016) is taken as the estimate of the systematic uncertainty from the continuum extrapolation.

  \item The global fit contains a $\Lambda_\text{QCD}/m_\Psi$ term to account for the twist-3 contribution.  In principle, higher-twist contributions are also present.   To estimate such systematic effects, one could add a $\Lambda_\text{QCD}/m_\Psi^2$ term to the global fit in Eq.~(\ref{eqn:global-fit}).  This changes the fit result to $\langle \xi^2 \rangle = 0.185 \pm 0.017$ which has a central value differing from that of the primary procedure by 0.025.  This is taken to be the systematic uncertainty from higher twist effects.

  \item As explained in Sec.~\ref{time-momentum}, at small values of $\tau = \tau_m - \tau_e$, the data are contaminated with uncontrolled lattice artifacts.  The primary fit omits the $\tau/a = 0$, 1, and 2 points, where such effects are the most significant and result in unacceptable $\chi^2$ values in the fits.  To analyse errors arising from the placement of this cut, one can exclude $\tau/a = 3$ from the fits, which gives a modified result of $\langle \xi^2 \rangle = 0.208 \pm 0.014$ and therefore a small systematic uncertainty from the difference in central values of 0.002.

  \item The Wilson coefficients $C_W$ are calculated in perturbation theory, and in this analysis, they are only computed to 1-loop order.  As an estimate of the magnitude of higher-loop corrections, one can perform this analysis at a larger renormalization scale of $\mu = 4$ GeV and then run back to $\mu = 2$ GeV using Eq.~(\ref{eq:Gegen_moment_one_loop_running}).  Such a procedure results in $\langle \xi^2 \rangle(\mu = 4\text{ GeV}) = 0.216 \pm 0.012$, which evolves to $\langle \xi^2 \rangle(\mu = 2\text{ GeV}) = 0.218 \pm 0.014$, giving a systematic uncertainty of 0.008 from the change in central value.
\end{itemize}

The above procedure for estimating systematic effects leads to a final value of $\langle \xi^2 \rangle (\mu = 2\text{ GeV}) = 0.210 \pm 0.013\text{ (stat.)} \pm 0.034\text{ (sys.)}$, which can be combined in quadrature to give $\langle \xi^2 \rangle (\mu = 2\text{ GeV}) = 0.210 \pm 0.036$ (total, exc. quenching).  The above error estimates are summarized in Table~\ref{tab:error-budget}.

\begin{table}
  \centering
    \begin{tabular}{l c} \hline\hline
    Source of error & Size \\ \hline
    Statistical & 0.013 \\ 
    Continuum extrapolation & 0.016 \\
    Higher-twist & 0.025 \\
    Excited-state contamination & 0.002 \\
    Unphysical $m_\pi$ & 0.014 \\
    Fit range & 0.002 \\
    Running coupling & 0.008 \\\hline
    \textbf{Total (exc.~quenching)} & \textbf{0.036} \\ \hline
  \end{tabular}
  \caption{The error budget for the computation of the second Mellin moment $\langle \xi^2 \rangle$ using the HOPE method, with the data processed in the time-momentum representation.}
  \label{tab:error-budget}
\end{table}

The dominant sources of uncertainty are from the continuum and higher-twist extrapolations.  In principle, both these extrapolations can be better controlled by including finer lattice spacings, which would also allow the inclusion of larger heavy-quark masses.  However, computations at finer lattices are expensive and therefore beyond the scope of this preliminary work.
The error from quenching is formally uncontrollable, although empirically it is a 10--20\% effect in many calculations.  To perform a precise comparison of this result to dynamical calculations would require redoing these calculations on dynamical ensembles.

\subsection{Determination of $f_\pi$}
The previous two subsections describe the determination of the second moment of the pion LCDA using the time-momentum analysis procedure.  To check the validity of the HOPE strategy, it is worth noting that the pion decay constant $f_\pi$ is computed as a byproduct of this analysis.  As is clear in the OPE formula, Eq.~(\ref{eq:2nd_Mellin_OPE_had_amp_target_mass}), $f_\pi$ is an overall normalization factor for the hadronic amplitude $V^{\mu\nu}$.

One can extrapolate the $f_\pi$ values computed at various heavy quark masses on the four ensembles to the continuum using the same procedure as for the extrapolation of $\langle \xi^2 \rangle$, giving a global fit value of $161 \pm 2$ MeV after removal of lattice discretization and higher-twist effects, where the error reflects statistical uncertainties only.  It should be noted that this measurement suffers from not only the systematic errors mentioned in the previous subsection but also additional uncertainties from the normalization constants $Z_A$ and $\tilde{b}_A$ to which the second moment is completely insensitive in the time-momentum analysis.\footnote{Since this normalization factor is only used in the determination of $f_\pi$ rather than in the computation of $\langle \xi^2 \rangle$ that is the main focus of the paper, this work uses the approximation $\tilde{b}_A a \tilde{m}_{ij} \approx b_A a m_{ij}$, which is correct up to a mass-independent $\mathcal{O}(a)$ term~\cite{Bhattacharya_2006}.  The value of $b_A$ was taken to be a constant value of 1.25 across all lattice spacings, which is consistent with the values quoted in Ref.~\cite{Bhattacharya_2006} at all lattice spacings measured in that work.}

On the other hand, $f_\pi$ can be directly measurable on the lattice via the axial-axial correlator at $0 \ll \tau \ll T$,
\begin{equation}
  \langle 0 | A_4(\tau) A_4(0) | 0 \rangle \sim \frac{(f_\pi Z_A)^2 m_\pi}{2} \left[e^{-m_\pi \tau} + e^{-m_\pi (T-\tau)}\right] \, ,
  \label{pion-decay}
\end{equation}
where $A_4$ is the local unsmeared axial current $\overline \psi \gamma_4 \gamma_5 \psi$ and the convention where $f_\pi \sim 130$ MeV at the physical pion mass is used.  As written, Eq.~(\ref{pion-decay}) contains $O(a)$ corrections, so the continuum extrapolation must include a term linear in $a$ rather than in $a^2$.  A low-statistics (one source per configuration) computation of $f_\pi$ at the four lattices gives a continuum value of $f_\pi = 157 \pm 6$ MeV.

\begin{figure}
  \centering
  \includegraphics[scale=0.5]{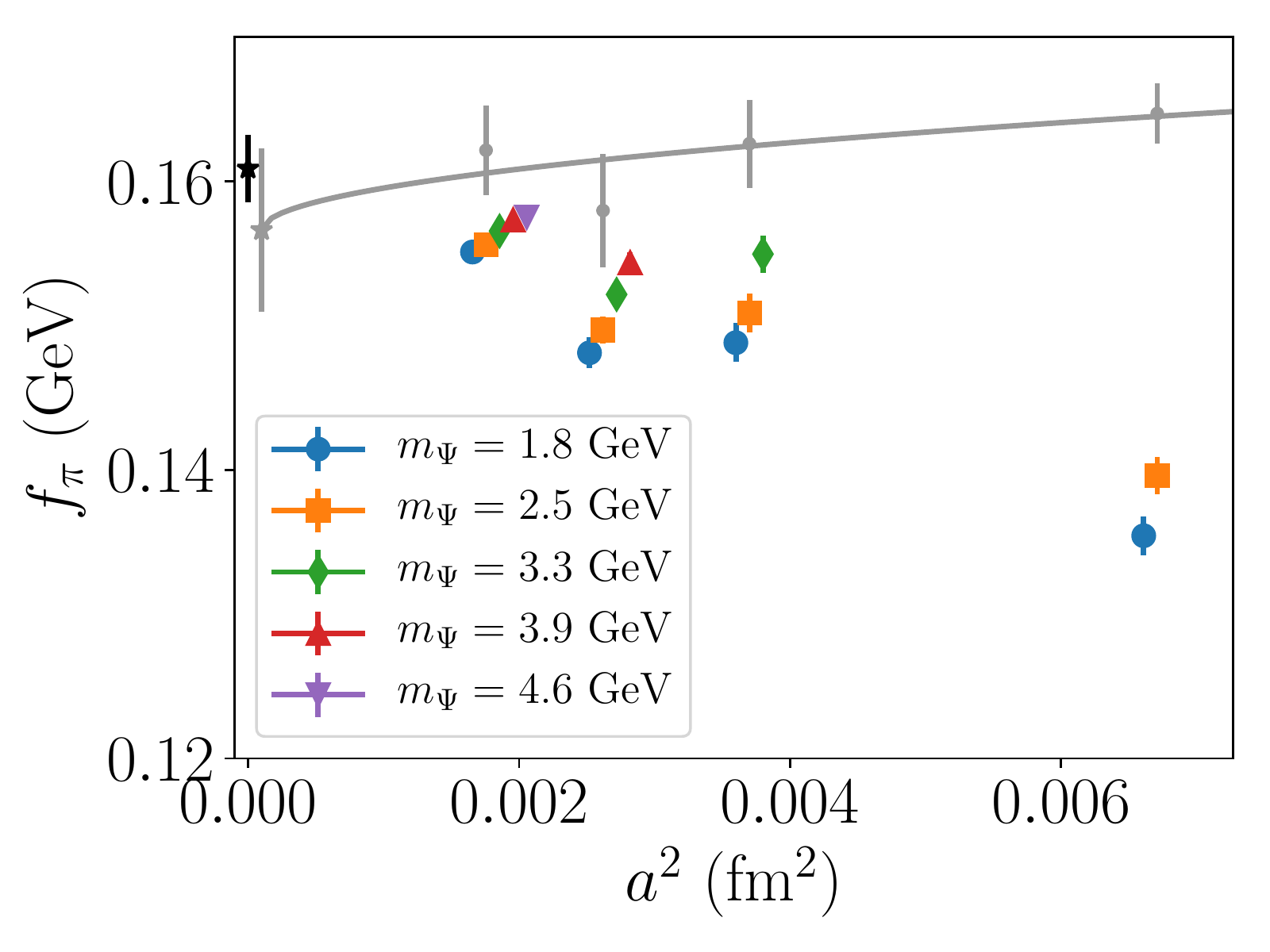}
  \caption{Extrapolation of $f_\pi$ to the continuum.  The coloured points at finite lattice spacing are the values of $f_\pi$ extracted from the four lattice spacings at various heavy quark masses as an intermediate step in the determination of $\langle \xi^2 \rangle$.  The black point is the global fit of $161 \pm 2$ MeV to the hadronic tensor data.  The gray points are direct computations of $f_\pi$ at the four lattice spacings.  These direct calculations suffer from $O(a)$ discretization errors, but an extrapolation linear in $a$ gives a continuum value of $158 \pm 5$ MeV, in good agreement with the value extracted from hadronic tensor measurements.}
  \label{fig:fpi-extrapolation}
\end{figure}

Despite the systematics that could affect the value extracted from the hadronic tensor measurement, the two determinations of $f_\pi$ are in good agreement (see Fig.~\ref{fig:fpi-extrapolation}).  While $f_\pi$ is not directly relevant to the calculation, this serves as a useful cross-check of the validity of the operator product expansion and, more generally, of this calculational method.

\subsection{Momentum-Space Analysis}
\label{sec:momanalysis}
A further check of the validity of the time-momentum representation method can be obtained by analysing the same data using a momentum-space analysis.  The starting point for the momentum-space analysis is the time-momentum representation ratio $R^{[\mu\nu]}(\tau,\mathbf{p},\mathbf{q};a)$ constructed in Eq.~\eqref{eq:r12}. The lattice-regularized data are converted to momentum-space via
\begin{equation}
V^{[\mu\nu]}(p,q;a)=a\sum_{\tau=-\tau_\text{max}}^{\tau_\text{max}} e^{i\tau q_4}R^{[\mu\nu]}(\tau,\mathbf{p},\mathbf{q};a)\, ,
\label{DFT}
\end{equation}
where $\tau_\text{max}$ was taken to be approximately $1\fm=5\gev^{-1}$.  By Fourier transforming the tree-level HOPE equation, it is possible to show that the numerical data decay exponentially in $\tau$ as
\begin{equation}
R^{[\mu\nu]}(\tau,\mathbf{p},\mathbf{q};a)\sim 
\frac{e^{-E_{\mathbf{q}}\tau}}{2E_{\mathbf{q}}}\, ,
\end{equation}
where $E_{\mathbf{q}}=\sqrt{m_\Psi^2+\mathbf{q}^2}$. Since $R^{[\mu\nu]}(\tau,\mathbf{p},\mathbf{q};a)$ exhibits exponential decay in $\tau$ with an exponent with magnitude greater than  approximately $2\gev$, the truncation in the sum in Eq.~(\ref{DFT}) is expected to be well controlled.
While the discrete Fourier transform formally only produces a discrete set of Fourier modes, in this work, interpolation between these Fourier modes is achieved by evaluating Eq.~(\ref{DFT}) for arbitrary $q_4$.  Note that the largest Fourier mode, $q_4^\text{max}$ must be taken sufficiently small to remain below the Nyquist frequency which corresponds to requiring $q_4^\text{max}<\pi/a$. For the ensemble with the coarsest lattice spacing ($a=0.0813\fm$) this results in the constraint that $q_4^\text{max}<7.5\gev$. In practice, data at momenta close to the Nyquist frequency may possess large lattice artifacts, and in this analysis $q_4^\text{max}=5\gev$ is chosen.

While data at non-zero $\tau$ are guaranteed to have a well-defined continuum limit, $\tau=0$ data contain additional UV divergences arising from the mixing of the current-current operator with lower-dimensional operators. After performing the Fourier transform, this divergence will appear as an additive shift in the numerical data. Thus in order to ensure that the hadronic amplitude considered in this work has a well-defined continuum limit (after finite, multiplicative renormalization), a single subtraction is first performed at fixed lattice spacing:
\begin{equation}
  V_\text{sub}^{[\mu\nu]}(p,q;a)=V^{[\mu\nu]}(p,q;a)-V^{[\mu\nu]}\left[p,(\mathbf{q},q_{4,\text{sub}});a\right]
\, ,
\end{equation}
where $q_{4,\text{sub}}$ is chosen to be $q_{4,\text{sub}}=q_4^\text{max}=5\gev$. This choice is informed by the desire to minimize the statistical noise introduced in this process. As a result of this subtraction, the matrix element may be expressed generically as 
\begin{equation}
V_\text{sub}^{[\mu\nu]}(p,q;a)=\sum_{n=0}^\infty a^n V_{(n),\text{sub}}^{[\mu\nu]}(p,q)\, ,
\end{equation}
where $V_{(0),\text{sub}}^{[\mu\nu]}(p,q)\equiv V_{\text{sub}}^{[\mu\nu]}(p,q)$ is the continuum hadronic matrix element. As emphasized previously, the simplicity of this continuum limit is one of the advantages of considering a current-current correlator. As argued in Sec.~\ref{sec:order-a}, it is possible to show that the use of the the Sheikholeslami-Wohlert (clover) improved action leads to the removal of $\mathcal{O}(a)$ corrections for the matrix element studied here (see Appendices~\ref{app:ca},~\ref{app:ca-prime} for details), so
\begin{equation}
V_\text{sub}^{[\mu\nu]}(p,q;a)=V_{\text{sub}}^{[\mu\nu]}(p,q)+a^2 V_{(2),\text{sub}}^{[\mu\nu]}(p,q)+\mathcal{O}(a^3)\, .
\end{equation}

\begin{figure}
\subfloat[][]{
\includegraphics[scale=0.5]{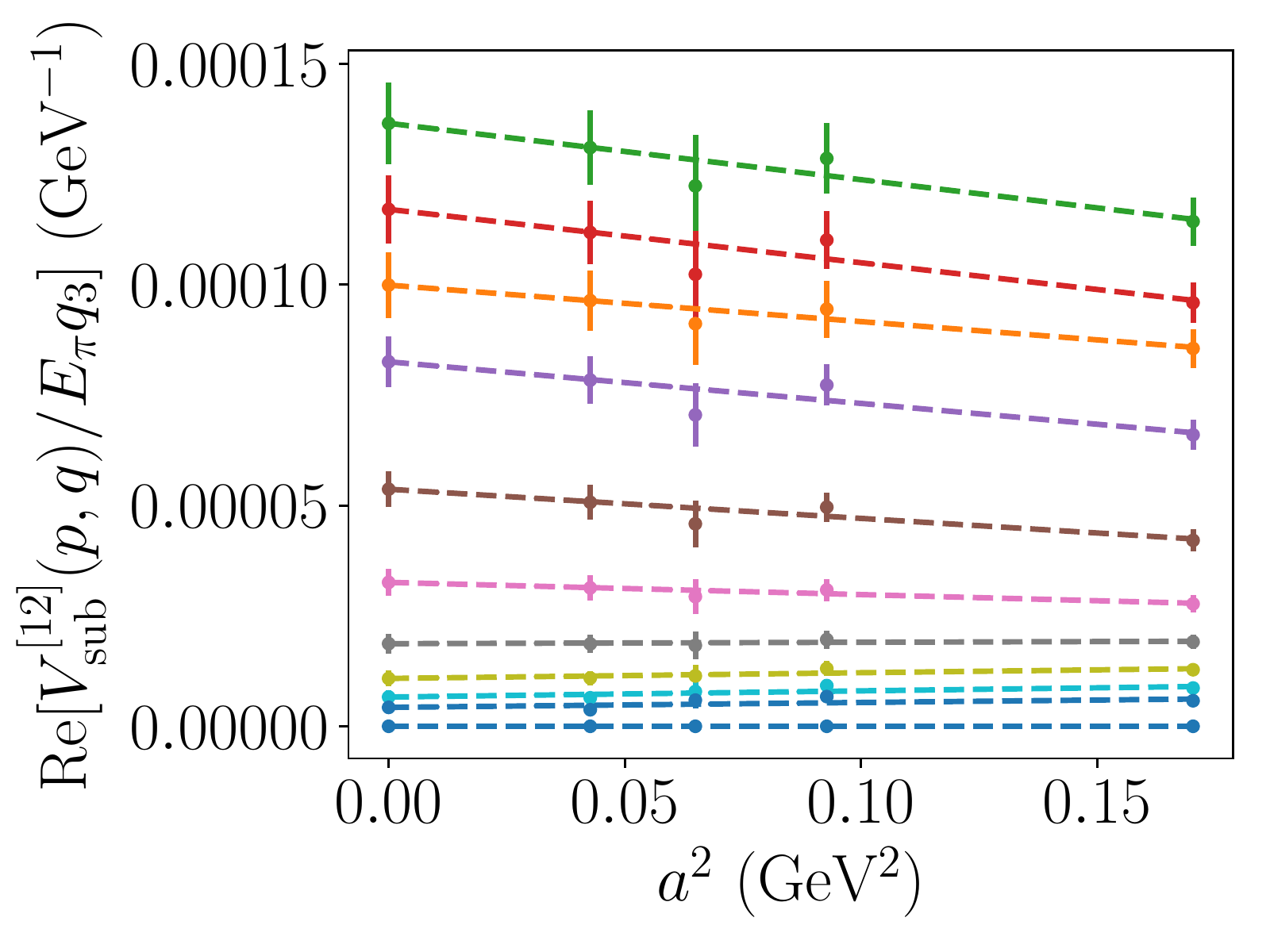}\label{fig:re_ctm_0}
}
\subfloat[][]{
\includegraphics[scale=0.5]{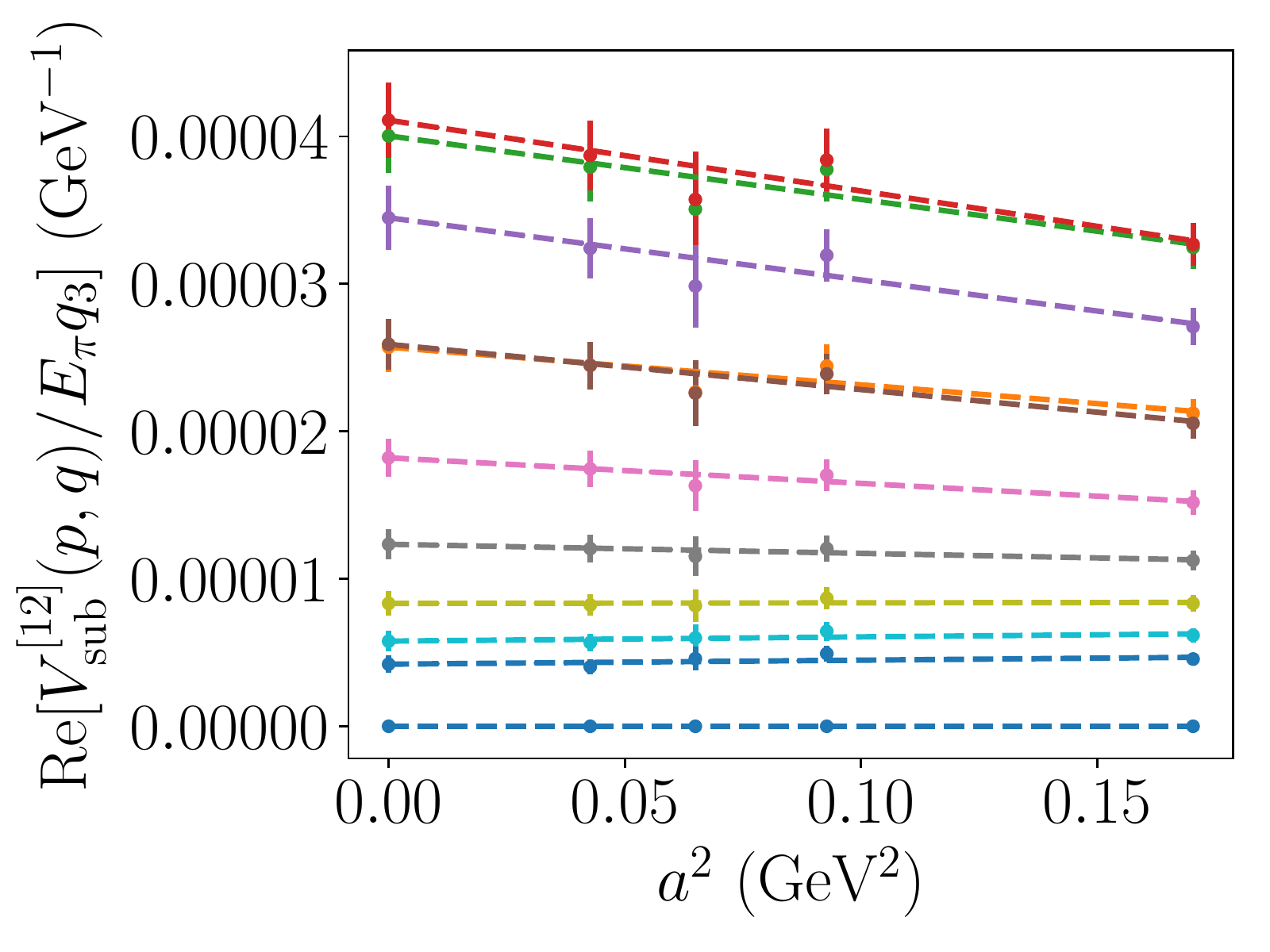}\label{fig:re_ctm_1}
}

\subfloat[][]{
\includegraphics[scale=0.5]{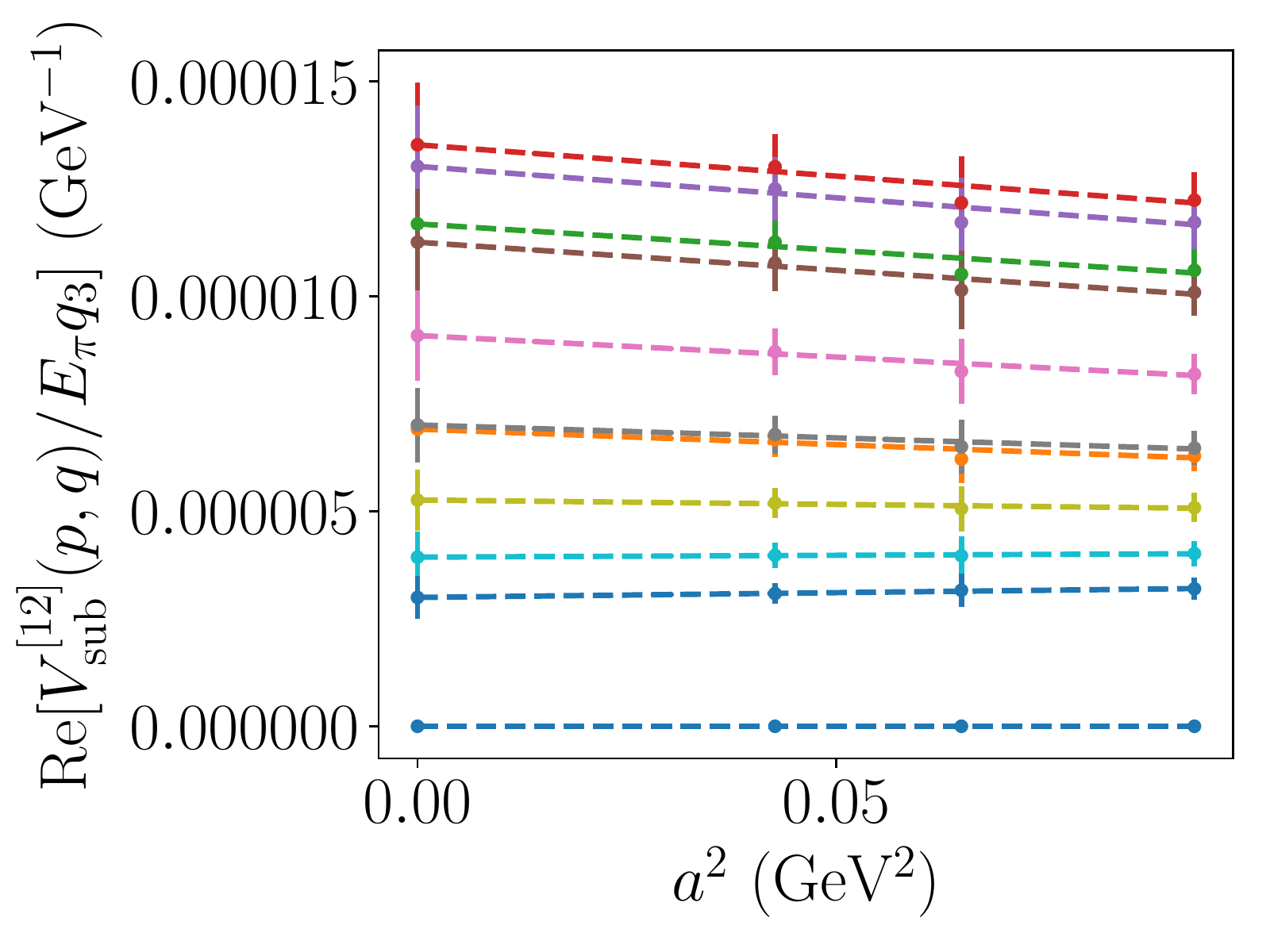}\label{fig:re_ctm_2}
}
\subfloat[][]{
\includegraphics[scale=0.5]{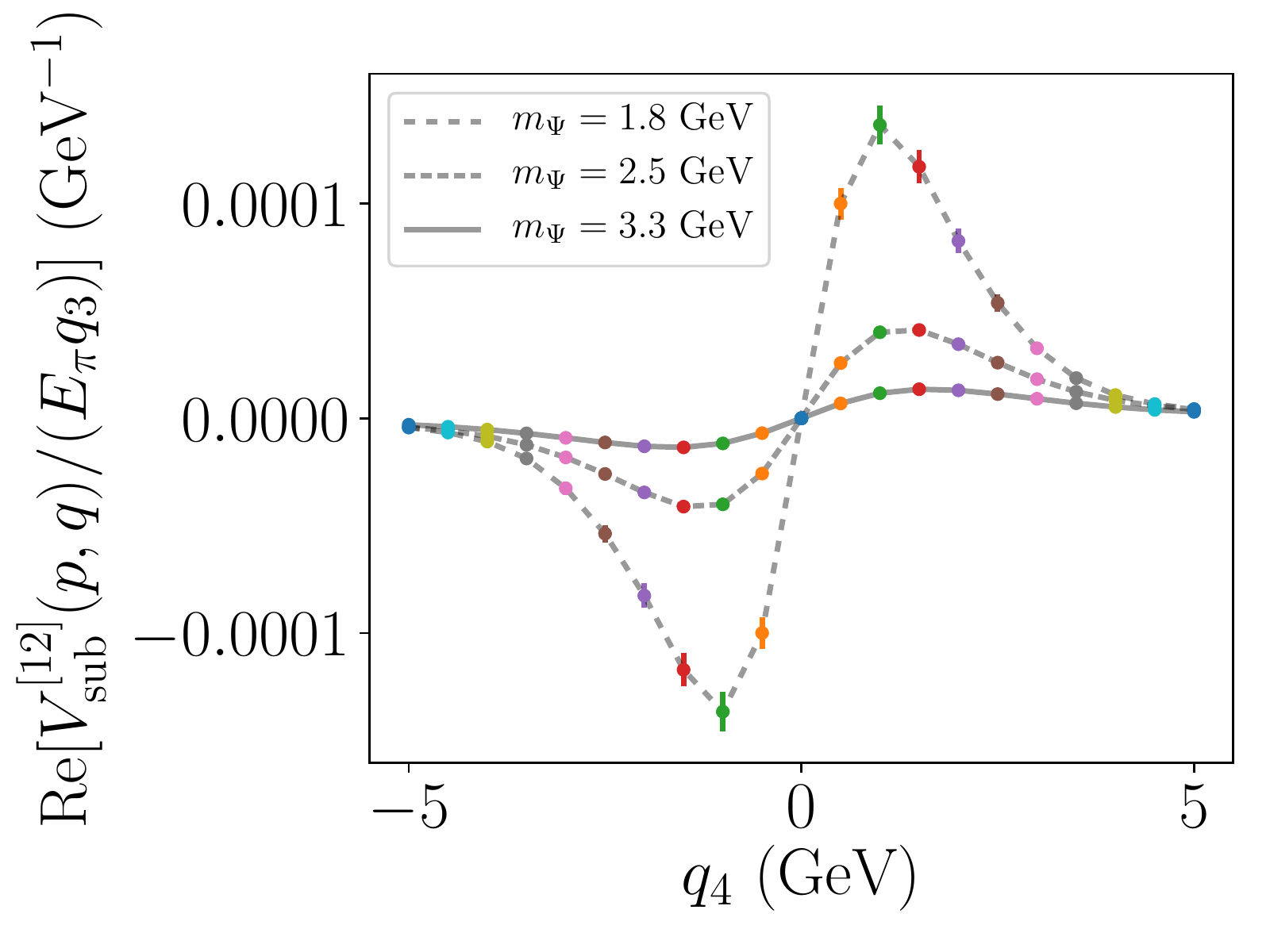}\label{fig:re_ctm_result}
}
\caption{Continuum extrapolation of real part of hadronic matrix element. Figs.~\ref{fig:re_ctm_0} - \ref{fig:re_ctm_2} show the pointwise continuum extrapolation for each value of $q_4$ for heavy quark masses of (a) $m_\Psi = 1.8\gev$, (b) $m_\Psi = 2.5\gev$, and (c) $m_\Psi = 3.3\gev$. The resulting hadronic amplitude in the continuum is shown in Fig.~\ref{fig:re_ctm_result}, where curves are included to guide the eye.}
\label{fig:real_part_continuum}
\end{figure}

\begin{figure}
\subfloat[][]{
\includegraphics[scale=0.5]{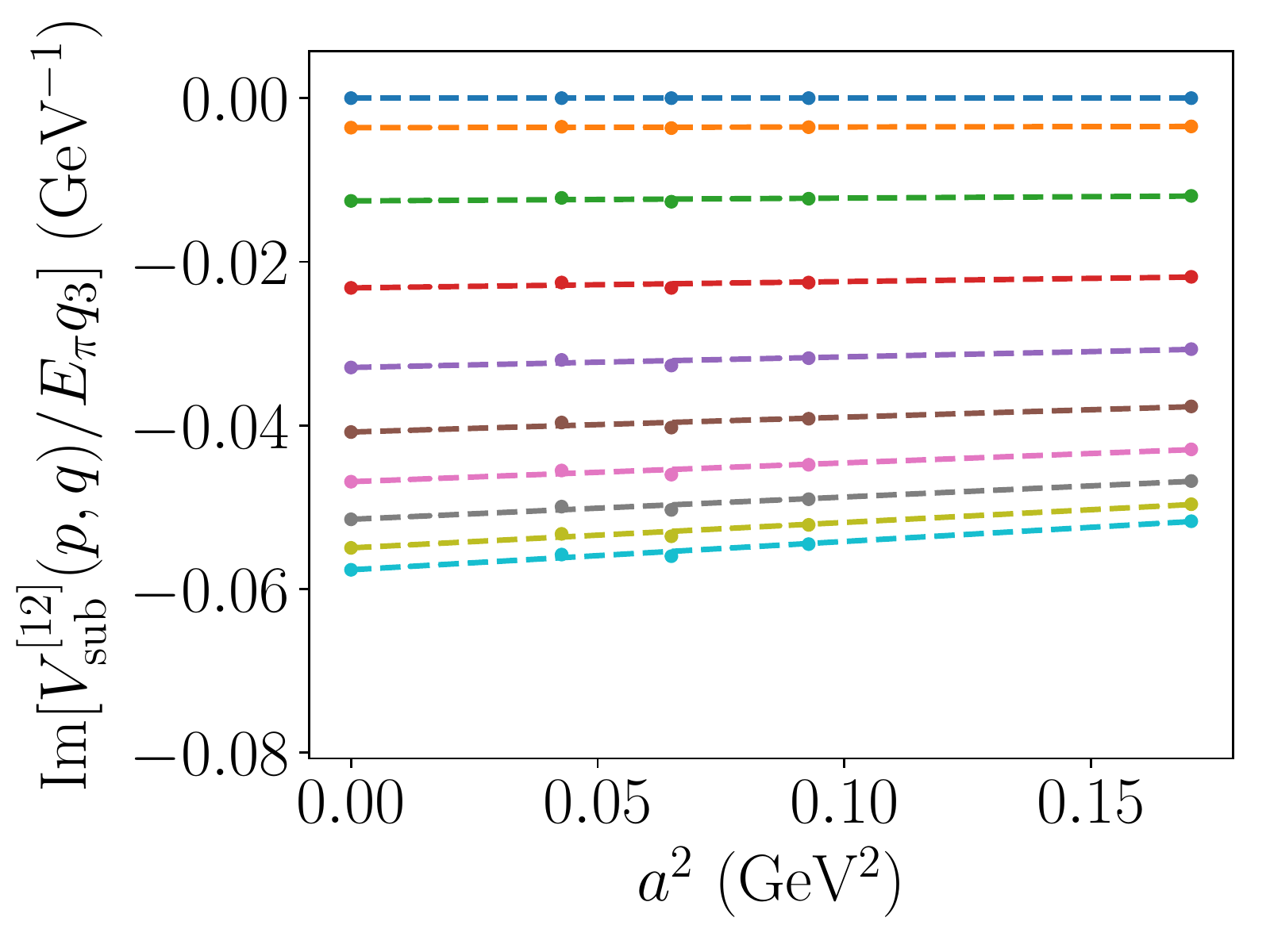}\label{fig:im_ctm_0}
}
\subfloat[][]{
\includegraphics[scale=0.5]{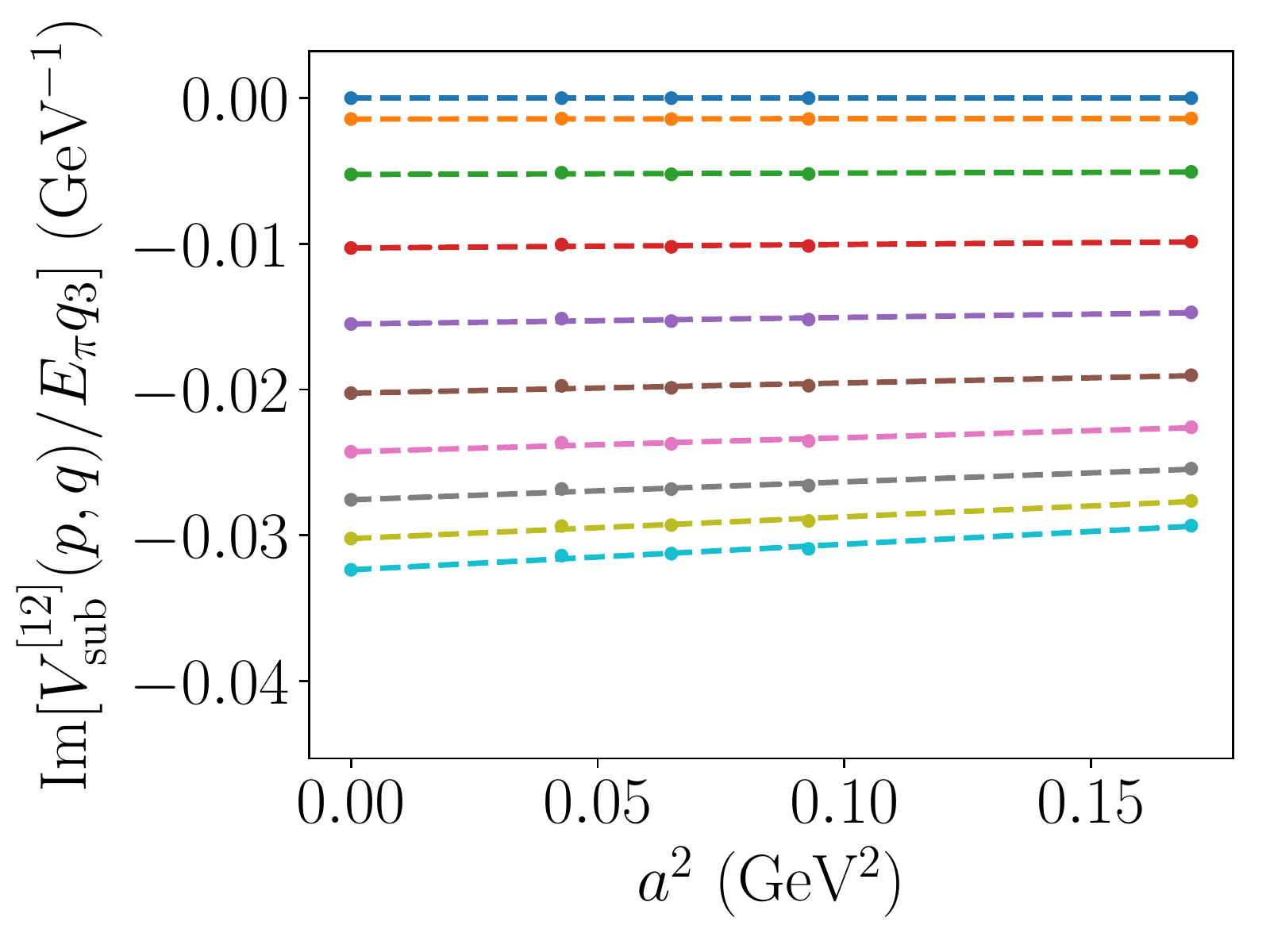}\label{fig:im_ctm_1}
}

\subfloat[][]{
\includegraphics[scale=0.5]{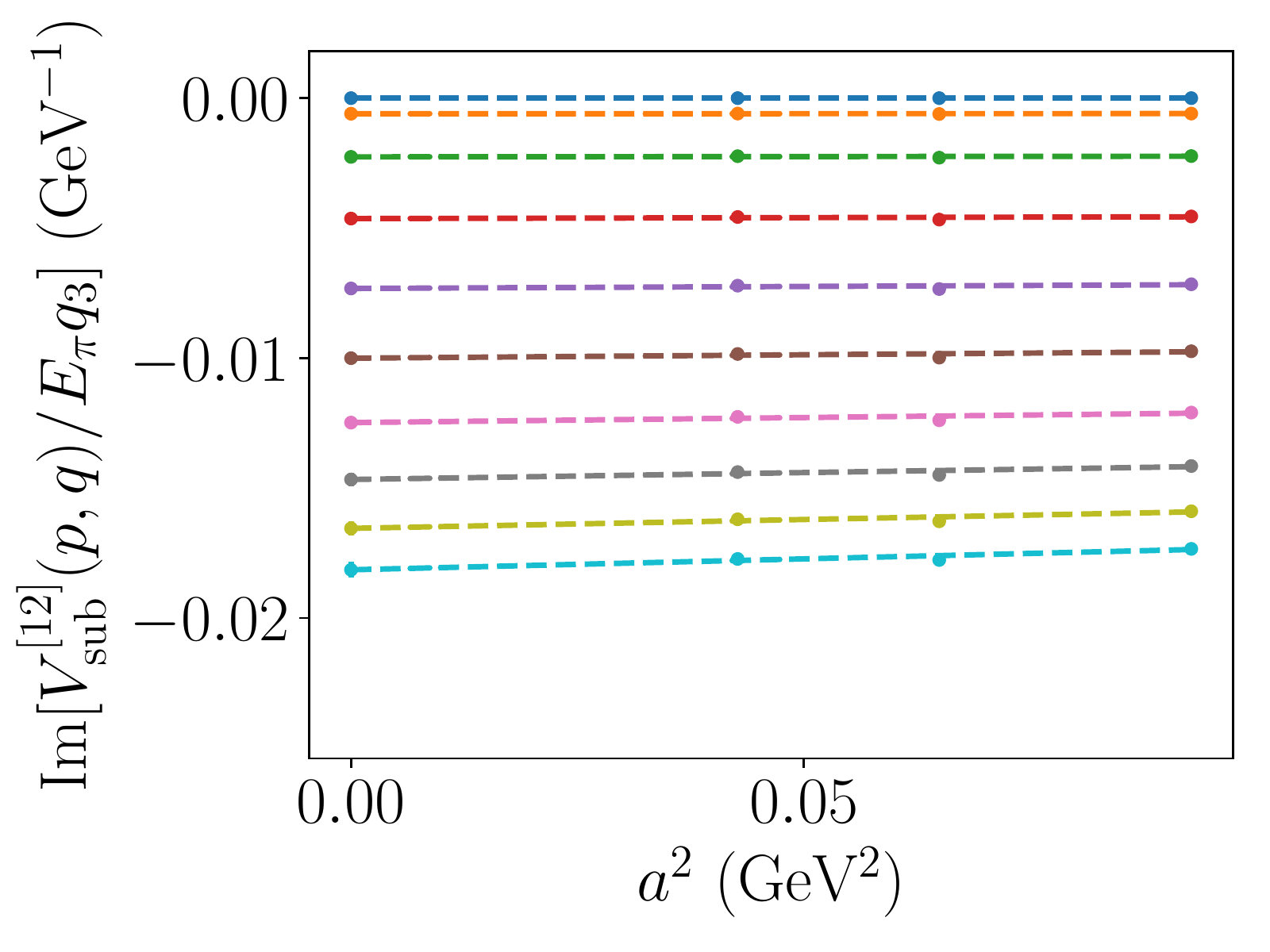}\label{fig:im_ctm_2}
}
\subfloat[][]{
\includegraphics[scale=0.5]{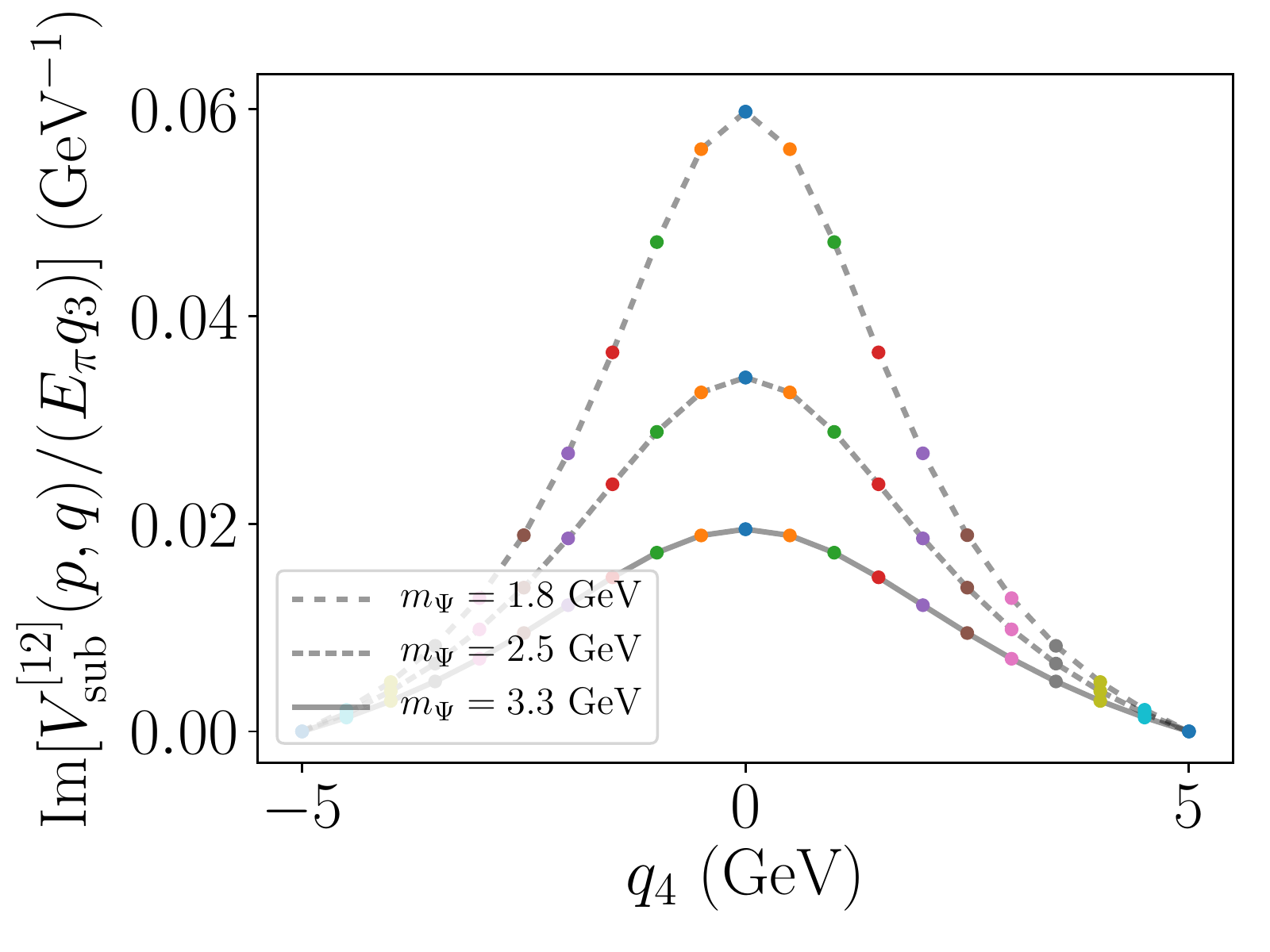}\label{fig:im_ctm_result}
}

\caption{Continuum extrapolation of imaginary part of hadronic matrix element. Figs.~\ref{fig:im_ctm_0} - \ref{fig:im_ctm_2} show the pointwise continuum extrapolation for each value of $q_4$ for heavy quark masses of (a) $m_\Psi = 1.8\gev$, (b) $m_\Psi = 2.5\gev$, and (c) $m_\Psi = 3.3\gev$. The resulting hadronic amplitude in the continuum is shown in Fig.~\ref{fig:im_ctm_result}, where curves are included to guide the eye.}
\label{fig:imag_part_continuum}
\end{figure}

It is important to note that the continuum extrapolation requires taking $a\to0$ along a line of constant physics. This is achieved by tuning the bare parameters to ensure certain physical quantities remain fixed as lattice spacing is varied, but there are some mistunings resulting from percent level inaccuracies in the tuning process. While such a mistuning appears as a relatively mild effect in the time-momentum representation analysis where it results in a variation of $f_\pi$, the momentum-space approach is more sensitive to any such mistuning, since it affects the continuum extrapolation of the hadronic matrix element $V^{[\mu\nu]}(p,q)$. Evidence of this mistuning can be seen in Fig.~\ref{fig:tuning}. It is important to note that the tuning of the light-quark sector parameters to their physical values is an issue for all methods, and is not unique to this approach. One can reduce the systematic error associated with this mistuning by considering a ratio which is less sensitive to the volume and pion mass dependence. Examining the HOPE, the leading volume and pion mass dependent quantities are found in the prefactor (for $\mu=1$, $\nu=2$)
\begin{equation}
V_\text{sub}^{[12]}(p,q)\propto \epsilon^{12\alpha\beta}p_\alpha q_\beta\propto (E_\pi q_3)\, ,
\end{equation}
for the special kinematics considered. In order to reduce the volume and pion mass dependence, the ratio $V_\text{sub}^{[12]}(p,q)/(E_\pi q_3)$ is formed and then continuum extrapolated. 

In contrast to the time-momentum representation analysis, a further cut must be made on the data included in the analysis. All data presented in the time-momentum representation analysis satisfy the constraint $am_\Psi<1.05$. This restriction is placed to ensure control over lattice artifacts. However, in addition to this constraint, the momentum-space analysis requires data at a sufficient number of lattice spacings for the continuum extrapolation to be performed. In particular, the model for the continuum extrapolation contains two free parameters, and thus the analysis must be limited to the subset of data where the the heavy quark mass satisfies the constraint $a m_\Psi<1.05$ for at least three lattice spacings. From Fig.~\ref{fig:masses-used}, this criterion constrains this analysis to make use of only the lightest three heavy quark masses.  The continuum extrapolations of the real and imaginary parts of the hadronic amplitude are shown in Figs.~\ref{fig:real_part_continuum},~\ref{fig:imag_part_continuum}.

\subsection{Extraction of Second Moment}

\begin{figure}
\includegraphics[scale=0.5]{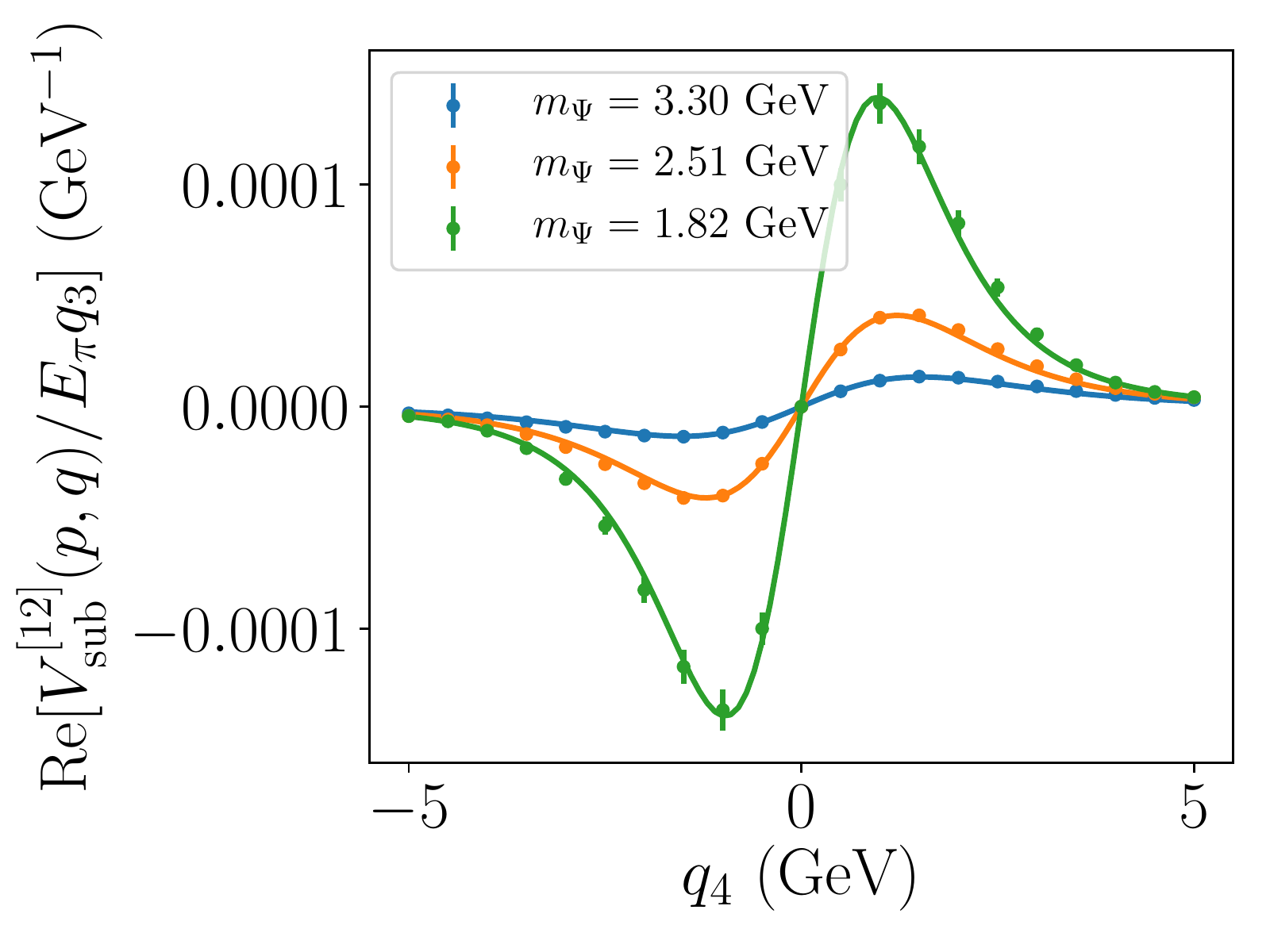}
\includegraphics[scale=0.5]{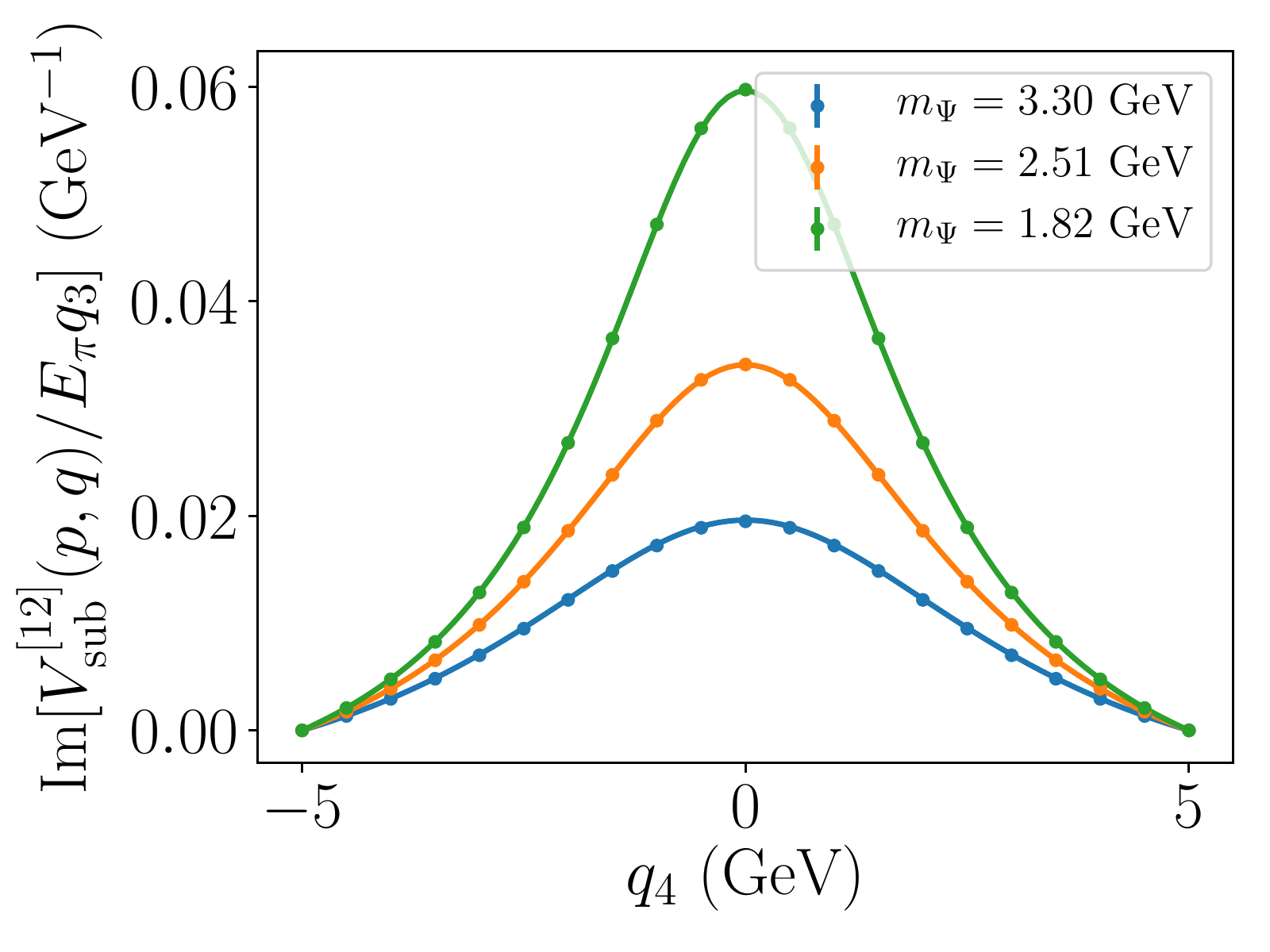}
\caption{Global fit of continuum matrix element data to the HOPE formula to extract $f_\pi$, $m_\Psi$ and $\expval{\xi^2}$.}\label{fig:mom_space_fit}
\end{figure}

Having extrapolated the real and imaginary parts of the hadronic matrix element to the continuum, a global fit is performed to the order-$\alpha_S$ HOPE formula given in Eq.~\eqref{eq:2nd_Mellin_OPE_had_amp_target_mass} plus an additional twist-three \textit{ansatz} which is used to control the leading higher-twist effects. The functional form of the model used in the extraction of the Mellin moments from the continuum data is
\begin{equation}
  \begin{split}
V^{[ 12]} (q,p,m_{\Psi})/(E_\pi q_3) = &-\frac{2 if_{\pi}}{\tilde{Q}^{2}}  \left \{
 C_{W}^{(0)} (\tilde{Q}^{2},
 \mu, \tau)  +  C_{W}^{(2)} (\tilde{Q}^{2},
 \mu, \tau)  \la \xi^{2}\ra  \left [ \frac{\zeta^{2} C_{2}^{2}
(\eta)}{12\tilde{Q}^{2}} \right ]\right \} \\
&+ \frac{2if_\pi \Lambda_\text{QCD}}{\tilde{Q}^3}\left\{b_0+b_2\left [ \frac{\zeta^{2} C_{2}^{2}
(\eta)}{12\tilde{Q}^{2}} \right ]\right\}\, .
\end{split}
\label{momentum-OPE}
\end{equation}
It is important to note that while the above parameterization of the twist-three piece is the most natural, other terms like $m_\Psi/\tilde{Q}^4$ are also possible. The resulting fit is shown in Fig.~\ref{fig:mom_space_fit}. As a result of this global analysis, the first two moments of the pion LCDA are found to be $f_\pi= 0.173 \pm 0.001\gev$, and $\expval{\xi^2}(\mu=2\gev^2)=0.210 \pm 0.013$, where the statistical uncertainty is determined from a bootstrap analysis of the numerical data. Since this a global fit, the three heavy quark masses are also determined to be $m_\Psi^{(1)}=1.82 \pm 0.02\gev$, $m_\Psi^{(2)}=2.52 \pm 0.02\gev$ and $m_\Psi^{(3)}=3.34 \pm 0.02\gev$, which are in good agreement with the heavy-quark masses determined at fixed lattice spacing in the time-momentum representation.

Systematic uncertainties in $\langle \xi^2 \rangle$ for the momentum-space analysis may also be estimated following the same procedure as described in the time-momentum representation analysis. Since both analyses use the same lattice data they share some sources of systematic error. Thus the excited state contamination is taken to be a $\sim 1\%$ effect, and as with the time-momentum representation analysis, finite-volume effects are taken to be negligible. Finally, as in the time-momentum representation analysis, the unphysically heavy pion mass is assumed to contribute a 5\% systematic error. In addition to these shared sources of systematic uncertainty, the momentum-space method has several additional sources of systematic error. These arise from the difference in the order of steps of the analysis, and are discussed below:
\begin{itemize}
\item Following the procedure employed above, $\mathcal{O}(a^3)$ effects are studied by making a more conservative cut on $am_\Psi$. In particular, the cut is chosen to be $am_\Psi<0.7$, which is consistent with the time-momentum representation. The resulting fit leads to a value for the second moment of $\expval{\xi^2}=0.222 \pm 0.068$. Taking the differences of central values leads to a systematic error of $0.012$.
\item The effects of higher-twist contributions may be studied by adding a twist-four \textit{ansatz} to the continuum HOPE formula in Eq.~(\ref{momentum-OPE}). The form chosen is
\begin{equation}
V_{\text{higher-twist}}^{[ 12]} (q,p,m_{\Psi})/(E_\pi q_3) = V^{[ 12]} (q,p,m_{\Psi})/(E_\pi q_3) + \frac{2if_\pi \Lambda_\text{QCD}^2}{\tilde{Q}^4}\left\{c_0+c_2\left [ \frac{\zeta^{2} C_{2}^{2}
(\eta)}{12\tilde{Q}^{2}} \right ]\right\}\, ,
\end{equation}
mirroring the choice for the twist-three term. The resulting value for the second moment is $\expval{\xi^2}=0.245 \pm 0.019$. This results in a systematic uncertainty of $0.035$.
\item The higher-loop corrections to the Wilson coefficients are studied by repeating the above analysis at a renormalization scale of $\mu=4\gev$, and then evolving back to $\mu = 2$ GeV using the renormalization-group evolution of the Gegenbauer moments given by Eq.~(\ref{eq:Gegen_moment_one_loop_running}).  This gives $\expval{\xi^2}(\mu=4\gev)=0.236\pm 0.016$. Running this to $2\gev$ results in $\expval{\xi^2}(\mu=2\gev)=0.239\pm 0.017$. Taking the difference between the evolved $\expval{\xi^2}$ and the original fitted $\expval{\xi^2}(\mu = 2\text{ GeV})$ gives a systematic uncertainty of $0.029$. 
\end{itemize}

A breakdown of the sources of systematic error described here is given in Table~\ref{tab:mom-error-budget}. This analysis of systematic errors leads to a final value for the second moment of $\expval{\xi^2}(\mu=2\gev)=0.210\pm0.013\text{(stat)}\pm 0.044\text{(syst)}$. These two sources of error may be added in quadrature to obtain the final result $\expval{\xi^2}(\mu=2\gev)=0.210\pm0.046$. Similarly to the time-momentum space analysis, the dominant source of systematic uncertainty arises from the higher-twist terms. This issue is made worse in momentum-space due to the additional cut on lattice data required for the continuum extrapolation of the hadronic amplitude. 

\begin{table}
  \centering
  \begin{tabular}{l c} \hline\hline
    Source of error & Size \\ \hline
    Statistical &  0.013\\ 
    Excited-state contamination & 0.002 \\ 
    Continuum extrapolation & 0.012 \\ 
    Higher-twist  &  0.035\\ 
    Running coupling & 0.029 \\ 
    Unphysical $m_\pi$ & 0.014 \\ \hline 
    \textbf{Total (exc.~quenching)} & \textbf{0.046} \\ \hline
  \end{tabular}
  \caption{The error budget for the computation of the second Mellin moment $\langle \xi^2 \rangle$ using the HOPE method, with the data analysed in the momentum-space approach.}
  \label{tab:mom-error-budget}
\end{table}

\subsection{Discussion of Results}

The ratio $R^{[\mu\nu]}(\tau,\mathbf{p},\mathbf{q};a)$ was analysed using two alternative approaches, termed the time-momentum representation (TMR) analysis and the momentum-space (Mom) analysis. The results of the second moment from these approaches are
\begin{align}
  \expval{ \xi^2 }_\text{TMR}(\mu=2\gev)&= 0.210 \pm 0.013\text{ (stat.)} \pm 0.034\text{ (sys.)} = 0.210 \pm 0.036 \, , \\
\expval{ \xi^2 }_\text{Mom}(\mu=2\gev)&=0.210 \pm 0.013\text{ (stat.)} \pm 0.044\text{ (sys.)} = 0.210 \pm 0.046 \, . 
\end{align}
The central values and statistical errors are the same in both approaches. The agreement of central values is the result of statistical coincidence; with a different choice of fit parameters this extrapolated central value is expected to vary. The equivalence of the statistical error is relatively unsurprising, since both approaches mostly share the same raw lattice data.

As a cross-check, the pion decay constant $f_\pi^\text{2pt}=0.158\pm0.005\gev$ was extracted from a conventional analysis of the axial-vector 2-point correlation function.  This is to be compared with the HOPE-derived values $f_\pi^\text{TMR}=0.161\pm0.002\text{ (stat.)}\gev$ and $f_\pi^\text{Mom}=0.173 \pm 0.001\text{ (stat.)}\gev$, with systematic uncertainties in $f_\pi$.  The systematic uncertainties in these determinations $f_\pi$ are likely comparable to the systematic uncertainty in $\langle \xi^2 \rangle$ (about 10-20\%), or perhaps slightly larger due to the added uncertainty in the normalization factors.

Examining both procedures allows the study of the advantages and shortcomings of both approaches and serves as a further cross-check of the analysis. The above equations show that the time-momentum representation approach results in a smaller systematic error than that of the momentum-space analysis. While the systematic uncertainty incurred from the truncation of the twist expansion is the largest systematic error in both analyses, the additional cut placed on the data in the momentum-space analysis results in the removal of data with the heaviest heavy quark masses. Since higher-twist corrections are suppressed by factors of $1/\tilde{Q}\sim1/m_\Psi$, this results in less control over the higher-twist effects.

Given the above considerations, the central value is chosen to be the more precise time-momentum representation analysis value of
\begin{equation}
\expval{ \xi^2 }(\mu=2\gev)= 0.210 \pm 0.036\,.
\end{equation}
This corresponds to a Gegenbauer moment of
\begin{equation}
  \phi_2 (\mu = 2\gev) = 0.03 \pm 0.11\,.
  \label{gegenbauer}
\end{equation}

\begin{figure}
\includegraphics[scale=0.5]{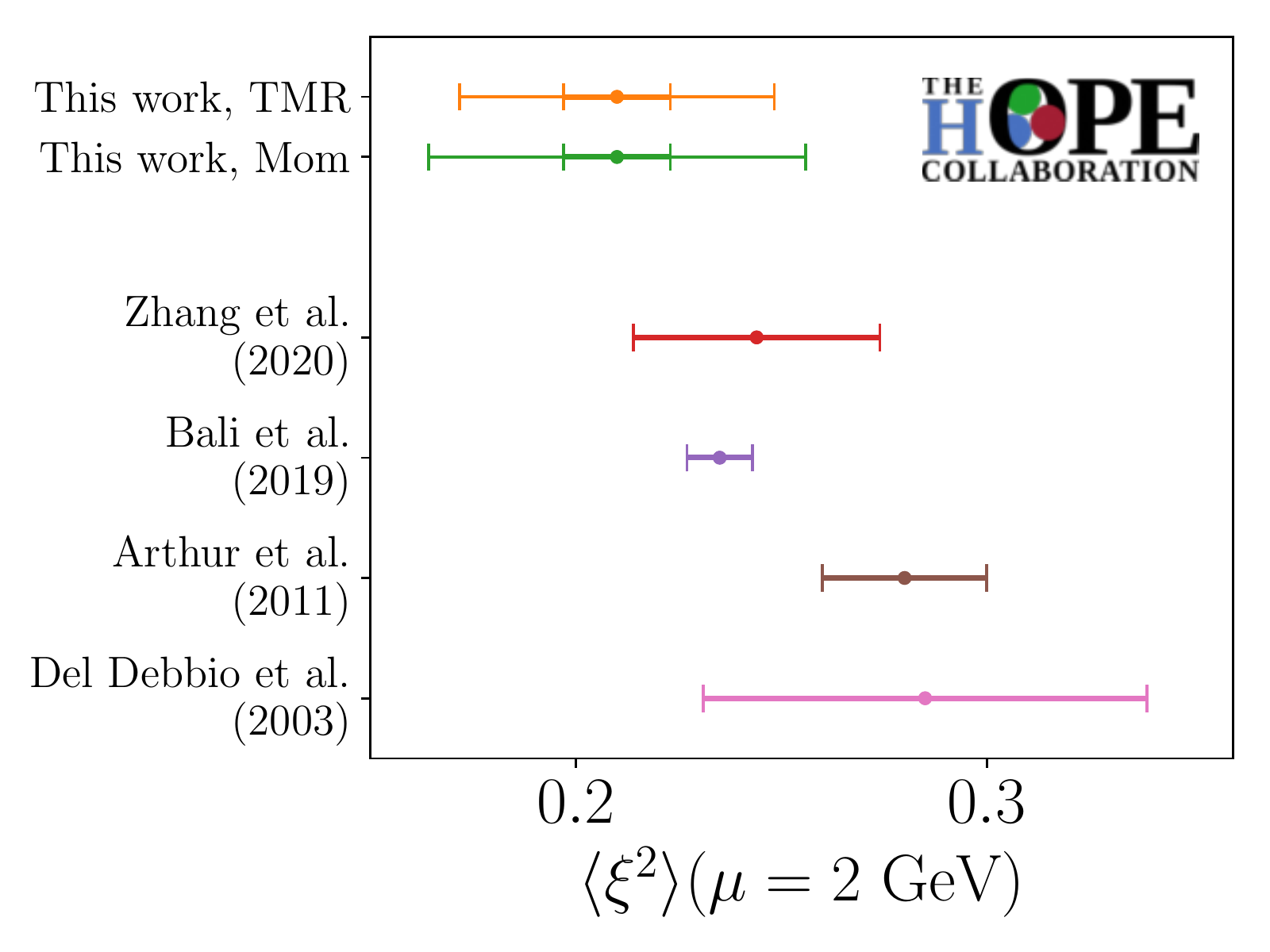}
\caption{Comparison of $\langle \xi^2 \rangle$ extracted from the time-momentum representation (TMR) and momentum-space analyses to various results in the literature.  Note that the values from this work and Del Debbio et al.~are in the quenched approximation, whereas the results from Zhang et al., Bali et al., and Arthur et al.~use dynamical QCD, and the error bars do not reflect the uncertainty from the quenched approximation.}\label{fig:comparison}
\end{figure}

Most previous lattice calculations have used local operators to compute $\langle \xi^2 \rangle$.  
In the quenched approximation, $\langle \xi^2 \rangle$ was previously computed to be $0.280 \pm 0.051$ at a renormalization scale of $\mu = 2.67$ GeV \cite{Del_Debbio_2003}.  Running this down to 2 GeV gives $\langle \xi^2 \rangle = 0.285 \pm 0.054$, which agrees with this quenched calculation, albeit with a large error bar.

More recent calculations with the local operator method have been performed in dynamical QCD, giving $\langle \xi^2 \rangle = 0.28 \pm 0.02$ \cite{Arthur_2011} and $\langle \xi^2 \rangle = 0.235 \pm 0.008$ \cite{RQCD:2019osh}, both at $\mu = 2$ GeV.  A separate approach is to proceed via the quasi-distribution amplitude (the distribution amplitude analogue of the quasi-PDF), which was used to give a result of $\langle \xi^2 \rangle = 0.244 \pm 0.030$ \cite{Zhang_2020}. These results are compared to the second moment determined in this work in Fig.~\ref{fig:comparison}. Formally, the uncertainty from quenching cannot be controlled, so a precise comparison of the results in this work to these dynamical calculations is not possible.  However, in practice, quenching errors are usually at the order of 10--20\%, and the calculation presented here agrees with the dynamical results within the listed uncertainties combined with this approximate quenching uncertainty.

\section{Conclusion and outlook}
\label{sec:conclusion}
Factorization theorems in QCD imply that the LCDA of the pion is relevant to a variety of experimental processes. Since the LCDA is a non-perturbative object, it is a quantity of importance for LQCD calculations. While direct computation of $\phi_{\pi} (\xi, \mu)$ is impossible in a Euclidean field theory, a range of different theoretical approaches which allow one to indirectly study the LCDA have been proposed and pursued. These methods include direct calculation of the local matrix elements~\cite{Kronfeld:1984zv, Martinelli:1987si,Del_Debbio_2003,Arthur_2011,Bali:2019dqc}, factorization apporaches like the pseudo-DA approach~\cite{,Ji:2013dva}, and a light-quark operator product expansion~\cite{Braun:2007wv}, have been used or proposed to this effect.

Knowledge of the first non-trivial Mellin moment of the pion $\expval{\xi^2}$  provides an important constraint on the shape of the LCDA. Due to the one-loop running behaviour of the Gegenbauer moments, one expects that the second moment is especially important for the shape of the LCDA at large enough renormalization scale. This quantity has previously been studied with the conventional approach of calculating the relevant matrix element of the local twist-two operator. As a result, this quantity is relatively well known and therefore provides a good test of the validity and applicability of the new methods.

This paper presents the first numerical study of the HOPE method to extract the second Mellin moment of the pion LCDA. This approach utilizes a quenched fictitious heavy-quark species which enables more control over higher-twist effects. After a discussion of the numerical calculation of the hadronic matrix element, two alternative approaches were explored for extracting the second Mellin moment of the pion LCDA from the numerical data.  These two approaches were termed the time-momentum analysis and the momentum-space analysis. Central to both analyses is the fact that the matrix element has a well-defined continuum limit after multiplicative renormalization. 

In the time-momentum analysis, the $\mathcal{O}(\alpha_s)$ formula of the HOPE is fit to the lattice data, and the resulting fit parameters are then extrapolated to the continuum. In the momentum-space analysis, the order of operations is reversed, and instead after Fourier transforming the lattice data, the correlators are extrapolated to the continuum before being compared with the $\mathcal{O}(\alpha_s)$ continuum HOPE. Both analyses produce results in good agreement with each other, and with other calculations in the literature. Due to the order of operations, more lattice data are included in the time-momentum analysis. This leads to an improved estimate of the statistical error in the second Mellin moment. As a result, the final value of $\langle \xi^2 \rangle$ determined in this work is
\begin{equation*}
  \langle \xi^2 \rangle(\mu=2\gev) = 0.210 \pm 0.036\,.
\end{equation*}
The uncertainty in this result is dominated by systematic effects, especially from higher-twist terms and the continuum extrapolation.  
From Table~\ref{tab:error-budget}, it is clear that reducing the systematic effect arising
from higher-twist contributions is the most important task for
improving these calculations.  For this purpose, it could be helpful to
adopt other heavy-quark formulations, such as that in
Ref.~\cite{El-Khadra:1996wdx}, for future lattice numerical calculations.

The results shown here demonstrate the viability of the HOPE approach to determine moments of light-cone quantities with comparable statistical precision to results from other methods. This paves the way for further investigations of the pion LCDA, including dynamical studies of the second moment using the HOPE method and a determination of higher Mellin moments. Early numerical studies have commenced, and preliminary results for the fourth moment are discussed in Ref.~\cite{Detmold:2021LattConf}. The success of this approach for the LCDA also suggests that the HOPE method can be applied to the study of other light cone quantities. Key objects of interest are the kaon LCDA which would allow the study of Mellin moments of a system with non-zero strangeness and pion PDF and helicity PDF, for which the Wilson coefficients have already been calculated to one loop order~\cite{Detmold:2021uru}.

\acknowledgements
The authors thank ASRock Rack Inc. for their support of the construction of an
Intel Kights Landing cluster at National Yang Ming Chiao Tung
University, where the numerical calculations were performed.  Help
from Balint Joo in tuning Chroma is acknowledged.
We thank M. Endres for providing the ensembles of gauge field configurations used in this work.
  CJDL and RJP are supported by the Taiwanese MoST Grant
No. 109-2112-M-009-006-MY3 and MoST Grant No. 109-2811-M-009-516. 
The work of IK is partially supported by the MEXT as ``Program for
Promoting Researches on the Supercomputer Fugaku'' (Simulation for
basic science: from fundamental laws of particles to creation of
nuclei) and JICFuS.
YZ is supported by the U.S. Department of Energy, Office of Science, Office of Nuclear Physics, contract no. DEAC02-06CH11357.
WD and AVG acknowledge support from the U.S.~Department of Energy (DOE) grant DE-SC0011090. WD is also supported within the framework of the TMD Topical Collaboration of the U.S.~DOE Office of Nuclear Physics, and by the SciDAC4 award DE-SC0018121.
WD is supported by the National Science Foundation
under Cooperative Agreement PHY-2019786 (The NSF AI Institute for Artificial Intelligence and Fundamental Interactions, http://iaifi.org/).
SM is partly supported by the LANL LDRD program.

\appendix

\label{app:order-a}
\section{Symmetry Constraints on Matrix Element}

\subsection{Lorentz Invariant Decomposition}\label{app:lor_invar_decomp}
The derivation of the $\mathcal{O}(a)$ improvement relies on the symmetry properties of the hadronic amplitude that is constructed. The momentum-space hadronic amplitude is defined as 
\begin{equation}
  V^{\mu\nu}(p,q)=\int d^4z \, e^{-iq\cdot z}\bra{\Omega}|\mathcal{T}\{J^\mu(z/2)J^\nu(-z/2)\}\ket{\pi(\mathbf{p})} \, .
  \label{hadronic-tensor}
\end{equation}
The time-momentum representation of this amplitude, $R^{\mu\nu}(\tau, \mathbf{p}, \mathbf{q})$, is
obtained by performing a Fourier transform in the Euclidean momentum $q_4$:
\begin{equation}
\begin{split}
  R^{\mu\nu}(\tau,\mathbf{p},\mathbf{q})&=\int \frac{dq_4}{(2\pi)} e^{-iq_4\tau}V^{\mu\nu}(p,q) \, .
\end{split}
\end{equation}
In momentum-space, the most general Lorentz covariant form of the amplitude, $V^{\mu\nu}(p,q)$, is
\begin{equation}
  V^{\mu\nu}(p,q)=a_1 p^\mu p^\nu +a_2 p^\mu q^\nu + a_3 q^\mu p^\nu +a_4 q^\mu q^\nu+ a_5 g^{\mu\nu} + a_6 \epsilon^{\mu\nu\alpha\beta}p_\alpha q_\beta \, ,
  \label{tensor-decomposition}
\end{equation}
where $a_{i}$ are scalar functions of the invariants $q^2$, $p^2$ and $p\cdot q$, that is,
\begin{equation}
  a_i=a_i(p^2,q^2,p\cdot q) \, .
\end{equation}
Applying a parity transformation $\mathcal{P}$ to the hadronic amplitude gives
\begin{equation}
\begin{split}
  V^{\mu\nu}(p,q)&=\int d^4z \, e^{-i{q}\cdot {z}}\bra{0}|\mathcal{P}^\dagger\mathcal{P}\mathcal{T}\{J^\mu(\mathbf{z}/2,z_4/2)\mathcal{P}^\dagger\mathcal{P}J^\nu(-\mathbf{z}/2,-z_4/2)\}{\mathcal{P}^\dagger\mathcal{P}}\ket{\pi(\mathbf{p})}
\\
&={-}(-1)^\mu(-1)^\nu\int d^4z e^{-i q \cdot z}\bra{0}|\mathcal{T}\{J^\mu(-\mathbf{z}/2,z_4/2)J^\nu(\mathbf{z}/2,-z_4/2)\}\ket{\pi(-\mathbf{p})}
\\
&={-}(-1)^\mu(-1)^\nu V^{\mu\nu}(\tilde{p},\tilde{q}) \, ,
\end{split}
\end{equation}
where for simplicity the notation $\tilde{k}=(-\mathbf{k},p^4)$ has been introduced and
\begin{equation}
(-1)^\mu\equiv
\begin{cases}
-1,~\text{for}~\mu=1,2,3
\\
\hphantom{-}1,~\text{for}~\mu=4
\end{cases} \, .
\end{equation}
Applying these transformations to the terms in the amplitude decomposition, Eq.~\ref{tensor-decomposition}, gives
\begin{equation}
\begin{split}
a_1 p^\mu p^\nu +a_2 p^\mu q^\nu &+ a_3 q^\mu p^\nu +a_4 q^\mu q^\nu+ a_5 g^{\mu\nu} + a_6 \epsilon^{\mu\nu\alpha\beta}p_\alpha q_\beta
\\
&=-(-1)^\mu(-1)^\nu a_1 \tilde{p}^\mu \tilde{p}^\nu -(-1)^\mu(-1)^\nu a_2 \tilde{p}^\mu \tilde{q}^\nu -(-1)^\mu(-1)^\nu  a_3 \tilde{q}^\mu \tilde{p}^\nu -(-1)^\mu(-1)^\nu a_4 \tilde{q}^\mu \tilde{q}^\nu
\\
&-(-1)^\mu(-1)^\nu a_5 g^{\mu\nu} -(-1)^\mu(-1)^\nu a_6 \epsilon^{\mu\nu\alpha\beta}\tilde{p}_\alpha \tilde{q}_\beta \, .
\end{split}
\end{equation}
Noting that $p^\mu =(-1)^\mu \tilde{p}^\mu$, one may conclude all terms but $a_6$ vanish, and thus the most general form of the amplitude is
\begin{equation}
  V^{\mu\nu}(p,q)=a_6(q^2,p\cdot q) \epsilon^{\mu\nu\alpha\beta}p_\alpha q_\beta \, .
\end{equation}

\subsection{Vanishing of $c_A$ Term}\label{app:ca}

The three-point correlation function is defined as
\begin{equation}
  G_3^{\mu\nu}(x,y)=\bra{\Omega}|\mathcal{T}\{J_{A}^\mu(x)J_{A}^\nu(y)\mathcal{O}_\pi^{\dagger}(0)\}\ket{\Omega} \, .
  \label{3-pt-correlation-function}
\end{equation}
where $J_{A}^\mu(x)$ is given in eq.~\eqref{eq:order-a_current}, and $\mathcal{O}_\pi^{\dagger}(0)=\overline{\psi}(0)\gamma_5\psi(0)=Z_\pi\overline{\psi}^{(0)}(0)\gamma_5\psi^{(0)}(0)$. Neglecting terms proportional to $c_A'$ in Eq.~(\ref{eq:order-a_current}), which will be considered in the next subsection,
\footnote{At $O(a)$, there are no cross-terms between the $c_A$ and $c_A'$ corrections, since these would be proportional to $a^2 c_A c_A'$, so these two effects can be studied separately.} 
one can expand the $O(a)$ improvements to the currents in Eq.~(\ref{3-pt-correlation-function}) as \cite{Bhattacharya_2006}
\begin{equation}
\begin{split}
G_3^{\mu\nu}(x,y)&=Z^2(a)Z_\pi\bra{\Omega}|T\bigg\{\bigg(\overline{\Psi}^{(0)}(x)\gamma^\mu\gamma^5 \psi^{(0)}(x)+\overline{\psi}^{(0)}(x)\gamma^\mu\gamma^5 \Psi^{(0)}(x)
\\
&+ac_A \partial^\mu \bigg\{\overline{\Psi}^{(0)}(x)\gamma^5 \psi^{(0)}(x)\bigg\}+ac_A \partial^\mu \bigg\{\overline{\psi}^{(0)}(x)\gamma^5 \Psi^{(0)}(x)\bigg\}\bigg)
\\
&\times\bigg(\overline{\Psi}^{(0)}(y)\gamma^\nu\gamma^5 \psi^{(0)}(y)+\overline{\psi}^{(0)}(y)\gamma^\nu\gamma^5 \Psi^{(0)}(y)
\\
&+ac_A \partial^\nu \bigg\{\overline{\Psi}^{(0)}(y)\gamma^5 \psi^{(0)}(y)\bigg\}+ac_A \partial^\nu \bigg\{\overline{\psi}^{(0)}(y)\gamma^5 \Psi^{(0)}(y)\bigg\}\bigg)
\\
&\times [\overline{\psi}^{(0)}(0)\gamma^5 \psi^{(0)}(0)]^{\dagger}\bigg\}\ket{\Omega}+\mathcal{O}(a^2) \, ,
\end{split}
\end{equation}
where $Z(a)=Z_A(1+\tilde{b}_Aa \tilde{m}_{ij})$, and $\tilde{m}_{ij}=(\tilde{m}_i+\tilde{m}_j)/2$. At order $a$, there are two terms proportional to $c_A$, one with the $c_A$ contribution originating from the $J^\mu$ current and the other with $c_A$ from the $J^\nu$ current.  To illustrate that terms proportional to $c_{A}$ vanish, it suffices to investigate one of these two terms.  For this purpose, contribution with $c_{A}$ coming from the improvement of $J^{\nu}$ is chosen:
\begin{equation}
G_{3,2}^{\mu\nu}(x,y)\equiv 
Z^2(a)Z_\pi ac_A \bra{\Omega}|T\bigg\{\bigg(\overline{\Psi}^{(0)}(x)\gamma^\mu\gamma^5 \psi^{(0)}(x)\bigg)\bigg(\partial^\nu \bigg\{\overline{\psi}^{(0)}(y)\gamma^5 \Psi^{(0)}(y)\bigg\}\bigg)
[\overline{\psi}^{(0)}(0)\gamma^5 \psi^{(0)}(0)]^{\dagger}\bigg\}\ket{\Omega}+\mathcal{O}(a^2) \, .
\end{equation}
The Fourier transform of $G_{3,2}^{\mu\nu}(x,y)$ is denoted $\tilde{G}_{3,2}^{\mu\nu}(p_1,p_2)$. Integrating by parts yields
\begin{equation}
\begin{split}
  \tilde{G}_{3,2}^{\mu\nu}(p_1,p_2)=Z^2(a)Z_\pi ac_A ip_2^\nu\int d^4x \, e^{-i p_1\cdot x } \, d^4y e^{-i p_2\cdot y} \bra{\Omega}|T\bigg\{\bigg(\overline{\Psi}^{(0)}(x)\gamma^\mu\gamma^5 \psi^{(0)}(x)\bigg)\bigg(\overline{\psi}^{(0)}(y)\gamma^5 \Psi^{(0)}(y)\bigg)
\\
\times
[\overline{\psi}^{(0)}(0)\gamma^5 \psi^{(0)}(0)]^{\dagger}\bigg\}\ket{\Omega} \, .
\end{split}
\label{matrix-element}
\end{equation}

Consider the most general form of the matrix element in Eq.~(\ref{matrix-element}). Since there is already one factor of $p_2^\nu$, the integral must transform as a vector under Lorentz boosts. Thus, the most general form of $C_{3,2}^{\mu\nu}(p_1,p_2)$ must be
\begin{equation}
  \tilde{G}_{3,2}^{\mu\nu}(p_1,p_2)=b_1(p_1^2,p_2^2,p_1\cdot p_2) p_1^\mu p_2^\nu +b_2(p_1^2,p_2^2,p_1\cdot p_2) p_2^\mu p_2^\nu \, .
\end{equation}
Using the parity constraints derived previously allows one to conclude that $C_{3,2}^{\mu\nu}(p_1,p_2)$ contains no terms which satisfy these symmetry properties, and therefore it must vanish. The other term proportional to $c_A$ (arising from the correction to $J^\mu$) vanishes analogously.  Thus, $c_A$ terms do not contribute to the $O(a)$ correction.

\subsection{Vanishing of $c_A'$ terms}\label{app:ca-prime}

Neglecting the $c_A$ and $Z_A$ corrections discussed above, the discretized version of the three-point correlator given in Eq.~\eqref{3pt-corr} is given by
\begin{equation}
  \begin{split}
    C_3^{\mu\nu} = \sum_{\mathbf{x}_e, \mathbf{x}_m} \langle &\left[ \overline \psi_u (x_m) \gamma^\nu \gamma^5 \Psi(x_m) - \frac{a}{4} c'_A \left( \overline \psi_u (x_m) \gamma^\nu \gamma^5 (\overrightarrow{D}_\Psi \Psi)(x_m) - (\overline \psi_u \overleftarrow{D}_\psi)(x_m) \gamma^\nu \gamma^5 \Psi(x_m) \right) \right] \\
    &\left[ \overline \Psi(x_e) \gamma^\mu \gamma^5 \psi_d(x_e) - \frac{a}{4}c'_A \left( \overline \Psi(x_e) \gamma^\mu \gamma^5 (\overrightarrow{D}_\psi \psi_d)(x_e) - (\overline \Psi \overleftarrow{D}_\Psi)(x_e) \gamma^\mu \gamma^5 \psi_d(x_e) \right) \right] \\
    &\left[ \overline \psi_d(0) \gamma^5 \psi_u(0) \right] \rangle e^{i\mathbf{p}_e \cdot \mathbf{x}_e} e^{i\mathbf{p}_m \cdot \mathbf{x}_m}
  \end{split}
  \label{contact-terms}
\end{equation}
where the $u$ and $d$ quarks are degenerate and are solely distinguished in order to avoid disconnected diagrams.

Expanding this out gives 
\begin{equation}
  \begin{split}
    C_3^{\mu\nu} &= \sum_{\mathbf{x}_e, \mathbf{x}_m} \left\langle \left[ \overline \psi_u (x_m) \gamma^\nu \gamma^5 \Psi(x_m) \right]\left[ \Psi(x_e) \gamma^\mu \gamma^5 \psi_d (x_e) \right]\left[ \overline \psi_d (0) \gamma^5 \psi_u(0) \right] \right\rangle e^{i\mathbf{p}_e \cdot \mathbf{x}_e} e^{i\mathbf{p}_m \cdot \mathbf{x}_m} \\
    &- a \sum_{\mathbf{x}_e, \mathbf{x}_m} \frac{c_A'}{4} \left\langle \left[ \overline \psi_u (x_m) \gamma^\nu \gamma^5 (\overrightarrow D_\Psi \Psi)(x_m) - (\overline \psi_u \overleftarrow D_\psi)(x_m) \gamma^\nu \gamma^5 \Psi (x_m) \right] \right. \\
    &\hspace{63pt} \left.\left[ \overline \Psi(x_e) \gamma^\mu \gamma^5 \psi_d(x_e) \right]\left[ \overline \psi_d (0) \gamma^5 \psi_u (0) \right] \right\rangle e^{i\mathbf{p}_e \cdot \mathbf{x}_e} e^{i\mathbf{p}_m \cdot \mathbf{x}_m} \\
    &- a \sum_{\mathbf{x}_e, \mathbf{x}_m} \frac{c_A'}{4} \left\langle \left[ \overline \psi_u (x_m) \gamma^\nu \gamma^5 \Psi(x_m) \right] \left[ \overline \Psi(x_e) \gamma^\mu \gamma^5 (\overrightarrow D_\psi \psi_d)(x_e) - (\overline \Psi \overleftarrow D_\Psi)(x_e) \gamma^\mu \gamma^5 \psi_d (x_e) \right] \right. \\
    &\hspace{63pt} \left.\left[ \overline \psi_d (0) \gamma^5 \psi_u (0) \right] \right\rangle e^{i\mathbf{p}_e \cdot \mathbf{x}_e} e^{i\mathbf{p}_m \cdot \mathbf{x}_m} \\
    &+ O(a^2) \, .
  \end{split}
  \label{expanded-contact-terms}
\end{equation}

The first line of Eq.~(\ref{expanded-contact-terms}) is the tree-level contribution studied above.  The second and third lines contain four very similar terms, the first of which is
\begin{equation}
  -a \sum_{\mathbf{x}_e, \mathbf{x}_m} \frac{c_A'}{4} \left\langle \left[ \overline \psi_u (x_m) \gamma^\nu \gamma^5 (\overrightarrow D_\Psi \Psi)(x_m) \right] \left[ \overline \Psi(x_e) \gamma^\mu \gamma^5 \psi_d(x_e) \right]\left[ \overline \psi_d (0) \gamma^5 \psi_u (0) \right] \right\rangle e^{i\mathbf{p}_e \cdot \mathbf{x}_e} e^{i\mathbf{p}_m \cdot \mathbf{x}_m} \, .
\end{equation}
Explicitly writing out all spin, colour, and spacetime indices, this becomes
\begin{equation}
  \begin{split}
- \frac{a c_A'}{4} \sum_{\mathbf{x}_e, \mathbf{x}_m, x} &\left\langle \left[ \overline \psi_u (x_m)_a^\alpha (\gamma^\nu \gamma^5)^{\alpha\beta} D_\Psi (x_m | x)_{ab}^{\beta\gamma} \Psi(x)_b^\gamma \right] \right. \\
&\hspace{6pt}\left. \left[ \overline \Psi(x_e)_c^\delta (\gamma^\mu \gamma^5)^{\delta \varepsilon} \psi_d (x_e)_c^\varepsilon \right] \left[ \overline \psi_d(0)_d^\zeta (\gamma^5)^{\zeta\eta} \psi_u(0)_d^\eta \right]  \right\rangle e^{i\mathbf{p}_e \cdot \mathbf{x}_e} e^{i\mathbf{p}_m \cdot \mathbf{x}_m} \, .
  \end{split}
  \label{explicit-indices}
\end{equation}
Performing the contractions with Wick's theorem gives
\begin{equation}
  \begin{split}
    \frac{ac_A'}{4} \sum_{\mathbf{x}_e, \mathbf{x}_m, x} &\left[ D_u^{-1}(0|x_m)_{da}^{\eta\alpha} \right] \left[ D_d^{-1}(x_e|0)_{cd}^{\varepsilon\zeta} \right] \left[ D_\Psi^{-1}(x|x_e)_{bc}^{\gamma\delta} \right] \left[ D_\Psi (x_m|x)_{ab}^{\beta\gamma} \right] \\
    &\left( \gamma^\nu \gamma^5 \right)^{\alpha\beta} \left( \gamma^\mu \gamma^5 \right)^{\delta\varepsilon} \left( \gamma^5 \right)^{\zeta\eta} e^{i\mathbf{p}_e \cdot \mathbf{x}_e} e^{i\mathbf{p}_m \cdot \mathbf{x}_m} \, .
  \end{split}
  \label{wick-contractions}
\end{equation}
Note that
\begin{equation}
  \sum_x D_\Psi(x_m|x)_{ab}^{\beta\gamma} D_\Psi^{-1}(x|x_e)_{bc}^{\gamma\delta} = \delta^{(4)}(x_m - x_e) \delta^{\beta\delta} \delta_{ac}
  \label{dirac-inverse}
\end{equation}
and thus these terms cannot contribute to this analysis, where contributions from $\tau \equiv \tau_m - \tau_e = 0$ have been explicitly removed.  However, for completeness, it should be noted that one can show that these terms are in fact zero.  With this result, equation (\ref{wick-contractions}) becomes
\begin{align}
  \frac{ac_A'}{4} &\sum_{\mathbf{x}_e, \mathbf{x}_m} \sum_{\mathbf{x}_e, \mathbf{x}_m} e^{i(\mathbf{p}_e + \mathbf{p}_m)\cdot \mathbf{x}_e}\left[ D_u^{-1}(0|x_m)_{da}^{\eta\alpha} \right] \left[ D_d^{-1}(x_e|0)_{cd}^{\varepsilon\zeta} \right] \left[ \delta^{(4)}(x_m-x_e) \delta^{\beta\gamma} \delta_{ac} \right] \left( \gamma^\nu \gamma^5 \right)^{\alpha\beta} \left( \gamma^\mu \gamma^5 \right)^{\delta\varepsilon} \left( \gamma^5 \right)^{\zeta\eta} \\
  &= \frac{ac_A'}{4} \delta(\tau_e - \tau_m) \sum_{\mathbf{x}_e} e^{i(\mathbf{p}_e + \mathbf{p}_m)\cdot \mathbf{x}_e} D^{-1}(0|x_e)_{da}^{\eta\alpha} D^{-1}(x_e|0)_{ad}^{\varepsilon\zeta} \left( \gamma^\nu \gamma^5 \right)^{\alpha\beta} \left( \gamma^\mu \gamma^5 \right)^{\beta\varepsilon} \left( \gamma^5 \right)^{\zeta\eta} \\
  &= \frac{ac_A'}{4} \delta(\tau_e - \tau_m) \sum_{\mathbf{x}_e} e^{i\mathbf{p}_\pi \cdot \mathbf{x}_e} \text{Tr}\left[ D^{-1}(0|x_e) \gamma^\nu \gamma^\mu D^{-1}(x_e|0) \gamma^5 \right] \, .
  \label{pion-b1-correlator}
\end{align}

Under charge conjugation, $\psi \rightarrow C^{-1} \overline \psi^T$ and $\overline \psi \rightarrow -\psi^T C$ (with $C = i \gamma^2 \gamma^4$), so
\begin{equation}
  D^{-1}(x|y) = \langle \psi(x) \overline \psi(y) \rangle \rightarrow C D^{-1}(y|x)^T C^{-1}
  \label{charge-conjugation}
\end{equation}
Further noting that $C\gamma^\mu C^{-1} = -(\gamma^\mu)^T$ and $C\gamma^5 C^{-1} = \left( \gamma^5 \right)^T$, charge conjugation sends the trace in Eq.~(\ref{pion-b1-correlator}) to
\begin{align}
  \text{Tr}\left[ C D^{-1}(x_e|0)^T C^{-1}\gamma^\nu \gamma^\mu C D^{-1}(0|x_e)^T C^{-1} \gamma^5 \right]
  &= \text{Tr}\left[ D^{-1}(x_e|0)^T (\gamma^\nu)^T (\gamma^\mu)^T D^{-1}(0|x_e)^T (\gamma^5)^T \right] \\
  &= \text{Tr}\left[ D^{-1}(0|x_e) \gamma^\mu \gamma^\nu D^{-1}(x_e|0) \gamma^5 \right] \\
  &= -\text{Tr}\left[ D^{-1}(0|x_e) \gamma^\nu \gamma^\mu D^{-1}(x_e|0) \gamma^5 \right]  \, .
  \label{flipped-sign}
\end{align}
Since QCD is invariant under charge conjugation, amplitudes must be similiarly invariant, so this trace must vanish.  The remaining $c_A'$ terms vanish similarly, so there is no need to include the corresponding operators in the definition of the axial currents $J_A^\mu$.

\addcontentsline{toc}{chapter}{Bibliography} 
\bibliographystyle{apsrev.bst} 
\bibliography{refs}

\end{document}